\documentclass[11pt]{article}
\pdfoutput=1
\usepackage{amsmath} % AMS Math Package
\usepackage{amsthm} % Theorem Formatting
\usepackage{amssymb}	% Math symbols such as \mathbb
\usepackage{bm}	% Boldfaced math style with \bm{XXX}
\usepackage{graphicx} % Allows for eps images
\usepackage[squaren,Gray]{SIunits}
\usepackage{multirow}
\usepackage{booktabs}
\usepackage{makecell}
\usepackage[textsize=tiny,backgroundcolor=white]{todonotes}
\usepackage{subfig}
\usepackage[export]{adjustbox}

\makeatletter
\g@addto@macro\bfseries{\boldmath}
\makeatother

\graphicspath{ {img/} }


\newcommand{\perc}{\%}

\let\baraccent=\= % rename builtin command \= to \baraccent
\renewcommand{\=}[1]{\stackrel{#1}{=}} % for putting numbers above =

\theoremstyle{definition}

\theoremstyle{remark}

\newcommand{\ddk}[1]{\frac{d^d k_{#1}}{(4\pi)^d}}

\newcommand\calo[1]{{\cal O}\hspace{-0.2em}\left(#1\right)}

\newcommand{\cala}{{\cal A}}
\newcommand{\bbH}{\ensuremath{b\bar{b}H}}
\newcommand{\ttH}{\ensuremath{t\bar{t}H}}
\newcommand{\bbphi}{\ensuremath{b\bar{b}\phi}}
\newcommand{\yt}{\ensuremath{y_t}}
\newcommand{\ytsq}{\ensuremath{y_t^2}}
\newcommand{\yb}{\ensuremath{y_b}}
\newcommand{\ybsq}{\ensuremath{y_b^2}}
\newcommand{\ybyt}{\ensuremath{y_b\, y_t}}

\newcommand{\pt}{\ensuremath{p^T}}
\newcommand{\pth}{\ensuremath{p^T_H}}
\newcommand{\ptb}{\ensuremath{p^T_b}}
\newcommand{\ptj}{\ensuremath{p^T_j}}
\newcommand{\ptbone}{\ensuremath{p^T_{b_1}}}

\newcommand{\Mbb}{\ensuremath{M(bb)}}
\newcommand{\Rbb}{\ensuremath{\Delta R(bb)}}
\newcommand{\MBB}{\ensuremath{M(BB)}}
\newcommand{\RBB}{\ensuremath{\Delta R(BB)}}

\newcommand{\citere}[1]{ref.\,\cite{#1}}
\newcommand{\citeres}[1]{refs.\,\cite{#1}}

\newcommand{\eqn}[1]{eq.~(\ref{#1})}

\newcommand{\fig}[1]{figure~\ref{#1}}
\newcommand{\Fig}[1]{Figure~\ref{#1}}
\newcommand{\figs}[1]{figures~\ref{#1}}
\newcommand{\tab}[1]{table~\ref{#1}}
\newcommand{\sct}[1]{section~\ref{#1}}

\usepackage[public]{jheppub}

\preprint{\vspace{-0.83cm}\begin{flushright} CERN-TH-2018-175\\ CP3-18-52\\ NIKHEF/2018-037 \end{flushright}\vspace{0.83cm}}

\title{\nobreak{Top-Yukawa}~contributions~to~bbH~production~at~the~LHC}

\author[a]{Nicolas Deutschmann,}
\author[b]{Fabio Maltoni,}
\author[c]{Marius Wiesemann,}
\author[d]{and Marco Zaro}

\affiliation[a]{Institute for Theoretical Physics, ETH Zurich, 8093 Zurich, Switzerland}
\affiliation[b]{Centre for Cosmology, Particle Physics and Phenomenology (CP3),\\
Université catholique de Louvain, B-1348 Louvain-la-Neuve, Belgium}
\affiliation[c]{Theoretical Physics Department, CERN, Geneva, Switzerland}
\affiliation[d]{Nikhef, Science Park 105, NL-1098 XG Amsterdam, The Netherlands}

\emailAdd{ndeutschmann@itp.phys.ethz.ch}
\emailAdd{fabio.maltoni@uclouvain.be}
\emailAdd{marius.wiesemann@cern.ch}
\emailAdd{m.zaro@nikhef.nl}

\abstract{We study the production of a Higgs boson in association with bottom quarks (\bbH) in hadronic collisions at the LHC, including the different contributions stemming from terms proportional to the top-quark Yukawa coupling (\ytsq), to the bottom-quark one (\ybsq), and to their interference (\ybyt).
Our results are accurate to next-to-leading order in QCD, employ the four-flavour scheme and
the (Born-improved) heavy-top quark approximation. We find that next-to-leading order corrections to the \ytsq{} component are sizable, making it the dominant production mechanism for associated \bbH{} production in the Standard Model and increasing its inclusive rate by almost a factor of two. By studying final-state distributions of the various contributions, we identify observables and selection cuts that can be used to select the various components and to improve the experimental sensitivity of \bbH{}  production on the bottom-quark Yukawa coupling. }


\begin{document}
\maketitle \flushbottom

\section{Introduction}
\label{sec:intro}

After the discovery of a scalar resonance with a mass of about $125\,\giga\electronvolt$ \cite{Aad:2012tfa,Chatrchyan:2012ufa},
the accurate determination of its couplings to Standard-Model (SM) particles has become
one of the major objectives of LHC Run II and beyond. Data collected at the LHC so
far supports the hypothesis that this resonance is the scalar boson predicted by the
Brout--Englert--Higgs mechanism of electroweak symmetry breaking \cite{Englert:1964et,Higgs:1964pj} as implemented in the SM \cite{Weinberg:1967tq}:
the Higgs couplings are universally set by the masses of the corresponding particles the Higgs boson interacts with. Global fits of various production
and decay modes of the Higgs boson \cite{Aad:2015gba,Aad:2015zhl,Khachatryan:2016vau,ATLAS:2018doi,CMS:2018hhg} constrain its couplings to third-generation fermions and to vector bosons to be within $10-20$\% of the values predicted by the SM. In particular, the recent measurement of Higgs production in association with a top-quark pair \cite{Sirunyan:2018hoz, Aaboud:2018urx} provides the first direct evidence of the coupling between the Higgs boson and the top quark, thereby proving that gluon--gluon Higgs production proceeds predominantly via top-quark loops. The coupling of the Higgs boson to $\tau$ leptons
has also been established at the 5$\sigma$ level for some time \cite{Aad:2015vsa,Sirunyan:2017khh},  while the Higgs coupling to bottom quarks has been observed only very recently~\cite{ATLAS:2018nkp}. By contrast, to date, we have no experimental confirmation that the Higgs boson couples to first-/second-generation fermions, nor about the strength of the Higgs self-interaction.

The ability to probe elementary couplings and to improve the experimental sensitivity strongly relies on precise theoretical predictions for both production and decay.
The bottom-quark Yukawa coupling ($\yb{}$) plays a rather special role in this context:
despite having a relatively low coupling strength with respect to the couplings to
vector bosons and top quarks, the $H\to b\bar b$ decay dominates the total decay width
in the SM for a Higgs-boson mass of about $\unit{125}{\giga\electronvolt}$ due to
kinematical and phase space effects.
The observation of this decay is, however, quite challenging because of
large backgrounds generated by QCD, especially in the gluon-fusion production
mode~\cite{atlas2016}, and has for now only been
searched for in  vector-boson fusion~\cite{TheCMSCollaboration2015,Aaboud2016} and
Higgsstrahlung~\cite{Chatrchyan2014,atlas2016}. The latter
is the most sensitive channel, yielding a signal strength for the decay branching ratio of
$\mu_{bb}=1.0\pm 0.2$~\cite{ATLAS:2018nkp}.  However, since the total Higgs width is dominated by $H \to b\bar b$, the corresponding branching ratio has a rather weak dependence on \yb{}. As a result, the sensitivity of processes involving Higgs decays to bottom quarks on this parameter is in fact rather low.

Studying production modes featuring a \bbH{} coupling is a promising alternative:
on the one hand, Higgs production in the SM (inclusive over any particles produced
in association) proceeds predominantly via the gluon-fusion process, where the Higgs--gluon coupling
is mediated by heavy-quark loops.
In particular, bottom-quark loops have a contribution of about $-6$\% to the inclusive cross section, which
can become as large as $-10$\% for Higgs bosons produced
at small transverse momentum~\cite{Mantler2013,Grazzini2013,Banfi2014,Bagnaschi2012,Frederix:2016cnl,Bagnaschi:2015bop,Mantler:2015vba,Harlander:2014uea}.
On the other hand, the associated production of a Higgs boson with
bottom quarks (\bbH{} production) provides direct access to the bottom-quark Yukawa coupling
already at tree-level~\cite{Raitio:1978pt}. It yields a cross section comparable to the one of the associated production with top quarks
(roughly $\unit{0.5}{\pico\barn}$ at $\unit{13}{\tera\electronvolt}$), which is about
$1\%$ of the fully-inclusive Higgs-production rate in the SM.
Furthermore, the inclusive rate decreases dramatically once conditions on the
associated $b$ jets are imposed to make it distinguishable from
inclusive Higgs-boson production.

The SM picture outlined above might be significantly modified by
beyond-SM effects: while a direct observation in the SM is challenging at the LHC,
\bbH{} production plays a crucial role in models with modified Higgs sectors.
In particular in a generic two Higgs-doublet-model (2HDM), or in a supersymmetric one
such as the MSSM, the bottom-quark Yukawa coupling can be significantly increased, promoting \bbphi{}
to the dominant Higgs production mode \cite{Rainwater2002,Dittmaier2004} in many
benchmark scenarios,  $\phi$ being any of the scalars or pseudo-scalars in such theories.
Given that a scalar sector richer than that of the SM has not yet been ruled out experimentally,
this is a fact that one must bear in mind, and that constitutes a strong motivation for theoretical
studies of scalar-particle production in association with bottom quarks.

The production of \bbH{} final states receives additional contributions from
the loop-induced gluon-fusion process (proportional to \ytsq{}; \yt{} being the top-quark Yukawa coupling), which in the SM is of similar size as \ybsq{} contributions, but
have rarely been studied in the literature.
In this paper, we consider Higgs production in association with bottom quarks
for all contributions proportional to \ybsq{} and \ytsq{} at NLO QCD, as well as their
interference terms proportional to \ybyt{}.
The \bbH{} process is particularly interesting also from a theoretical viewpoint in many respects.
First, as for all mechanisms that feature bottom quarks at the level
of the hard process, there are two schemes applicable to performing the computation.
These so-called four-flavour scheme (4FS) and five-flavour scheme (5FS) reflect the
issue that arise from different kinematic regimes, where either the mass of the
bottom-quark can be considered a hard scale or bottom quarks are treated on the same
footing as the other light quarks. Hence, the bottom-quark is considered to be massive
in the 4FS, while its mass can be set to zero in the 5FS. The advantages of either scheme in
the context of \bbH{} production have been discussed in detail in \citere{Wiesemann2015}.
We employ the 4FS throughout this paper, owing to its superior description of
differential observables related to final-state bottom quarks and
the definition of bottom-flavoured jets, which is particularly striking in fixed-order computations.
Another theoretical motivation lies in the nature of the loop-induced gluon-fusion process
that leads to the contributions proportional to \ytsq{}. Being dominated by kinematical configurations where
the Higgs boson recoils against a gluon which splits into a bottom-quark pair, this collider
process features the cleanest and most direct access to $g\to b\bar{b}$ splittings.
Thus, as a bonus, our computation also allows us to study the effect of NLO corrections on such
splittings.

Given that the NLO QCD corrections to \bbH{} production for \ybsq{} contributions
(and the LO \ybyt{} terms) were studied in great detail in \citere{Wiesemann2015},
including the effect of parton showers, we focus here
on the computation of NLO QCD corrections to the terms proportional to \yt{} and analyse their  behaviour with respect to the \ybsq{} contribution.
We note that our computation of NLO corrections to the  \ybyt{} and \ytsq{} terms employs an
effective field theory, where the top quark is integrated out from the theory and the Higgs directly couples to gluons, to which we refer as Higgs Effective Field Theory (HEFT). Besides a detailed description of the application of this approach to our problem, we will show that this approximation is quite accurate in the bulk of the phase space region which is relevant for this study.

Before introducing our calculation in the next section, we briefly summarise the status of
the results for \bbH{} production available in the literature. As far as 4FS computations
are concerned, seminal NLO QCD fixed-order parton-level predictions were obtained in \citeres{Dittmaier:2003ej,Dawson:2003kb}, and later updated to the case of MSSM-type couplings \cite{Dawson:2005vi}, and to SUSY-QCD corrections in the MSSM~\cite{Liu:2012qu,Dittmaier:2014sva}. Part of the 
NLO electroweak corrections were also obtained recently in \citere{Zhang:2017mdz}.
The presentation of differential results in these papers is very limited as the focus is on the total cross section. Given that computations in the 5FS are technically much simpler, far more results in this scheme exist in the literature: the total cross section are known at NLO \cite{Dicus:1998hs,Balazs:1998sb} since a long time and even  NNLO QCD \cite{Harlander:2003ai} predictions were among the first computations at this level of accuracy ever achieved relevant for LHC phenomenology.
Parton-level distributions were obtained at NLO for $H$+$b$ and $H$+jet production \cite{Campbell:2002zm,Harlander:2010cz}, and at NNLO for jet rates \cite{Harlander:2011fx} and fully differential distributions \cite{Buehler:2012cu}.
The analytical transverse-momentum spectrum of the Higgs boson was studied
up to $\mathcal{O}(\alpha_s^2)$ in \citere{Ozeren:2010qp}, while
analytically resummed NLO+NLL and NNLO+NNLL results were presented in \citere{Belyaev:2005bs} and \citere{Harlander:2014hya}, respectively.\footnote{Even the ingredients
for the full N$^3$LO prediction are already available \cite{Ahmed:2014pka,Gehrmann:2014vha}; their combination is far from trivial though.}
NLO+PS predictions for both the 4FS and the 5FS were presented for the first time in \citere{Wiesemann2015},
including a comprehensive comparison of the two schemes and the discussion
several differential distributions with NLO QCD accuracy.
Other NLO+PS results were later obtained in {\sc Powheg} \cite{Jager:2015hka} and {\sc Sherpa} \cite{Krauss:2016orf}. At the level of the total cross section advancements have been made by first understanding the differences between results obtained in the two schemes \citeres{Maltoni:2012pa,Lim:2016wjo} and then by consistently
combining state-of-the-art 4FS and 5FS predictions in \citeres{Forte:2015hba,Forte:2016sja,Bonvini:2015pxa,Bonvini:2016fgf}.

The paper is organised as follows. In \sct{sec:calculation} our computation is described
in detail. We first discuss the various contributions
to the \bbH{} cross section (\sct{sec:structure}), then introduce the HEFT
approximation to determine the \ytsq{} terms (\sct{sec:heft}) and
finally perform a comprehensive validation of the HEFT approximation for
the \ytsq{} cross section (\sct{sec:approx});
phenomenological results are presented in \sct{sec:results} --- see in particular \sct{sec:inputs}
for the input parameters, \sct{sec:predictions} for SM results,
\sct{sec:yb} for how to obtain the best sensitivity to extract \yb{} in the measurements, and
\sct{sec:gbb} for our analysis on NLO corrections to $g\to b\bar{b}$ splitting.
We conclude in \sct{sec:summary} and collect relevant technical information in the appendices.

\section{Outline of the calculation}\label{sec:calculation}

\subsection{Coupling structure of the \bbH{} cross section}
\label{sec:structure}

\begin{figure}[!h]
  \centering
  \subfloat[$\cala_b^{(0)}$]
  {\includegraphics[height=.17\textwidth,valign=b]{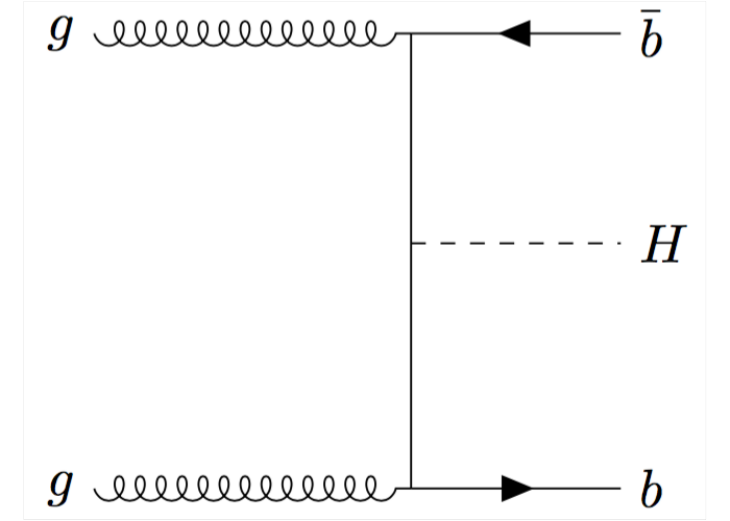}\label{fig:hbblo}}
  \subfloat[$\cala_b^{(1V)}$]
  {\includegraphics[height=.17\textwidth,valign=b]{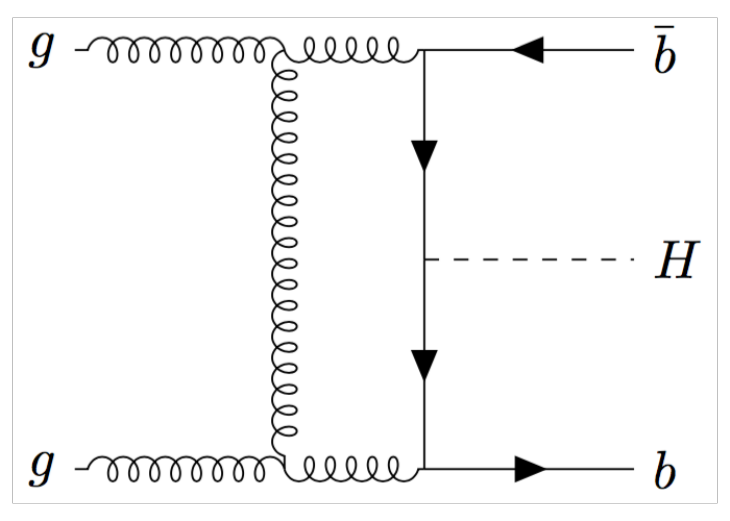}\label{fig:hbbnlov}}
  \subfloat[$\cala_b^{(1R)}$]
  { \includegraphics[height=.17\textwidth,valign=b]{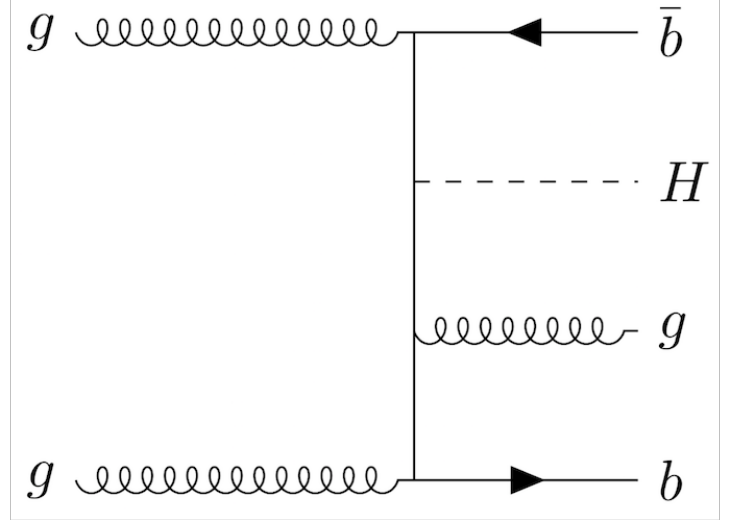}\label{fig:hbbnlor}}
  \subfloat[$\cala_t^{(0)}$]
  {\includegraphics[height=.17\textwidth,valign=b]{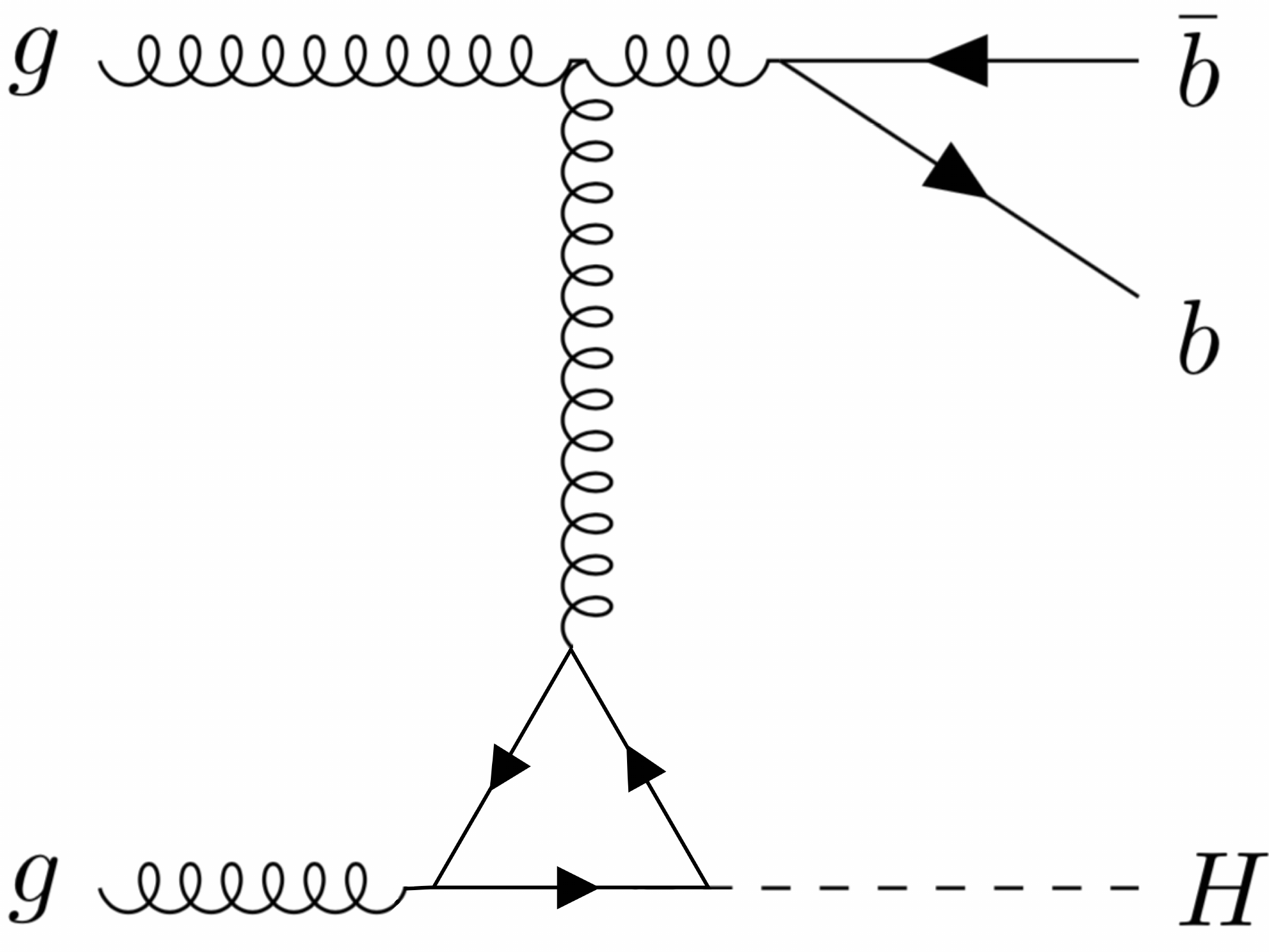}\label{fig:hbbnlot}}
  \caption{\label{fig:bbh}Examples of Feynman diagrams for $b \bar b H$ production at LO and at NLO, which contain virtual and real diagrams proportional to \yb{},
  and virtual diagrams with a top loop proportional to \yt{}. The corresponding amplitudes are named $\cala_b^{(0)}$, $\cala_b^{(1V)}$, $\cala_b^{(1R)}$ and $\cala_t^{(0)}$.}
  \label{fig:hbblonlo}
\end{figure}

\begin{figure}[!t]
  \centering
  \subfloat[$\cala_t^{(1V)}$]{\includegraphics[width=.27\textwidth]{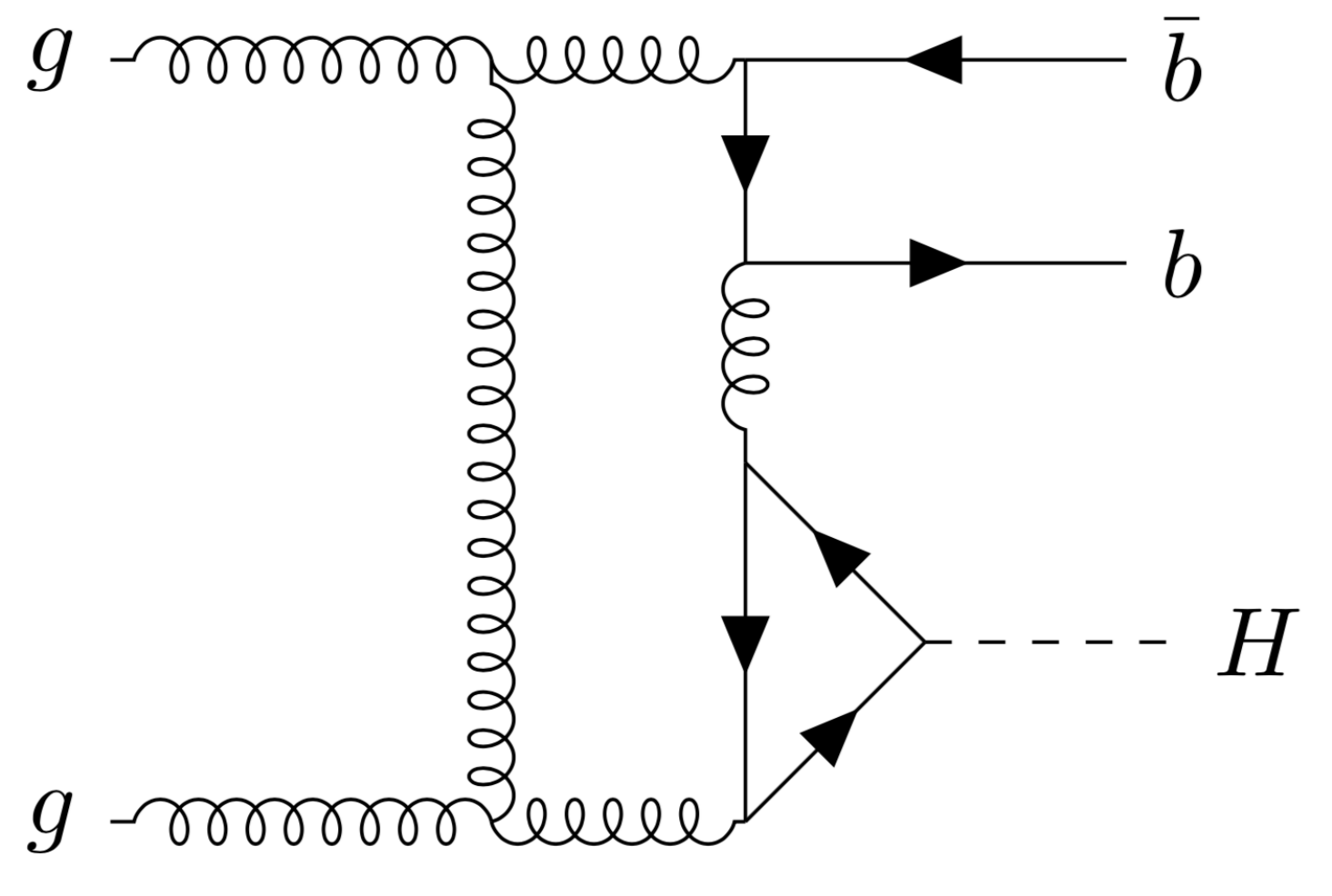}\label{fig:hbbnlovtop}}\qquad
  \subfloat[$\cala_t^{(1R)}$]{\includegraphics[width=.25\textwidth]{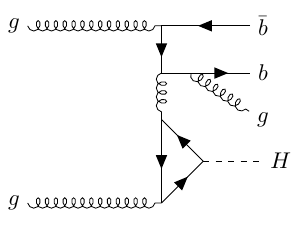}\label{fig:hbbnlortop}}
  \caption{\label{fig:bbHtop} Examples of virtual, $\cala_t^{(1V)}$, and real emission, $\cala_t^{(1R)}$, diagrams contributing to associated $b \bar b H$ production at
  $\calo{\ytsq{}\,\alpha_s^5}$, and
  $\calo{\ybyt{}\,\alpha_s^4}$ through their interference with $\cala_b^{(0)}$ and $\cala_b^{(1R)}$.}
  \label{fig:hbbnlotop}
\end{figure}

The leading contribution to the associated production of a Higgs boson with bottom quarks
in the 4FS starts at $\calo{\alpha_s^2}$ in QCD perturbation theory, and is mediated by the
bottom-quark Yukawa coupling. Hence, the coupling structure of the LO process is
$\ybsq\,\alpha_s^2$. A sample Feynman diagram is shown in \fig{fig:hbblo}.
At the next order in $\alpha_s$ the typical one-loop (\fig{fig:hbbnlov}) and real-emission (\fig{fig:hbbnlor}) diagrams are included, and yield a contribution of $\calo{\ybsq\,\alpha_s^3}$.
At the same order in $\alpha_s$ additional one-loop diagrams appear featuring a closed top-quark loop which the 
Higgs boson couples to (\fig{fig:hbbnlot}).
These diagrams introduce for the first time a dependence on top-quark Yukawa coupling in the \bbH{}
cross section and lead to contributions of $\calo{\ybyt\,\alpha_s^3}$ through their interference with \yb{} diagrams as shown in \fig{fig:hbblo}. At the next order in $\alpha_s$, the square of these
\yt{} amplitudes yields a contribution that starts at $\calo{\ytsq\,\alpha_s^4}$. Thus, it
is suppressed by two powers of $\alpha_s$ with respect to the first non-zero contribution to \bbH{} production of
$\calo{\ybsq\,\alpha_s^2}$ and  could be formally considered a NNLO contribution.
However, it is easy to understand that a na\"ive power counting just based on the single parameter $\alpha_s$ is not suitable for describing \bbH{} production, since the strong hierarchy between the top-quark and the bottom-quark Yukawa couplings in the SM is such that  $\ybsq\,\alpha_s^2$ terms turn out to be of a similar size as the $\ytsq\,\alpha_s^4$ contributions. In this respect, one also expects
that $\alpha_s$ corrections to the \ytsq{} contributions might turn out to be important,
which are of $\calo{\ytsq\,\alpha_s^5}$ and
formally part of the N$^3$LO corrections with respect to the leading $\calo{\ybsq\,\alpha_s^2}$ terms. They enter via virtual and real diagrams of the type shown in \fig{fig:hbbnlovtop} and
 in \fig{fig:hbbnlortop}, respectively.

Collecting all relevant terms at different orders in $\alpha_s$, one can express the cross section as
\begin{equation}
\begin{split}
    {\rm d} \sigma
    &\propto \alpha_s^2\, y_b^2 \left|\cala_b^{(0)}\right|^2\\
    & + \alpha_s^3 \left[ y_b^2 \left( 2 \text{Re} \left( \cala_b^{(0)} \cala_b^{(1V)} \right) + \int {\rm d} \Phi \left|\cala_b^{(1R)}\right|^2 \right) + y_b y_t 2 \text{Re} \left( \cala_b^{(0)} \cala_t^{(0)} \right)  \right]\\
    &+ \alpha_s^4 \left[ y_t y_b\, 2 \text{Re}\left( \cala_b^{(0)} \cala_t^{(1V)}+  \cala_b^{(1V)} \cala_t^{(0)}+ \int {\rm d} \Phi\cala_b^{(1R)} \cala_t^{(1R)} \right)+y_t^2\left|\cala_t^{(0)}\right|^2\right]\\
    &+ \alpha_s^5\, y_t^2 \left[ 2 \text{Re} \left(\cala_t^{(1V)} \cala_t^{(0)}\right) + \int {\rm d}\Phi \left|\cala_t^{(1R)}\right|^2 \right]\\
    &+ \mathcal O \left(\alpha_s^4\, y_b^2\right)+ \mathcal O \left(\alpha_s^5\, y_b y_t\right)+ \mathcal O \left(\alpha_s^6\, y_t^2\right)\,,
 \end{split}
\label{eq:hbbxsec1}
\end{equation}
where the $\cala_q^{(i)}$
amplitudes are introduced with the respective sample diagrams in \figs{fig:bbh}--\ref{fig:bbHtop},
and ${\rm d}\Phi$ denotes the appropriate phase space of the extra real emission
with all relevant factors in each case.
An equivalent, yet more appropriate and transparent way of organising the computation above is to consider a double coupling expansion in terms of $y_b$ and $y_t$ and then to systematically include $\alpha_s$ corrections to each of these terms.\footnote{This is possible because QCD corrections do not
induce any other coupling combinations on top of $\ybsq$, $\ytsq$, $\ybyt$.} Up to NLO, the cross section can be written as
\begin{equation}
\begin{split}
    {\rm d} \sigma &    = y_b^2\,\alpha_s^2\left( \Delta_{\ybsq}^{(0)} + \alpha_s \Delta_{\ybsq}^{(1)} \right) +y_t y_b\, \alpha_s^3\left( \Delta_{\ybyt}^{(0)} + \alpha_s \Delta_{\ybyt}^{(1)} \right) + y_t^2 \,\alpha_s^4\left( \Delta_{\ytsq}^{(0)} + \alpha_s \Delta_{\ytsq}^{(1)} \right)\, .
\end{split}
\label{eq:hbbxsec2}
\end{equation}
It is trivial to see that eq.~(\ref{eq:hbbxsec1}) and eq.~(\ref{eq:hbbxsec2}) feature exactly the same terms. In this formulation QCD corrections to \ybsq{}, \ybyt{} and \ytsq{} terms, the $\Delta_x^{(i)}$ contributions, are manifestly gauge invariant and can be calculated independently of each other at
LO and NLO.    All the coefficients up to $\calo{\alpha_s^3}$
($\Delta_{\ybsq}^{(0)}$, $\Delta_{\ybsq}^{(1)}$, and $\Delta_{\ybyt}^{(0)}$) were
determined and studied already in~\citere{Wiesemann2015}.  Our focus here is therefore on the calculation of  the contributions involving \yt{} in the 4FS, {\it i.e.}, $\Delta_{\ybyt}^{(1)}$,  $\Delta_{\ytsq}^{(0)}$, and $\Delta_{\ytsq}^{(1)}$.\footnote{The contributions $\Delta_{\ytsq}^{(0)}$ ($\Delta_{\ytsq}^{(1)}$) are implicitly included in the computation of gluon--gluon fusion at NNLO (N$^3$LO) in the 5FS~\cite{Anastasiou:2002yz,Anastasiou:2015ema}. These calculations, however, cannot provide information on final states specifically containing $b$ quarks.}

\subsection{HEFT approximation in \bbH{} production}
\label{sec:heft}

NLO corrections to the contributions proportional to the top-quark Yukawa coupling
require the computation of two-loop $2\to 3$ amplitudes with internal massive fermion
lines, see \fig{fig:hbbnlovtop}. The evaluation of such diagrams is beyond current
technology. Hence, in this section, we introduce the heavy top-mass approximation
that can be employed for the computation of these amplitudes, and we rearrange the SM cross
section discussed in \sct{sec:structure} in the HEFT.

In this effective theory, the top quark is integrated out and yields effective point-like 
interactions between the Higgs boson and gluons, described by the effective Lagrangian
\begin{align}
\mathcal{L}=-\frac{1}{4}\, C_1\;H\,G_{\mu\nu}^a\,G^{a, \mu\nu},
\end{align}
where $H$ denotes the field associated to the physical Higgs boson, $v$ is the vacuum expectation value of the Higgs field, $G_{\mu\nu}^a$ is the gluon field strength tensor and $C_1$ is
the Wilson coefficient that can be expressed in terms of SM parameters. The HEFT Lagrangian yields a point-like interaction for both $Hgg$ and $Hggg$ vertices. At leading order in the strong coupling, we have
\begin{align}
C_1 = - \frac{1}{v}\left[y_t \frac{v}{\sqrt{2}m_t} \right]\left(\frac{\alpha_s}{3\pi}+{\cal O} \left(\alpha_s^2\right) \right),
\end{align}
where the dependence on the top Yukawa coupling, which is explicit in the SM, is cancelled by the power supression of the loop integral, making the term in brackets exactly equal to 1. By matching the amplitude for the process $H\to gg$ in the HEFT and the SM 
at higher orders in perturbation theory, the expansion of $C_1$ in $\alpha_s$ 
can be determined \cite{Gerlach:2018hen}, which will then depend on the renormalization schemes adopted in the HEFT and in the SM. We discuss the details of the renormalization procedure in Appendix~\ref{app:renormalization}.

The HEFT approximation has been used successfully to compute a number of observables in the Higgs sector, with the gluon-fusion cross section through $\text{N}^3\text{LO}$ as the most notable example~\cite{Mistlberger:2018etf,Anastasiou:2016cez,Anastasiou:2015ema}. By substituting top loops with a point-like coupling, the HEFT allows for significant simplifications of Higgs-related observables at the price of a limited range of applicability: the approximation is expected to break down when one of the scales appearing in the process, and in particular in the massive loop integrals,
becomes comparable with the top-quark mass.
The case at hand corresponds to $H$+jet ($H$+$g$) production with
$g\to b\bar{b}$ splitting either in the initial or in the final state. It has been shown that the HEFT
provides an excellent approximation in that case as long as the scales of the process remain
moderate~\cite{Harlander:2012hf,Neumann:2014nha}, for example as long as the Higgs
transverse momentum (\pth{}) is below $\sim 150\,\giga\electronvolt$.
In section~\ref{sec:approx}, we provide a detailed assessment of the goodness of the heavy top-mass approximation. As we will show, the heavy-top mass approximation works extremely well (with differences from the full computation below 10\%) as long as the probed momentum scales (Higgs
or leading b-jet transverse momentum, or invariant mass of the b-jet pair) do not exceed $200\,\giga\electronvolt$.

\begin{figure}[t!]
  \centering
  \subfloat[$\hat{\cala}_t^{(0)}$]{\includegraphics[width=.25\textwidth]{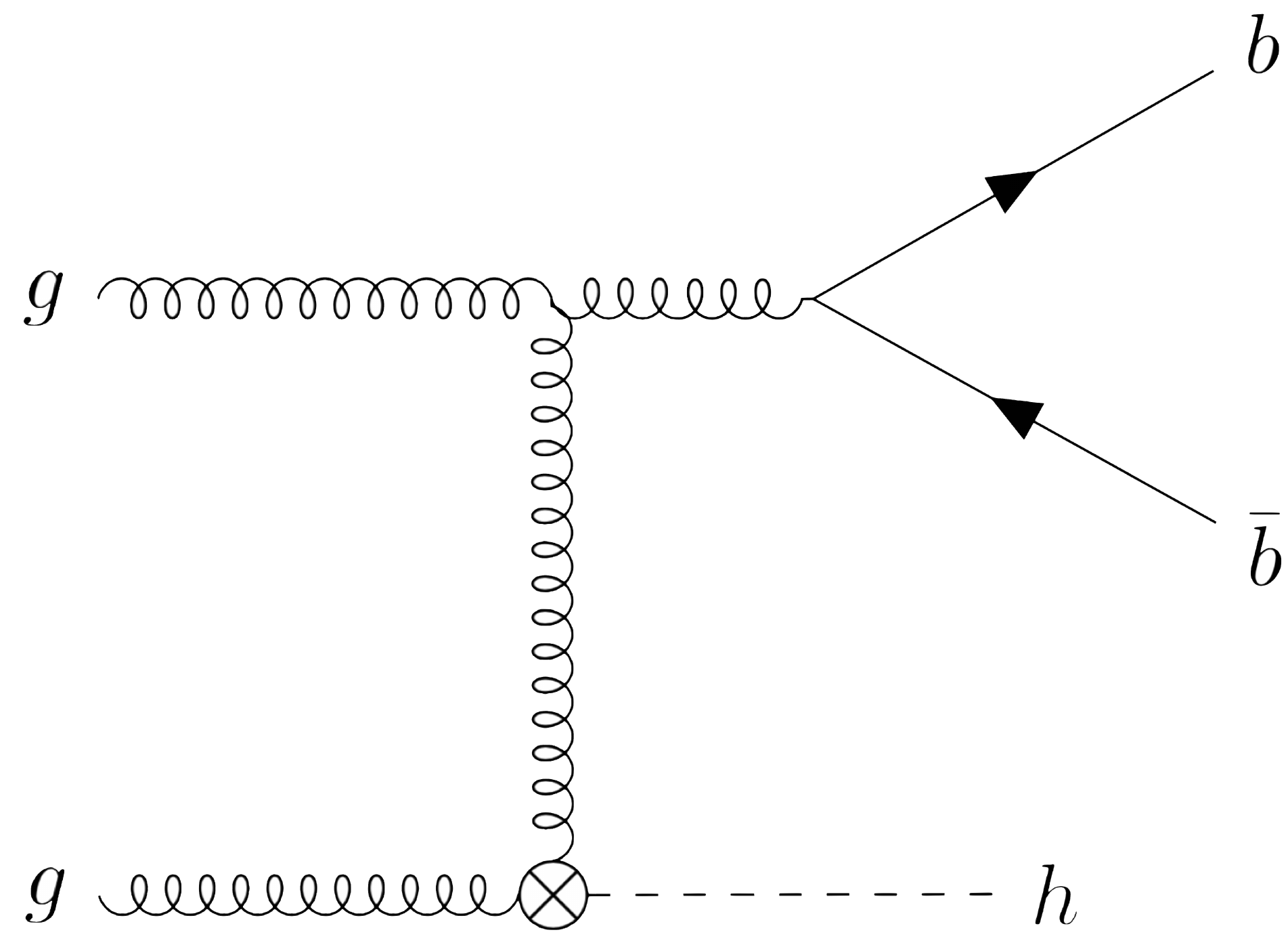}}\qquad
  \subfloat[$\hat{\cala}_t^{(1V)}$]{\includegraphics[width=.25\textwidth]{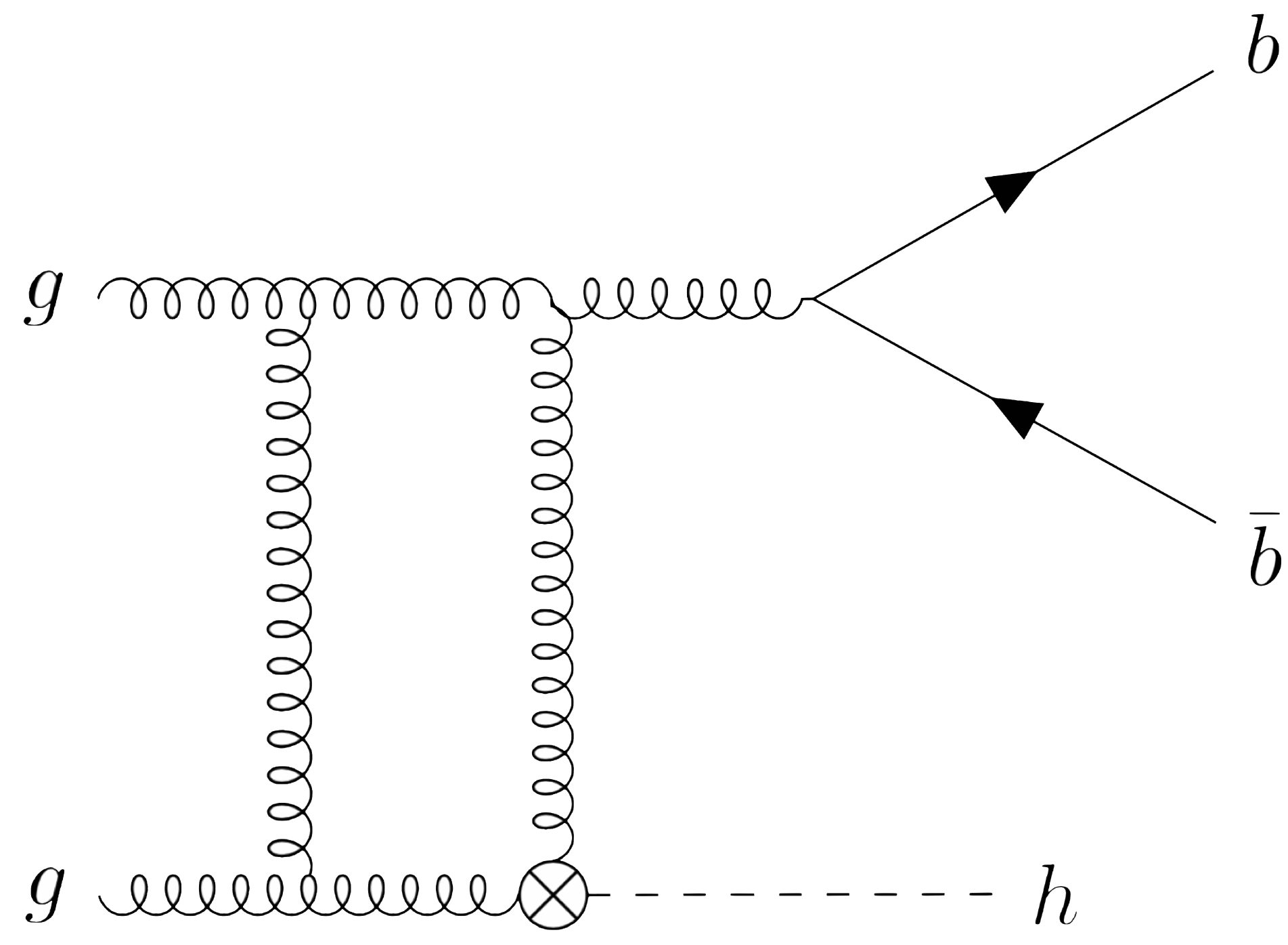}\label{fig:hbbeftnlo}}\qquad
  \subfloat[$\hat{\cala}_t^{(1R)}$]{\includegraphics[width=.25\textwidth]{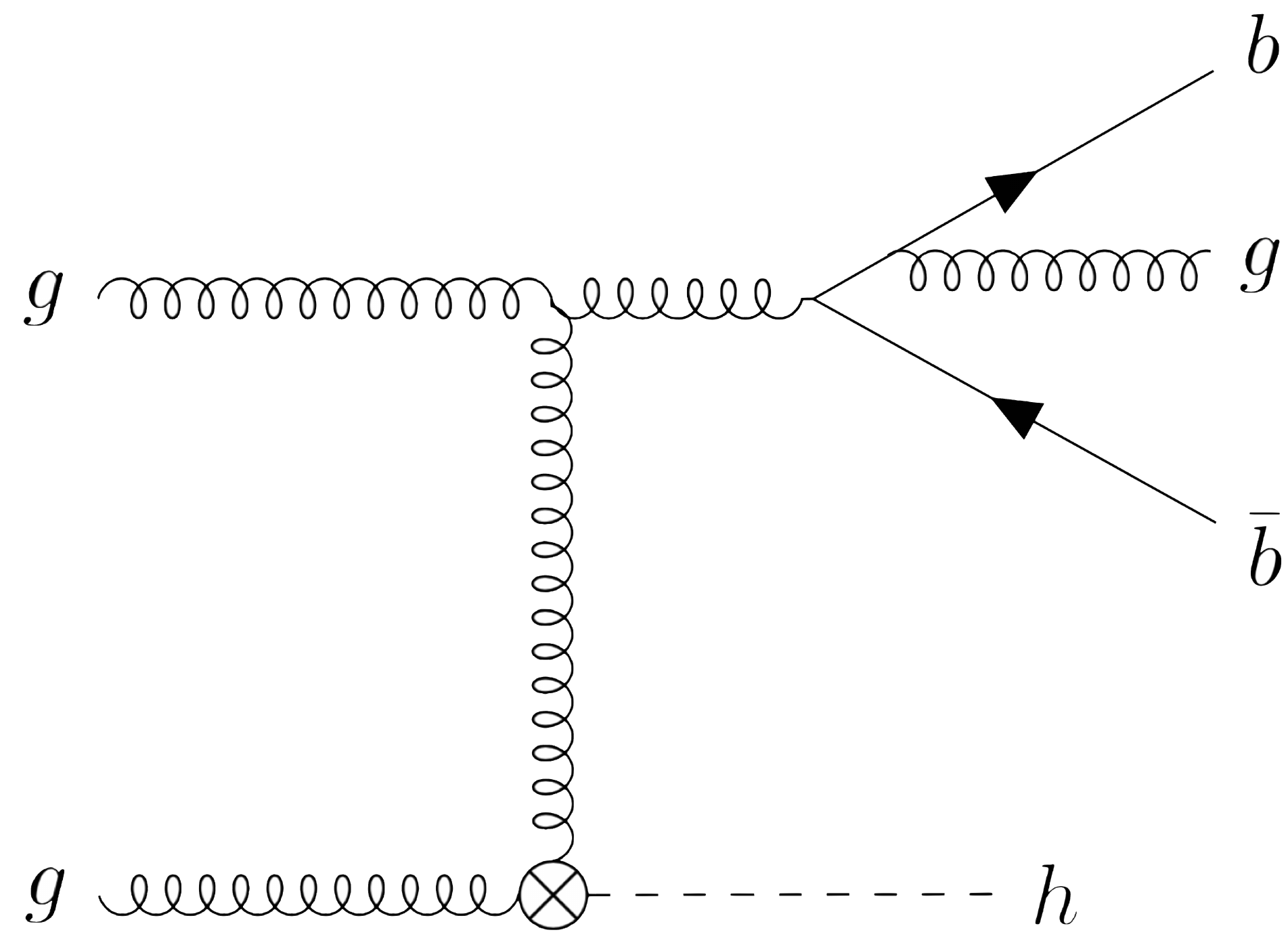}\label{fig:hbbeftreal}}
  \caption{\label{fig:heft}Examples of Born-level, virtual and real-emission diagrams for the \ytsq{} contribution to
  $b \bar b H$ production in the heavy-top quark approximation.}
  \label{fig:hbbeft}
\end{figure}

Working in the HEFT allows us to avoid the computation of the highly complicated
amplitudes in \fig{fig:hbbnlot} and \fig{fig:bbHtop}, and evaluate instead the diagrams shown in \fig{fig:heft},
which have a much lower complexity, being at most at the one-loop level.
In addition, the HEFT has been implemented in a Universal Feynrules Output (UFO) model~\cite{Artoisenet:2013puc,Demartin:2014fia}, 
and this calculation can  be performed using existing automated Monte Carlo tools. Nonetheless, present implementations neglect power-suppressed corrections to SM parameters generated by the heavy-top  mass approximation, which play a crucial role in the case at hand.
In particular the bottom-quark Yukawa coupling must be corrected in the following way:
\begin{equation}
  y_b^\text{HEFT} = y_b + y_t \alpha_s^2\frac{m_b}{m_t} \delta y_b\,,
\end{equation}
which generates additional terms of $\calo{\ybyt \alpha_s^4}$ and $\calo{\ytsq \alpha_s^5}$
entering the \bbH{} cross section at the perturbative order we are interested in. The exact expression
for $\delta y_b$ can be obtained from  \eqn{eq:ybheftvssm}. 
As a result, we insert  $\yb\to y_b^\text{HEFT}$ into \eqn{eq:hbbxsec2} to yield the \bbH{}
cross section in the HEFT and rearrange it as follows:
\begin{equation}
\begin{split}
      {\rm d} \sigma^\text{HEFT} &=
y_b^2\,\alpha_s^2\left( \Delta_{\ybsq}^{(0)} + \alpha_s \Delta_{\ybsq}^{(1)} \right)\\
&+y_t y_b\,\alpha_s^3\left( \hat{\Delta}_{\ybyt}^{(0)} + \alpha_s \hat{\Delta}_{\ybyt}^{(1)} +2\,\alpha_s \frac{m_b}{m_t}\delta y_b \Delta_{\ybsq}^{(0)} \right)\\
&+ y_t^2\,\alpha_s^4\left( \hat{\Delta}_{\ytsq}^{(0)} + \alpha_s \hat{\Delta}_{\ytsq}^{(1)}
+\alpha_s \frac{m_b}{m_t}\delta y_b \hat{\Delta}_{\ybyt}^{(0)} \right)\,,
  \end{split}
\label{eq:hbbxsecheft}
\end{equation}
where top-quark loops have been replaced by the HEFT contact interaction in quantities with a 
hat\footnote{In practice we do not replace loops explicitly in the SM calculation: our 
    UFO model contains all gauge-invariant leading power interactions of the HEFT and we 
    generate the HEFT amplitude independently using \textsc{MadGraph5\_aMC@NLO}~\cite{Alwall:2014hca}. The expressions for the $\hat{\Delta}$ terms contains all tree-level, real-emission and virtual diagrams that are relevant after replacing $C_1$ by its matched expression to order ${\cal O}(y_t\alpha_s^2)$. We articifically separate the insertion of the matched expression for the bottom Yukawa in  \eqn{eq:hbbxsecheft} to attract the reader's attention to the effect of this power suppressed correction, which would be missed if one only replaced top loops by contact operators -- even when consistently including higher orders of the matched Wilson coefficient.}. In 
    this cross section, the only contribution that could not be directly calculated using automated tools is the power-suppressed bottom-quark Yukawa correction. This was derived using a low-energy theorem, at ${\cal O}(\alpha_s^2)$ in~\citeres{PhysRevLett.78.594,CHETYRKIN199719}, which is the order needed for this calculation, and further improved to ${\cal O}(\alpha_s^4)$ in~\citere{Chetyrkin:1997un}. We rederive the ${\cal O}(\alpha_s^2)$ coefficient in the Appendix~\ref{app:matching} by an explicit two-loop matching calculation and find:
\begin{align}
y_b^\text{HEFT} = y_b^\text{SM}+y_t\left(\frac{\alpha_s}{\pi}\right)^2 \frac{m_b}{m_t}C_F \left(\frac{5}{24}-\frac{1}{4}\log\left(\frac{\mu_R^2}{m_t^2}\right)\right)\,,
\label{eq:ybheftvssm}
\end{align}
in agreement with the existing literature, where $\alpha_s$ and $y_b^\text{SM}$ are understood to be renormalised in the $\overline{\text{MS}}$ scheme at a scale $\mu_R$, while $m_b$, $m_t$ and $y_t$ are renormalised on-shell. Note that the renormalization scheme of SM parameters affects the matching coefficient $\delta y_b$ only at higher orders, neglected in the present calculation.

We have implemented by hand  this modification in the HEFT model at NLO.
This enables a complete calculation of the QCD corrections $\hat{\Delta}_{\ytsq}^{(1)}$ and $\hat{\Delta}_{\ybyt}^{(1)}$ in the heavy-top mass approximation  in a fully automated way. We therefore can employ \textsc{MadGraph5\_aMC@NLO}~\cite{Alwall:2014hca} to perform the calculation of the \bbH{} cross section in the 4FS at parton level.
We use the recently-released version capable of computing a mixed-coupling
expansion~\cite{Frederix:2018nkq} of the cross section in order to compute
all six contributions (\ybsq{}, \ybyt{} and \ytsq{} both at LO and NLO)
with the appropriate $\overline{\text{MS}}$ renormalisation of \yb{} simultaneously.\footnote{A similar computation was performed in the context of charged-Higgs production in the intermediate-mass range~\cite{Degrande:2016hyf}.}

Besides computing \eqn{eq:hbbxsecheft} in the HEFT, we also calculate the LO \ytsq{}
contributions in the full theory in order to rescale the \ytsq{} contributions and to provide the best approximation
of the \bbH{} cross section in \eqn{eq:hbbxsec2}. We refer to this approach as the Born-improved HEFT (BI-HEFT) in the following:
\begin{equation}
    \sigma_{y_t^2}^{\text{BI-HEFT}} \equiv \sigma_{y_t^2}^{\text{HEFT}} \times
        \frac{\sigma_{y_t^2}^{\text{SM, LO}}}{\sigma_{y_t^2}^{\text{HEFT, LO}}}\,.
\label{eq:bi-heft}
\end{equation}
For differential distributions, \eqn{eq:bi-heft} is applied bin-by-bin.

\subsection{Assessment of the HEFT approximation}
\label{sec:approx}
In this section we assess the accuracy of the heavy-top quark approximation. To
this end, we compare the LO \ytsq{} cross section
$\sigma_{\ytsq}^\text{SM}$ against its approximation in the HEFT $\sigma_{\ytsq}^\text{HEFT}$.
We use the same input parameters as for our phenomenological results in \sct{sec:results},
and refer to \sct{sec:inputs} for details. We perform a validation for both the
inclusive cross section and differential distributions. Since the topology of the process
at LO is very similar to that of the $H$+jet process, we expect the HEFT to provide
a good description in the relevant phase-space regions, in particular concerning the shapes
of distributions. We stress again that in our best prediction, the BI-HEFT, we use
the HEFT only to determine the radiative corrections in terms of the NLO $K$-factor. Total cross section
and kinematic distributions, obtained in the HEFT, are reweighted (bin-by-bin) by a factor equal
to the ratio between the full theory and the HEFT, both evaluated at LO. This has been shown
to be an excellent approximation for $H$+jet production as long as the relevant scales do
 not  become too large~\cite{Harlander:2012hf,Neumann:2014nha}.

We start by reporting the result for the inclusive cross section:
\begin{align}
  \sigma_{\ytsq}^\text{SM} &= \unit{0.375}{\pico\barn}\,, &   \sigma_{\ytsq}^\text{HEFT} &= \unit{0.358}{\pico\barn}\,.
\end{align}
The results lie within $5\%$ of each other. Considering that the perturbative uncertainties
are one order of magnitude larger, we conclude that the inclusive cross section is well described by
the heavy-top quark approximation. Furthermore, the accuracy of the BI-HEFT result
can be assumed to be considerably better than this value, since top-mass effects are included at LO by the rescaling in \eqn{eq:bi-heft}.
As in the case of $H$+jet production, the dominant configurations are with the Higgs at low transverse momentum,
which explains the quality of the approximation.

\begin{figure}[tp]
\begin{center}
\subfloat[]{\includegraphics[trim = 13mm 8.2cm 2cm 2cm, width=.245\textheight]{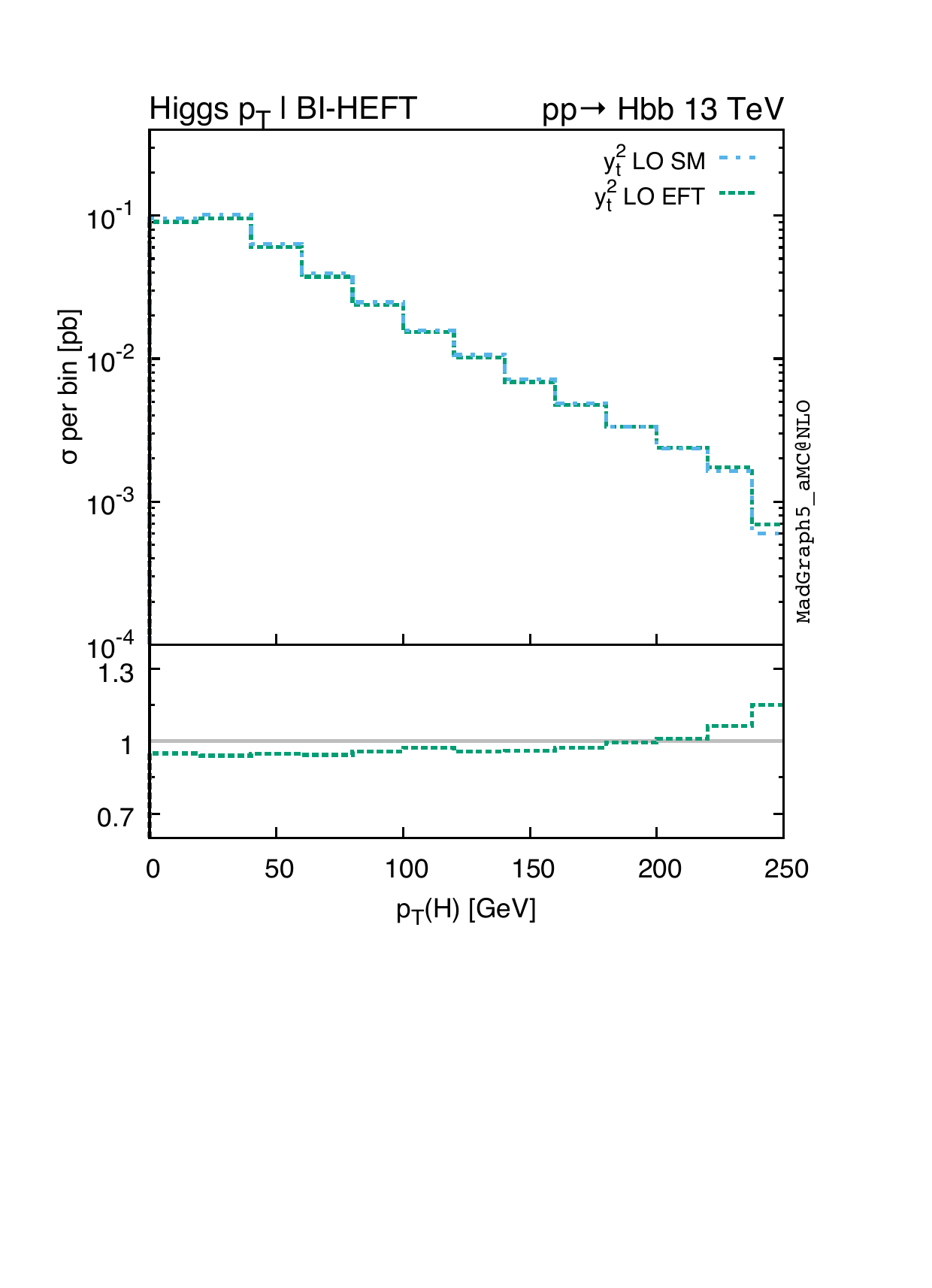}\label{fig:approx-pth}}
\subfloat[]{\includegraphics[trim = 13mm 8.2cm 2cm 2cm, width=.245\textheight]{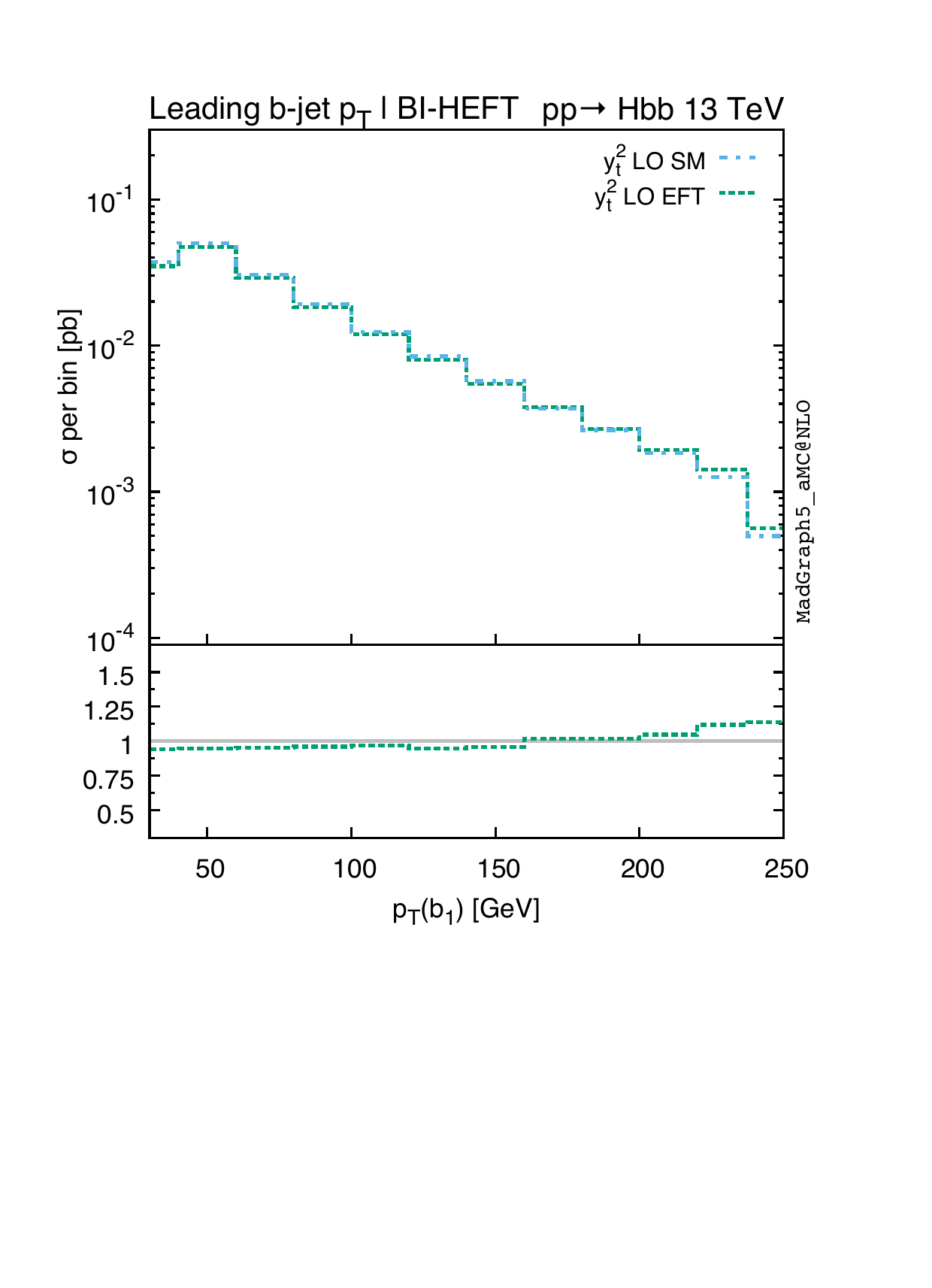}\label{fig:approx-ptb1}}
\subfloat[]{\includegraphics[trim = 13mm 8.2cm 2cm 2cm, width=.245\textheight]{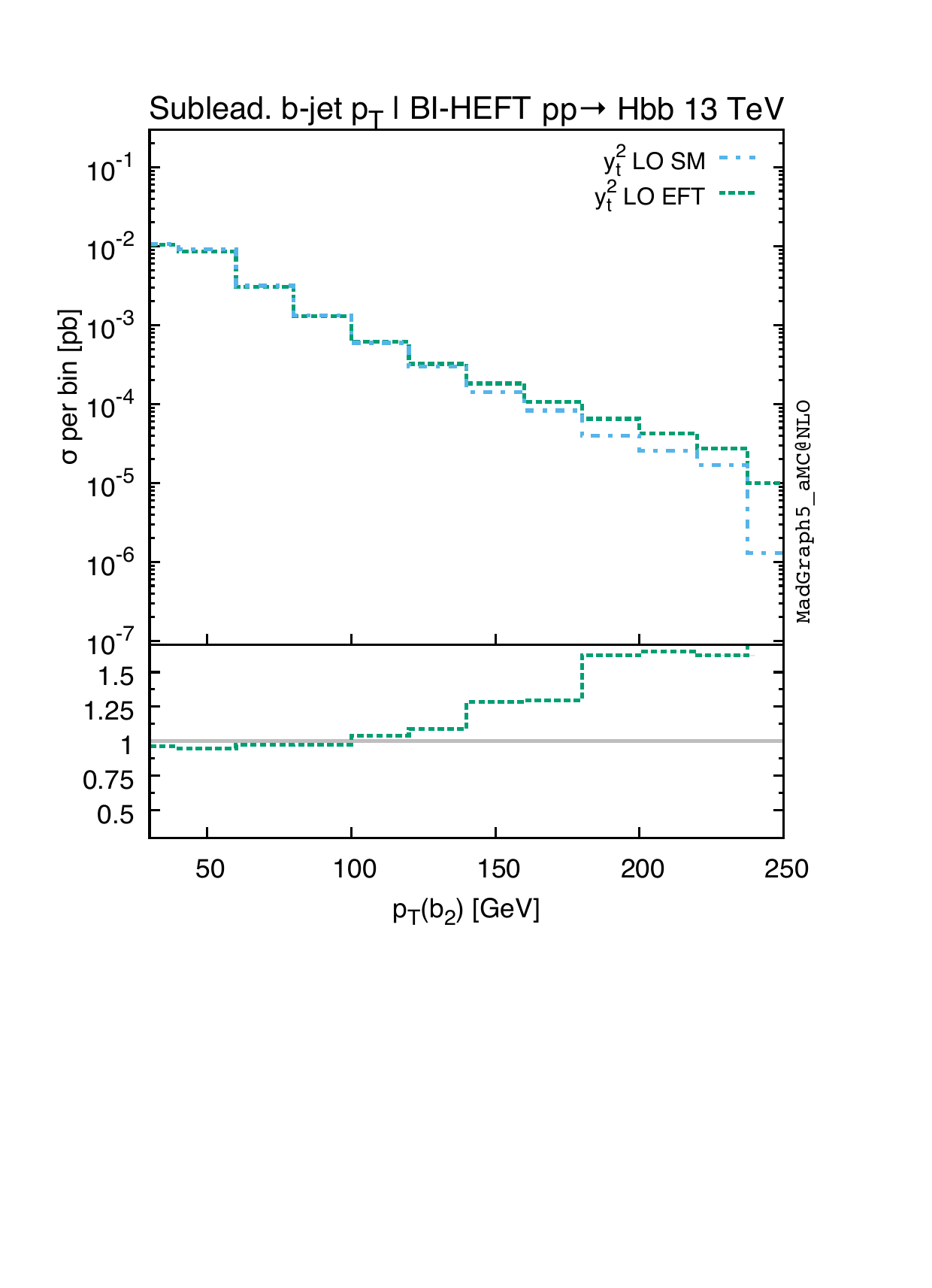}\label{fig:approx-ptb2}} \\
\subfloat[]{\includegraphics[trim = 13mm 8.2cm 2cm 2cm, width=.245\textheight]{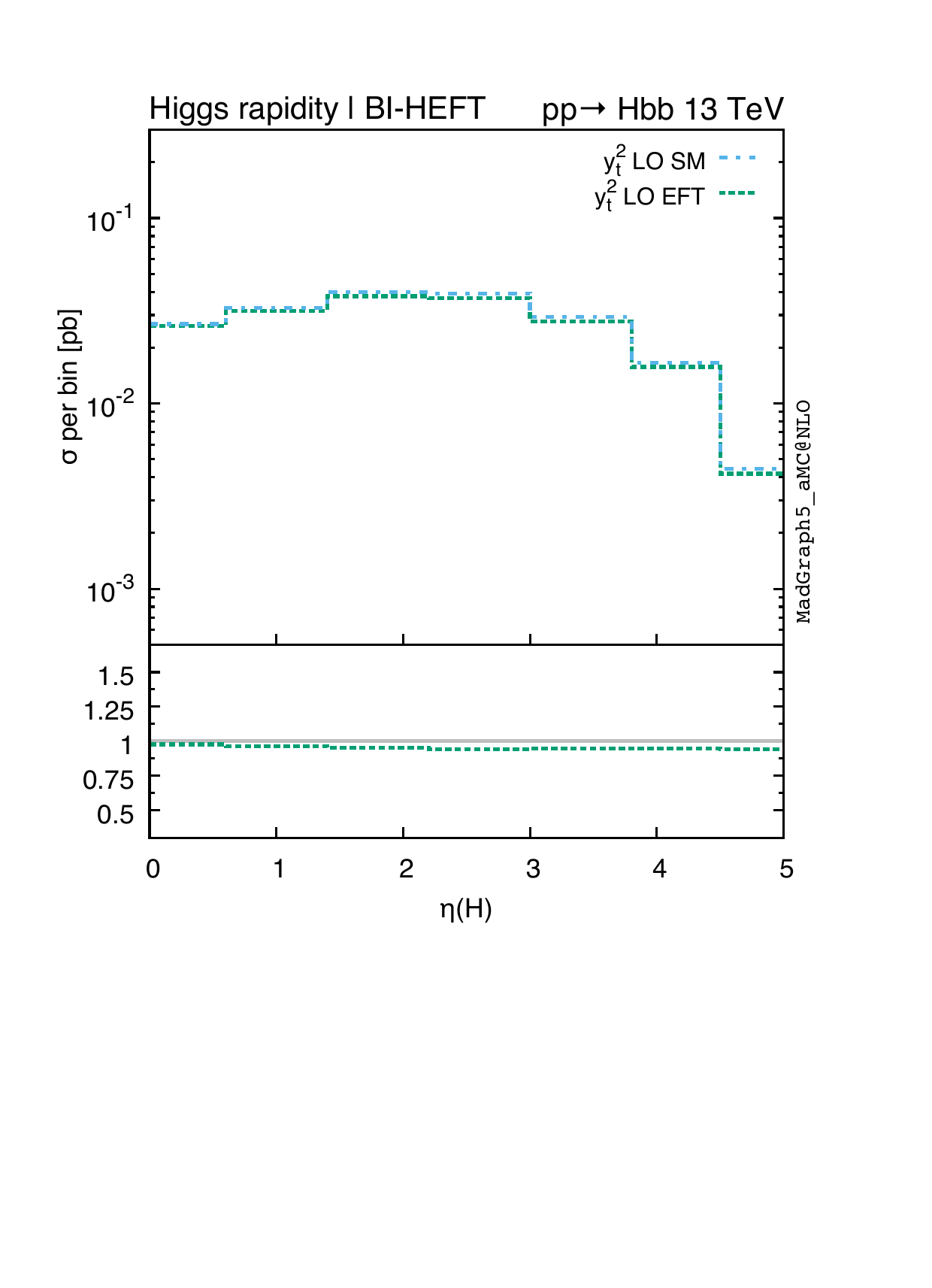}\label{fig:approx-etah}}
\subfloat[]{\includegraphics[trim = 13mm 8.2cm 2cm 2cm, width=.245\textheight]{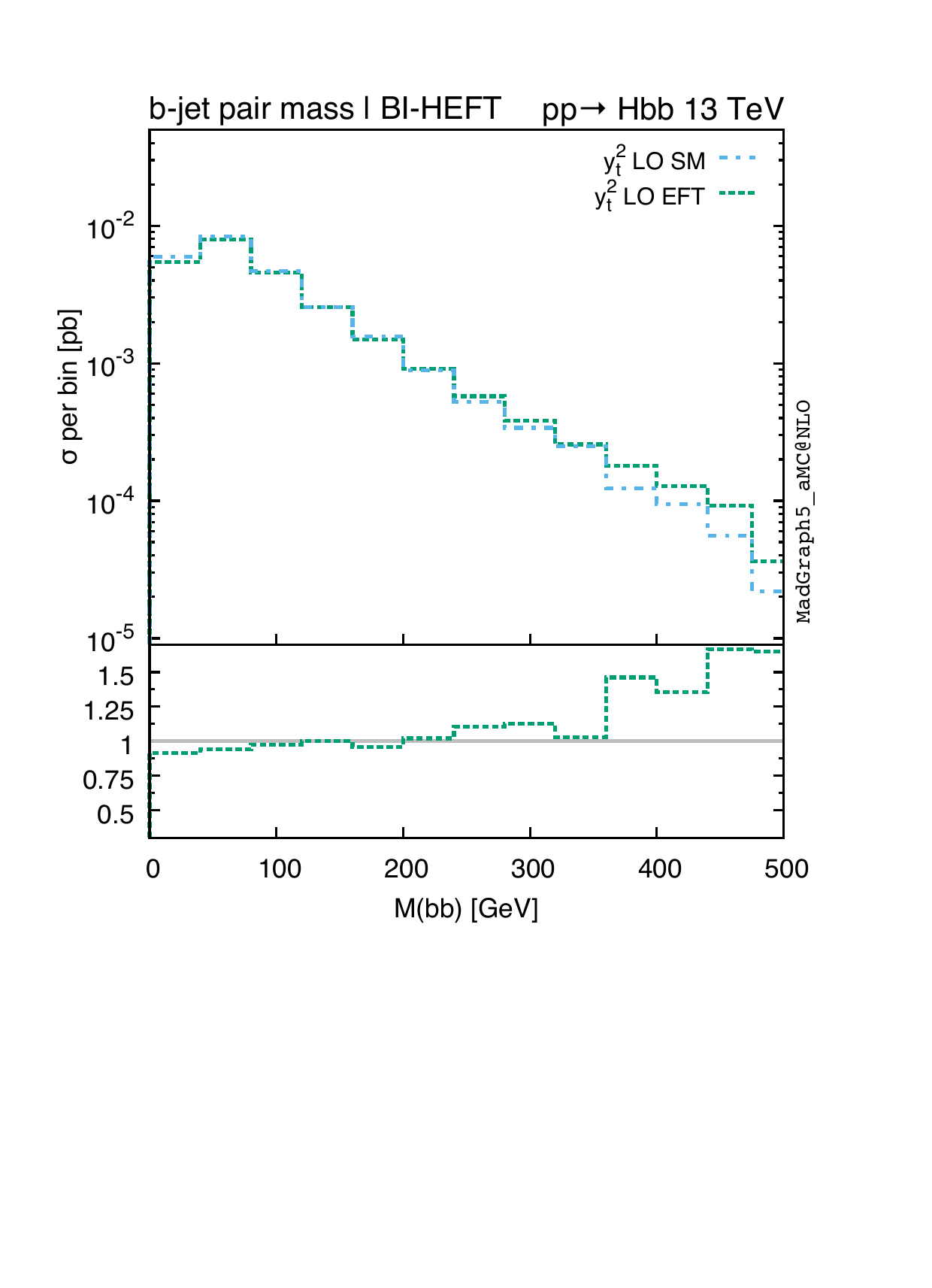}\label{fig:approx-mbb}}
\subfloat[]{\includegraphics[trim = 13mm 8.2cm 2cm 2cm, width=.245\textheight]{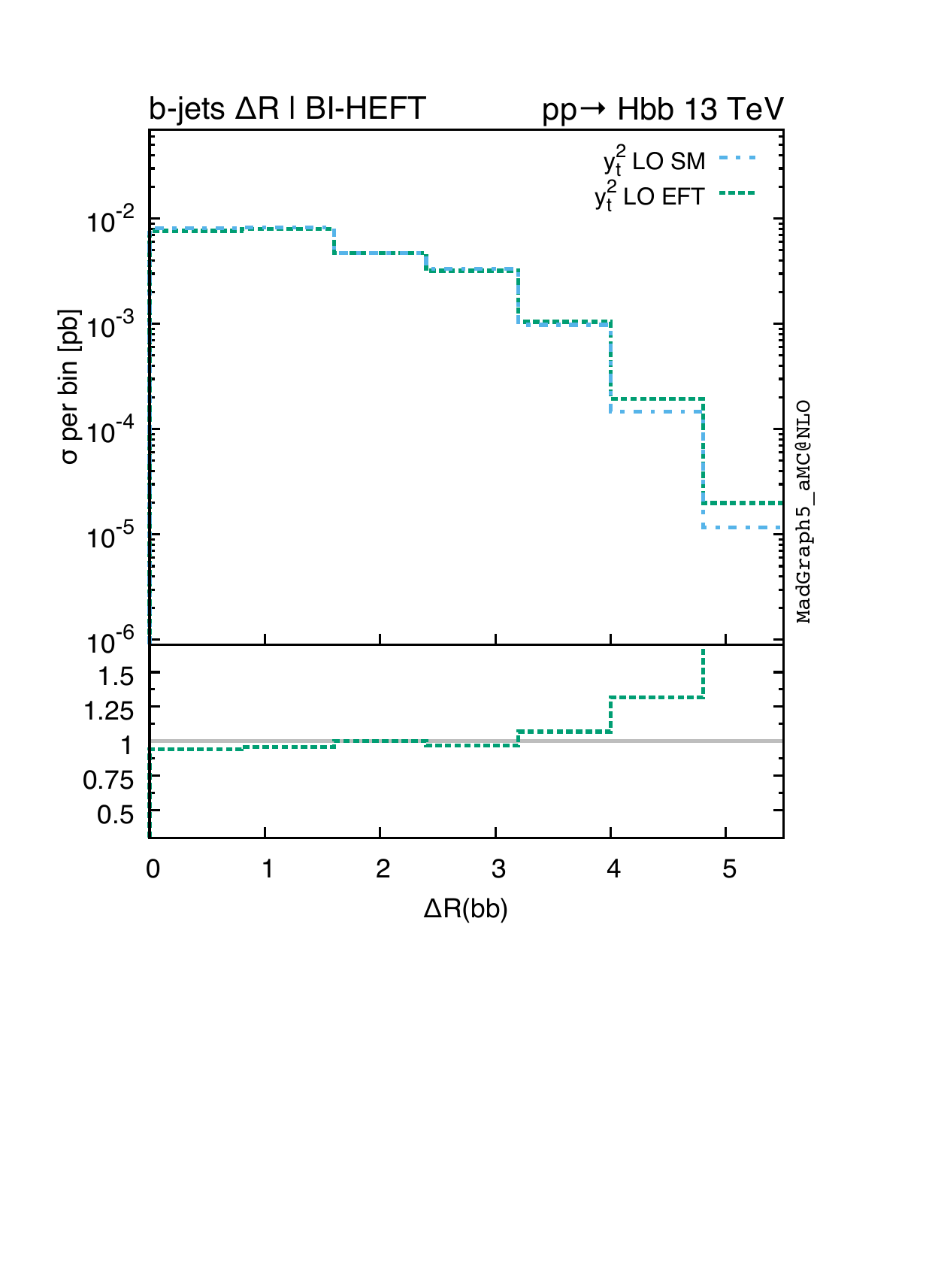}\label{fig:approx-drbb}}
\caption[]{\label{fig:approx}{
Comparison of LO predictions in the SM and the HEFT for various observables:
the transverse-momentum of the Higgs boson (\ref{fig:approx-pth}), of the leading (\ref{fig:approx-ptb1}), and of the subleading $b$ jet (\ref{fig:approx-ptb2}), the rapidity of the Higgs boson (\ref{fig:approx-etah}), the invariant mass of the $b$-jet pair (\ref{fig:approx-mbb}), and their
distance in the $\eta$--$\phi$ plane (\ref{fig:approx-drbb}); the lower insets show the ratio of the two predictions.
}}
\end{center}
\end{figure}

Let us now turn to differential cross sections in \fig{fig:approx}.
The main frame shows the SM (blue dash-dotted) and HEFT (green dotted) predictions.
The lower inset shows their bin-by-bin ratios. The first three plots, \figs{fig:approx-pth}-\ref{fig:approx-ptb2} feature
the transverse-momentum spectra of the Higgs boson, the leading and the subleading
$b$ jet respectively. As expected, we find that the HEFT provides a
good description of the SM result, especially in terms of shapes. Only at large
transverse momentum the two curves start deviating with the HEFT result becoming harder.
This happens after transverse momenta of $\sim 200\,\giga\electronvolt$ for the Higgs and the leading $b$ jet,
and a bit earlier for the second-hardest $b$ jet.

The Higgs rapidity distribution in \fig{fig:approx-etah} is hardly affected by the HEFT
approximation, with the HEFT/SM ratio being essentially flat. Also for
the invariant mass of the two $b$ jets in \fig{fig:approx-mbb}, the heavy-top quark result
provides a good description as long as $\Mbb \lesssim 200\,\giga\electronvolt$.
Finally, for the separation in the $\eta$--$\phi$ plane between the two $b$ jets, shown
in \fig{fig:approx-drbb} , the
agreement between HEFT and SM is very good up to $\Delta R = 4$. Above this value, the distribution
is dominated by large invariant-mass pairs, and the HEFT/SM ratio follows what happens for the invariant-mass
distribution.

Overall, the heavy-top quark approximation used in the HEFT results works extremely well for this process over a large fraction of the phase space and in particular where the majority of events are produced. For the goals of our study, it is especially important to verify that the comparison of the angular separation of the $b$ jets and of their invariant mass is well reproduced, as it indicates that we can safely explore the regime in which the two bottom jets merge into a single one.
This regime is particularly interesting to study for the \ytsq{} terms as we will see in \sct{sec:results}.
Furthermore, there is a reasonable range of $b$-jet transverse momentum where the process is correctly described, so that we can trust the prediction to study the impact of $b$-jet requirements on the relative importance of the \ybsq{} and \ytsq{} contributions.
It should be noted, however, that in the two $b$-jet configuration, the HEFT prediction is rather poor over a larger range of transverse momenta for the subleading $b$ jet.
Nevertheless, this is not expected to have an impact on our phenomenological study in the upcoming section.

\section{Phenomenological results}
\label{sec:results}

In this section we present differential results for \bbH{} production at the 13 TeV LHC including all contributions
 proportional to \ybsq{}, \ybyt{}, and \ytsq{} at NLO QCD, see \eqn{eq:hbbxsec2}.
We analyze the importance of radiative corrections and the
relative size of the three contributions.
Although we work in the SM, thanks to the separation of the cross section by the Yukawa coupling structure,
our predictions are directly applicable to 2HDM-type extensions of the SM (for $b\bar{b}\phi$ with a neutral Higgs boson $\phi\in\{h,H,A\}$)
by an appropriate rescaling of the top and bottom-quark Yukawa.
Even for the MSSM such rescaling has been shown to be an excellent approximation of the complete result~\cite{Dittmaier:2006cz,Dawson:2011pe}.

\subsection{Input parameters}\label{sec:inputs}

Our predictions are obtained in the four-flavour scheme throughout. We use the corresponding $n_f=4$
NNPDF 3.1~\cite{Ball:2017nwa} sets of parton densities at NLO with the corresponding running and $\alpha_s$ values.\footnote{More precisely,
    {\tt NNPDF31\_nlo\_as\_0118\_nf\_4}  ({\tt lhaid=320500} in LHAPDF6~\cite{Buckley:2014ana})
corresponding to $\alpha_s(m_Z)=0.118$. We would like to stress that two versions of this set 
exist, flagged by the {\tt DataVersion} entry of the {\tt NNPDF31\_nlo\_as\_0118\_nf\_4.info} file. In particular 
we used {\tt DataVersion=1}. The newer version has some mild but visible effects on the cross sections quoted in 
\tab{tab:rates}. For instance, the inclusive prediction for the $y_b^2$ term at LO becomes 0.247 pb, and 0.373 pb at NLO. We 
thank Ayan Paul and Zhuoni Quian for pointing this issue after the publication of this article, and Stefano Carrazza for clarifications.
}
The central values of the renormalisation ($\mu_R$) and factorisation ($\mu_F$) scales are set on an event-to-event basis to
\begin{align}\label{eq:scales}
    \mu_R = \mu_F = H_T/4 = \frac{1}{4}\sum_i \sqrt{(p^T_i)^2 + m_i^2},
\end{align}
where the index $i$ runs over all the final-state particles, possibly including the extra parton from the real
emission. Scale uncertainties are computed without extra runs using
a reweighting technique~\cite{Frederix:2011ss}, and
correspond to independent (nine-point) variations in the range
$H_T/8\le\mu_R, \mu_F\le H_T/2$.
Internal masses are set to their on-shell values $m_H=125\,\giga\electronvolt$, $m_t = 172.5\,\giga\electronvolt$ and $m_b=4.92\,\giga\electronvolt$.
The top-quark Yukawa is renormalised on-shell; for the bottom-quark Yukawa, instead, we
compute $m_b(\mu_R)$ by adopting the $\overline{\textrm{MS}}$ scheme, with a four-loop evolution~\cite{Marquard:2015qpa,Kataev:2015gvt} from
$m_b(m_b)=4.18\,\giga\electronvolt$ up to the central value of the renormalisation scale, and two-loop running for the scale variations,
as recommended by the LHC Higgs cross section working group~\cite{deFlorian:2016spz}.

Jets are reconstructed with the anti-$k_T$ algorithm~\cite{Cacciari:2008gp}, as implemented in {\sc FastJet}~\cite{Cacciari:2011ma},
with a jet radius of $R = 0.4$, and subject to the condition $\ptj{}> 30\,\giga\electronvolt$  and $\eta_j <2.5$. Results with a larger jet radius, $R=1$,
are available in appendix~\ref{app:r1}.
Bottom-quark flavoured jets ($b$ jets) are defined to include at least one bottom quark among the jet constituents.
A $b$ jet containing a pair of bottom quarks is denoted as a $bb$ jet. Within a fixed-order
computation, we will use the word $B$ hadrons to identify bottom quarks (the notation
$B$ will refer to bottom quarks, while the notation $b$ to bottom-tagged jets). Bottom-quark observables
are infrared safe owing to the finite bottom-quark mass in the 4FS. At variance with the case of $b$ jets, no cut is imposed on $B$ hadrons.

\begin{table}[!th]
\vspace{1.3cm}
\begin{center}
\resizebox{\columnwidth}{!}{%
    \begin{tabular}{c c| ccc}
         &  &  LO (acceptance) &  NLO (acceptance) & $K$ \\
 \hline\hline
\multirow{5}{*}{ $ \rm{inclusive} $ } 
 & ${y_t^2}_{\textrm{HEFT}}$  &  $ \phantom{-} 3.58 \cdot 10^{ -1 }\,\,{}^{+   74 \perc} _{ -39 \perc} $  (100\perc) &  $ \phantom{-} 8.80 \cdot 10^{ -1 }\,\,{}^{+   47 \perc} _{ -31 \perc} $ (100\perc) &   2.5  \\
 & ${y_t^2}_{\textrm{BI-HEFT}}$  &  $ \phantom{-} 3.75 \cdot 10^{ -1 }\,\,{}^{+   74 \perc} _{ -39 \perc} $  (100\perc) &  $ \phantom{-} 9.22 \cdot 10^{ -1 }\,\,{}^{+   47 \perc} _{ -31 \perc} $ (100\perc) &   2.5  \\
 & $y_b y_t$  &  $ -3.82 \cdot 10^{ -2 }\,\,{}^{+   65 \perc} _{ -36 \perc} $  (------) &  $ -7.37 \cdot 10^{ -2 }\,\,{}^{+   39 \perc} _{ -27 \perc} $ (------) &   1.9  \\
 & $y_b^2$  &  $ \phantom{-} 2.63 \cdot 10^{ -1 }\,\,{}^{+   57 \perc} _{ -34 \perc} $  (100\perc) &  $ \phantom{-} 4.05 \cdot 10^{ -1 }\,\,{}^{+   21 \perc} _{ -21 \perc} $ (100\perc) &   1.5  \\
 & $y_b^2 + y_b y_t + {y_t^2}_{\textrm{BI-HEFT}}$  &  $ \phantom{-} 6.00 \cdot 10^{ -1 }\,\,{}^{+   67 \perc} _{ -37 \perc} $  (100\perc) &  $ \phantom{-} 1.25 \cdot 10^{ 0 \phantom{-} }\,\,{}^{+   38 \perc} _{ -28 \perc} $ (100\perc) &   2.1  \\

 \hline\hline
\multirow{5}{*}{ $ \ge 1 b $ } 
 & ${y_t^2}_{\textrm{HEFT}}$  &  $ \phantom{-} 1.70 \cdot 10^{ -1 }\,\,{}^{+   72 \perc} _{ -39 \perc} $  (47\perc) &  $ \phantom{-} 3.67 \cdot 10^{ -1 }\,\,{}^{+   39 \perc} _{ -29 \perc} $ (42\perc) &   2.2  \\
 & ${y_t^2}_{\textrm{BI-HEFT}}$  &  $ \phantom{-} 1.76 \cdot 10^{ -1 }\,\,{}^{+   72 \perc} _{ -39 \perc} $  (47\perc) &  $ \phantom{-} 3.81 \cdot 10^{ -1 }\,\,{}^{+   39 \perc} _{ -29 \perc} $ (41\perc) &   2.2  \\
 & $y_b y_t$  &  $ -1.15 \cdot 10^{ -2 }\,\,{}^{+   62 \perc} _{ -35 \perc} $  (------) &  $ -1.63 \cdot 10^{ -2 }\,\,{}^{+   16 \perc} _{ -19 \perc} $ (------) &   1.4  \\
 & $y_b^2$  &  $ \phantom{-} 6.02 \cdot 10^{ -2 }\,\,{}^{+   52 \perc} _{ -31 \perc} $  (23\perc) &  $ \phantom{-} 8.49 \cdot 10^{ -2 }\,\,{}^{+   13 \perc} _{ -16 \perc} $ (21\perc) &   1.4  \\
 & $y_b^2 + y_b y_t + {y_t^2}_{\textrm{BI-HEFT}}$  &  $ \phantom{-} 2.25 \cdot 10^{ -1 }\,\,{}^{+   67 \perc} _{ -37 \perc} $  (37\perc) &  $ \phantom{-} 4.50 \cdot 10^{ -1 }\,\,{}^{+   35 \perc} _{ -27 \perc} $ (36\perc) &   2.0  \\

 \hline\hline
\multirow{5}{*}{ $ \ge 2 b $ } 
 & ${y_t^2}_{\textrm{HEFT}}$  &  $ \phantom{-} 2.48 \cdot 10^{ -2 }\,\,{}^{+   72 \perc} _{ -39 \perc} $  (6.9\perc) &  $ \phantom{-} 4.86 \cdot 10^{ -2 }\,\,{}^{+   33 \perc} _{ -27 \perc} $ (5.5\perc) &   2.0  \\
 & ${y_t^2}_{\textrm{BI-HEFT}}$  &  $ \phantom{-} 2.56 \cdot 10^{ -2 }\,\,{}^{+   72 \perc} _{ -39 \perc} $  (6.8\perc) &  $ \phantom{-} 5.02 \cdot 10^{ -2 }\,\,{}^{+   33 \perc} _{ -27 \perc} $ (5.4\perc) &   2.0  \\
 & $y_b y_t$  &  $ -6.95 \cdot 10^{ -4 }\,\,{}^{+   62 \perc} _{ -35 \perc} $  (------) &  $ -5.24 \cdot 10^{ -4 }\,\,{}^{+    5 \perc} _{ -53 \perc} $ (------) &   0.8  \\
 & $y_b^2$  &  $ \phantom{-} 5.07 \cdot 10^{ -3 }\,\,{}^{+   51 \perc} _{ -31 \perc} $  (1.9\perc) &  $ \phantom{-} 5.92 \cdot 10^{ -3 }\,\,{}^{+    1 \perc} _{ -12 \perc} $ (1.5\perc) &   1.2  \\
 & $y_b^2 + y_b y_t + {y_t^2}_{\textrm{BI-HEFT}}$  &  $ \phantom{-} 3.00 \cdot 10^{ -2 }\,\,{}^{+   69 \perc} _{ -38 \perc} $  (5.0\perc) &  $ \phantom{-} 5.56 \cdot 10^{ -2 }\,\,{}^{+   30 \perc} _{ -26 \perc} $ (4.4\perc) &   1.9  \\

 \hline\hline
\multirow{5}{*}{ $ \ge 1 bb $ } 
 & ${y_t^2}_{\textrm{HEFT}}$  &  $ \phantom{-} 3.84 \cdot 10^{ -2 }\,\,{}^{+   70 \perc} _{ -38 \perc} $  (11\perc) &  $ \phantom{-} 7.86 \cdot 10^{ -2 }\,\,{}^{+   36 \perc} _{ -28 \perc} $ (8.9\perc) &   2.0  \\
 & ${y_t^2}_{\textrm{BI-HEFT}}$  &  $ \phantom{-} 4.12 \cdot 10^{ -2 }\,\,{}^{+   70 \perc} _{ -38 \perc} $  (11\perc) &  $ \phantom{-} 8.43 \cdot 10^{ -2 }\,\,{}^{+   36 \perc} _{ -28 \perc} $ (9.1\perc) &   2.0  \\
 & $y_b y_t$  &  $ -7.91 \cdot 10^{ -5 }\,\,{}^{+   89 \perc} _{ -45 \perc} $  (------) &  $ \phantom{-} 2.02 \cdot 10^{ -4 }\,\,{}^{+  132 \perc} _{ -54 \perc} $ (------) &  -2.5  \\
 & $y_b^2$  &  $ \phantom{-} 3.37 \cdot 10^{ -4 }\,\,{}^{+   57 \perc} _{ -34 \perc} $  (0.1\perc) &  $ \phantom{-} 2.53 \cdot 10^{ -4 }\,\,{}^{+    4 \perc} _{ -48 \perc} $ (0.1\perc) &   0.7  \\
 & $y_b^2 + y_b y_t + {y_t^2}_{\textrm{BI-HEFT}}$  &  $ \phantom{-} 4.15 \cdot 10^{ -2 }\,\,{}^{+   70 \perc} _{ -38 \perc} $  (6.9\perc) &  $ \phantom{-} 8.48 \cdot 10^{ -2 }\,\,{}^{+   36 \perc} _{ -28 \perc} $ (6.8\perc) &   2.0  \\

 \hline\hline
\multirow{5}{*}{ $  p^T_H > 50\,\giga\electronvolt  $ } 
 & ${y_t^2}_{\textrm{HEFT}}$  &  $ \phantom{-} 1.38 \cdot 10^{ -1 }\,\,{}^{+   73 \perc} _{ -39 \perc} $  (39\perc) &  $ \phantom{-} 3.77 \cdot 10^{ -1 }\,\,{}^{+   52 \perc} _{ -33 \perc} $ (43\perc) &   2.7  \\
 & ${y_t^2}_{\textrm{BI-HEFT}}$  &  $ \phantom{-} 1.42 \cdot 10^{ -1 }\,\,{}^{+   73 \perc} _{ -39 \perc} $  (38\perc) &  $ \phantom{-} 3.87 \cdot 10^{ -1 }\,\,{}^{+   52 \perc} _{ -33 \perc} $ (42\perc) &   2.7  \\
 & $y_b y_t$  &  $ -7.43 \cdot 10^{ -3 }\,\,{}^{+   62 \perc} _{ -35 \perc} $  (------) &  $ -9.66 \cdot 10^{ -3 }\,\,{}^{+   10 \perc} _{ -17 \perc} $ (------) &   1.3  \\
 & $y_b^2$  &  $ \phantom{-} 3.20 \cdot 10^{ -2 }\,\,{}^{+   53 \perc} _{ -32 \perc} $  (12\perc) &  $ \phantom{-} 5.54 \cdot 10^{ -2 }\,\,{}^{+   24 \perc} _{ -21 \perc} $ (14\perc) &   1.7  \\
 & $y_b^2 + y_b y_t + {y_t^2}_{\textrm{BI-HEFT}}$  &  $ \phantom{-} 1.66 \cdot 10^{ -1 }\,\,{}^{+   70 \perc} _{ -38 \perc} $  (28\perc) &  $ \phantom{-} 4.33 \cdot 10^{ -1 }\,\,{}^{+   49 \perc} _{ -32 \perc} $ (35\perc) &   2.6  \\

 \hline\hline
\multirow{5}{*}{ $  p^T_H >  100\,\giga\electronvolt  $ } 
 & ${y_t^2}_{\textrm{HEFT}}$  &  $ \phantom{-} 5.03 \cdot 10^{ -2 }\,\,{}^{+   73 \perc} _{ -39 \perc} $  (14\perc) &  $ \phantom{-} 1.43 \cdot 10^{ -1 }\,\,{}^{+   53 \perc} _{ -33 \perc} $ (16\perc) &   2.8  \\
 & ${y_t^2}_{\textrm{BI-HEFT}}$  &  $ \phantom{-} 4.98 \cdot 10^{ -2 }\,\,{}^{+   73 \perc} _{ -39 \perc} $  (13\perc) &  $ \phantom{-} 1.41 \cdot 10^{ -1 }\,\,{}^{+   53 \perc} _{ -33 \perc} $ (15\perc) &   2.8  \\
 & $y_b y_t$  &  $ -1.35 \cdot 10^{ -3 }\,\,{}^{+   63 \perc} _{ -36 \perc} $  (------) &  $ -1.20 \cdot 10^{ -3 }\,\,{}^{+    2 \perc} _{ -32 \perc} $ (------) &   0.9  \\
 & $y_b^2$  &  $ \phantom{-} 5.65 \cdot 10^{ -3 }\,\,{}^{+   54 \perc} _{ -33 \perc} $  (2.1\perc) &  $ \phantom{-} 9.86 \cdot 10^{ -3 }\,\,{}^{+   24 \perc} _{ -21 \perc} $ (2.4\perc) &   1.7  \\
 & $y_b^2 + y_b y_t + {y_t^2}_{\textrm{BI-HEFT}}$  &  $ \phantom{-} 5.42 \cdot 10^{ -2 }\,\,{}^{+   72 \perc} _{ -39 \perc} $  (9.0\perc) &  $ \phantom{-} 1.50 \cdot 10^{ -1 }\,\,{}^{+   51 \perc} _{ -33 \perc} $ (12\perc) &   2.8  \\

 \hline\hline
\multirow{5}{*}{ $  p^T_H > 150\,\giga\electronvolt  $ } 
 & ${y_t^2}_{\textrm{HEFT}}$  &  $ \phantom{-} 2.10 \cdot 10^{ -2 }\,\,{}^{+   74 \perc} _{ -39 \perc} $  (5.9\perc) &  $ \phantom{-} 6.16 \cdot 10^{ -2 }\,\,{}^{+   53 \perc} _{ -33 \perc} $ (7.0\perc) &   2.9  \\
 & ${y_t^2}_{\textrm{BI-HEFT}}$  &  $ \phantom{-} 1.95 \cdot 10^{ -2 }\,\,{}^{+   74 \perc} _{ -39 \perc} $  (5.2\perc) &  $ \phantom{-} 5.73 \cdot 10^{ -2 }\,\,{}^{+   53 \perc} _{ -33 \perc} $ (6.2\perc) &   2.9  \\
 & $y_b y_t$  &  $ -3.18 \cdot 10^{ -4 }\,\,{}^{+   64 \perc} _{ -36 \perc} $  (------) &  $ -1.97 \cdot 10^{ -4 }\,\,{}^{+   11 \perc} _{ -84 \perc} $ (------) &   0.6  \\
 & $y_b^2$  &  $ \phantom{-} 1.40 \cdot 10^{ -3 }\,\,{}^{+   55 \perc} _{ -33 \perc} $  (0.5\perc) &  $ \phantom{-} 2.51 \cdot 10^{ -3 }\,\,{}^{+   25 \perc} _{ -22 \perc} $ (0.6\perc) &   1.8  \\
 & $y_b^2 + y_b y_t + {y_t^2}_{\textrm{BI-HEFT}}$  &  $ \phantom{-} 2.06 \cdot 10^{ -2 }\,\,{}^{+   73 \perc} _{ -39 \perc} $  (3.4\perc) &  $ \phantom{-} 5.96 \cdot 10^{ -2 }\,\,{}^{+   53 \perc} _{ -33 \perc} $ (4.8\perc) &   2.9  \\

    \end{tabular}}
    \caption{\label{tab:rates} Cross sections (in pb) for different $b$-jet multiplicities or minimum \pth{} cuts.}
\end{center}
\end{table}

\subsection{Predictions for \bbH{} production in the SM}\label{sec:predictions}

We start by discussing integrated cross sections in \tab{tab:rates}, both fully inclusive and within cuts.  As far as the latter are concerned, we have considered
various possibilities: the requirement that there be at least one or two $b$ jet(s); that there be at least one jet containing a pair of bottom quarks ($bb$ jets); and that the
transverse momentum of the Higgs boson be larger than $50\,\giga\electronvolt$, $100\,\giga\electronvolt$, and $150\,\giga\electronvolt$ (boosted scenarios), for simplicity without any requirement on $b$ jets.
The residual scale uncertainties are computed by varying the scales as indicated in \sct{sec:inputs}.
We present separately the results for terms proportional to $\ybsq$, $\ytsq$, and $\ybyt$. The $\ytsq$ contributions are provided in two approximations:
using the HEFT, on the one hand, and our BI-HEFT prediction, computed by
rescaling the HEFT result at NLO by the LO evaluated in the full theory,
on the other hand. For completeness, we also quote the BI-HEFT prediction for the sum of all individual contributions. Besides LO and NLO of the cross
sections we also provide the NLO/LO $K$-factor to assess the importance of QCD corrections.
Inside the bracket after the LO and NLO cross sections we quote the acceptance of the
respective scenario, defined as the ratio of the cross section within cuts divided by the
inclusive one. We refrain from quoting the acceptance for the \ybyt{} interference terms since
this quantity is meaningless on its own.
The conclusions that can be drawn from the table are the following:
\begin{itemize}
\item Already at LO the \ytsq{} terms yield a significant contribution to the SM \bbH{} cross section.
Due to sizable QCD corrections to the \ytsq{} terms ($K\approx 2.5$),
the inclusive NLO cross section is a factor of three larger
after including the loop-induced gluon-fusion component
than when considering only \ybsq{} contributions.
Hence, the \bbH{} cross section in the SM is substantially larger than generally assumed from \ybsq{} computations,
which could make its observation much easier. At the same time, however, the sensitivity to the extraction of the bottom-quark Yukawa coupling
is diminished. Below, we discuss how suitable phase-space cuts can be used to enhance the \ybsq{} over the \ytsq{} contributions and to retain
sensitivity to the extraction of \yb{}.
\item The relative size of \ytsq{} contributions further increases when considering the various scenarios with additional phase-space cuts.
The reason is that the loop-induced gluon-fusion component generates harder ($b$-)jet activity and the cuts  favour hard configurations. For example, tagging one $b$ jet has the effect of decreasing the \ybsq{} NLO cross section by
$-79$\%, while for \ytsq{} it is only $-59$\%, and the \ytsq{} NLO cross section is four times as large as \ybsq{}
in the $\ge 1b$-jet scenario, to be compared to the factor of two in the inclusive case. Tighter $b$-jet requirements or the \pth{} requirements only
have the effect of further increasing the relative size of the \ytsq{} contributions.
\item It is interesting to notice that the $\ge 1bb$ jet category, which requires one jet containing two bottom quarks, receives contributions
essentially only from \ytsq{} terms. This can be understood easily: a major part of the events for the loop-induced gluon fusion component
features the Higgs recoiling against a hard gluon, which splits into a $b\bar{b}$-pair. The two bottom quarks in these configurations are
boosted and generally close together, which makes it more likely for them to end up inside the same jet.
\item Two opposed effects render the measurement of the bottom-quark Yukawa coupling in \bbH{} production complicated:
as pointed before the relative size of terms proportional to \ybsq{} decreases as soon as $b$ jets are tagged. Nonetheless, tagging
at least one of the $b$ jets is essential to distinguish \bbH{} production from inclusive Higgs production, which predominantly proceeds
via gluon fusion.\footnote{Inclusive Higgs production may be used to extract \yb{} only through the measurement of the inclusive Higgs
transverse-momentum spectrum at very small \pth{}, see~\citere{Bishara:2016jga} for example.} Therefore, it is necessary to select suitable phase-space
requirements which increase the relative \ybsq{} contribution even in presence of at least one $b$ jet without loosing too much statistics.
Given our findings for the $\ge 1bb$-jet category, one can require at least one $b$ jet and veto all $bb$ jets. This decreases
the $\ge 1b$-jet rate for \ytsq{} by roughly $20$\%, while having a negligible effect on the \ybsq{} rate.
Below, we study differential distributions in order to find further requirements to enhance the relative size of the \ybsq{} contributions.
\item The LO contribution of the mixed \ybyt{} terms is negative in all scenarios, as has been observed already in~\citere{Wiesemann2015}. At NLO it yields a positive contribution only to the $\ge 1bb$-jet scenario. The
NLO/LO $K$-factor of the \ybyt{} term strongly depends on the scenario under consideration,
which is expected given its interference-type contribution.
By and large the impact of the \ybyt{} terms is minor though, reaching at most a few percent at NLO.
\item Overall, QCD corrections have an even larger impact on
the \ytsq{} terms than on the \ybsq{} terms, but they are quite sizable in either case. This can be understood
as follows: as pointed out in~\citere{Wiesemann2015} potentially large logarithmic terms of $\log(m_b^2/m_H^2)$ enter
the perturbative expansion of \ybsq{} contributions in the 4FS and cause large perturbative corrections.
Contributions proportional to \ytsq{}, on the other hand, feature a logarithmic enhancement of $g\to b\bar{b}$
splittings. These logarithms are taken into account for the first time up to NLO QCD
in this paper, and yield an
important correction to the cross section.
\item Given the large QCD corrections, it is not surprising that perturbative uncertainties estimated
from scale variations are relatively large as well. As expected they are largest for \ytsq{} terms.
The inclusion of NLO corrections reduces the uncertainties significantly, but they are still at
the level $30$\% to $40$\%. Their main source is again the logarithmic enhancement of the individual
contributions pointed out above.
\end{itemize}

\begin{figure}[tp]
\begin{center}
\subfloat[]{\includegraphics[trim = 13mm 8.2cm 2cm 2cm, width=.33\textheight]{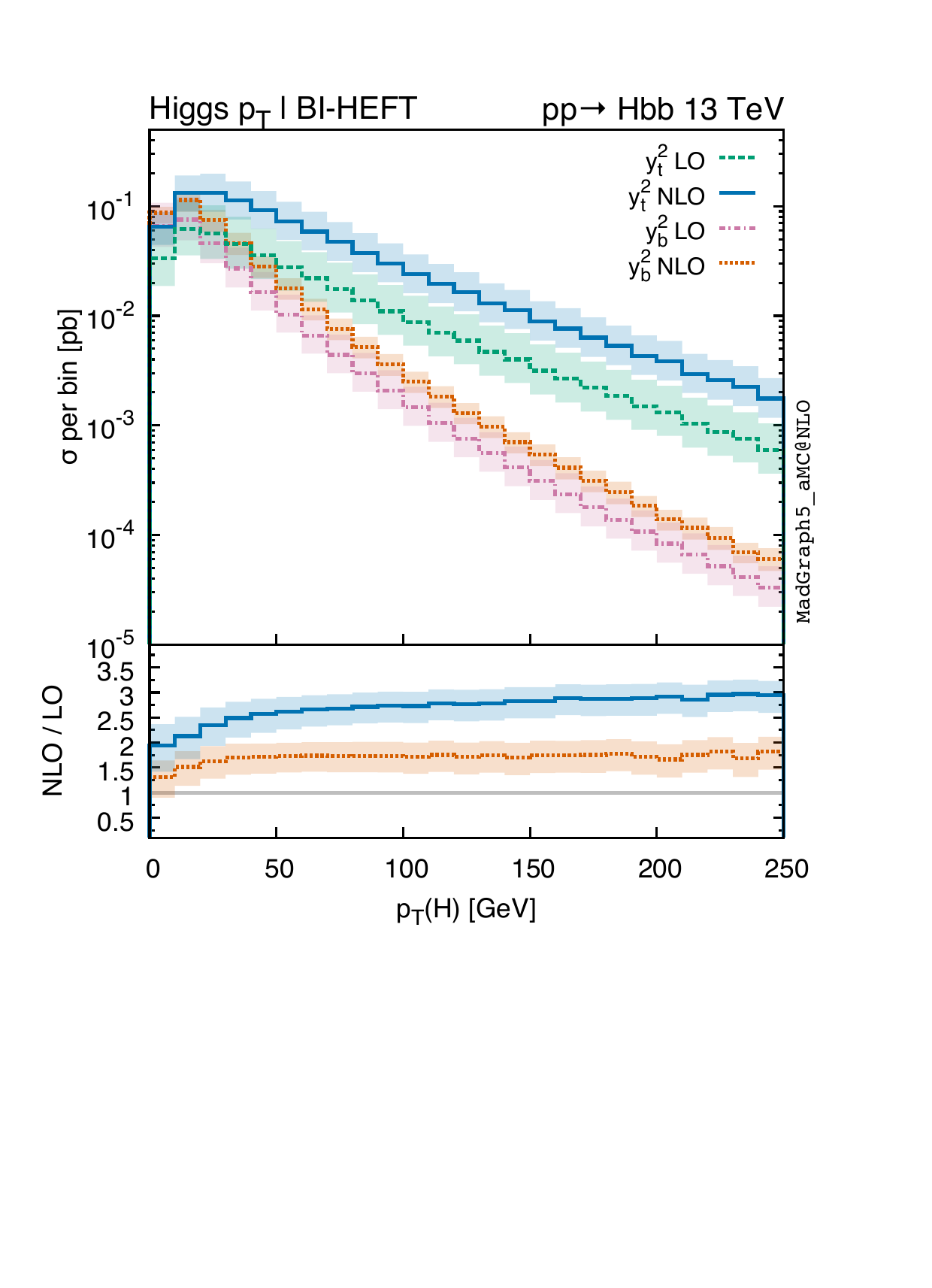}\label{fig:pt-h}}\hspace*{0.5cm}
\subfloat[]{\includegraphics[trim = 13mm 8.2cm 2cm 2cm, width=.33\textheight]{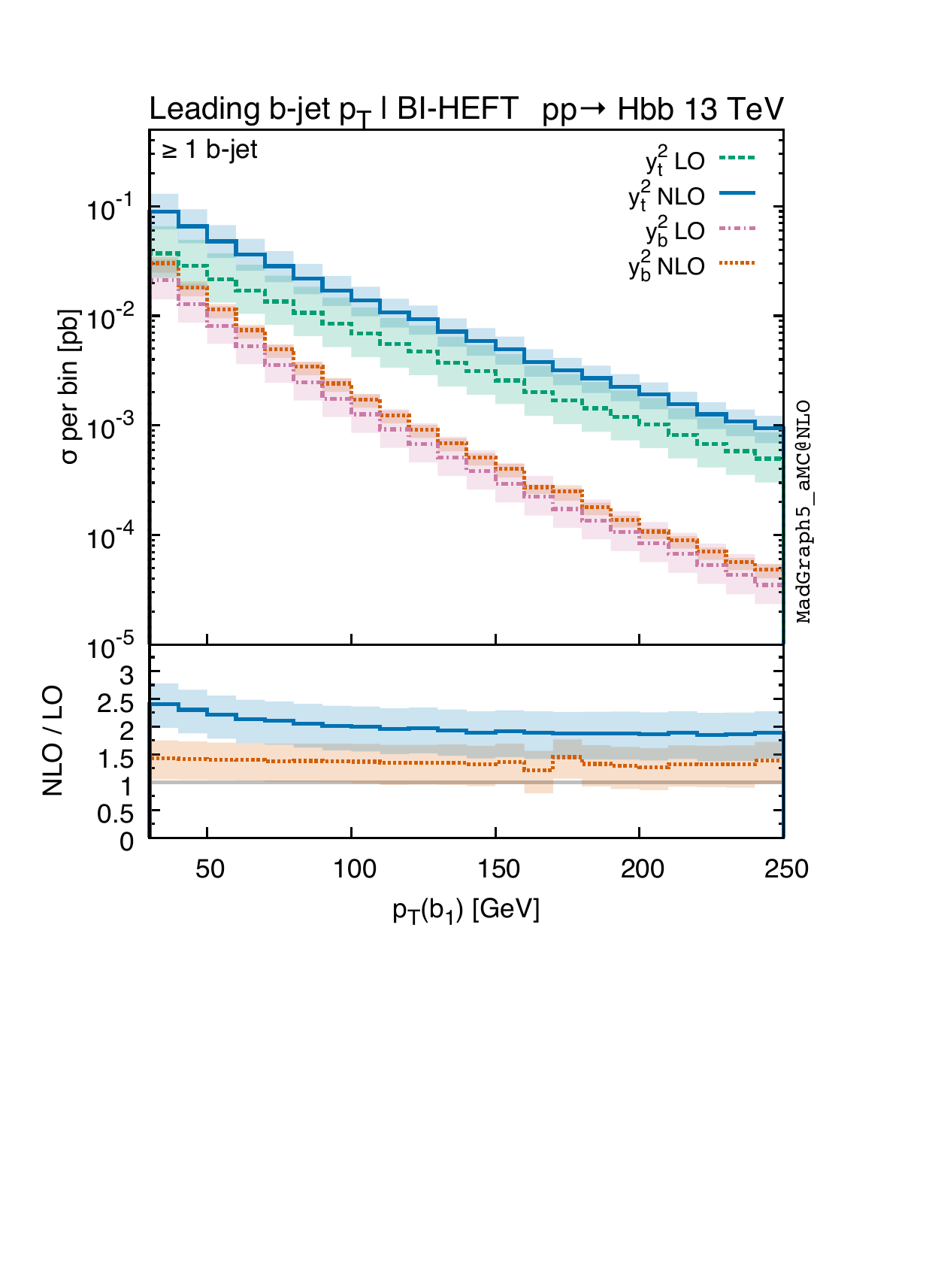}\label{fig:pt-b1}}
\caption[]{\label{fig:pt}{Distributions in the transverse momentum of the Higgs boson (\ref{fig:pt-h})
and of the hardest $b$ jet (\ref{fig:pt-b1}). See the text for details.}}
\end{center}
\begin{center}
\subfloat[]{\includegraphics[trim = 13mm 8.2cm 2cm 2cm, width=.33\textheight]{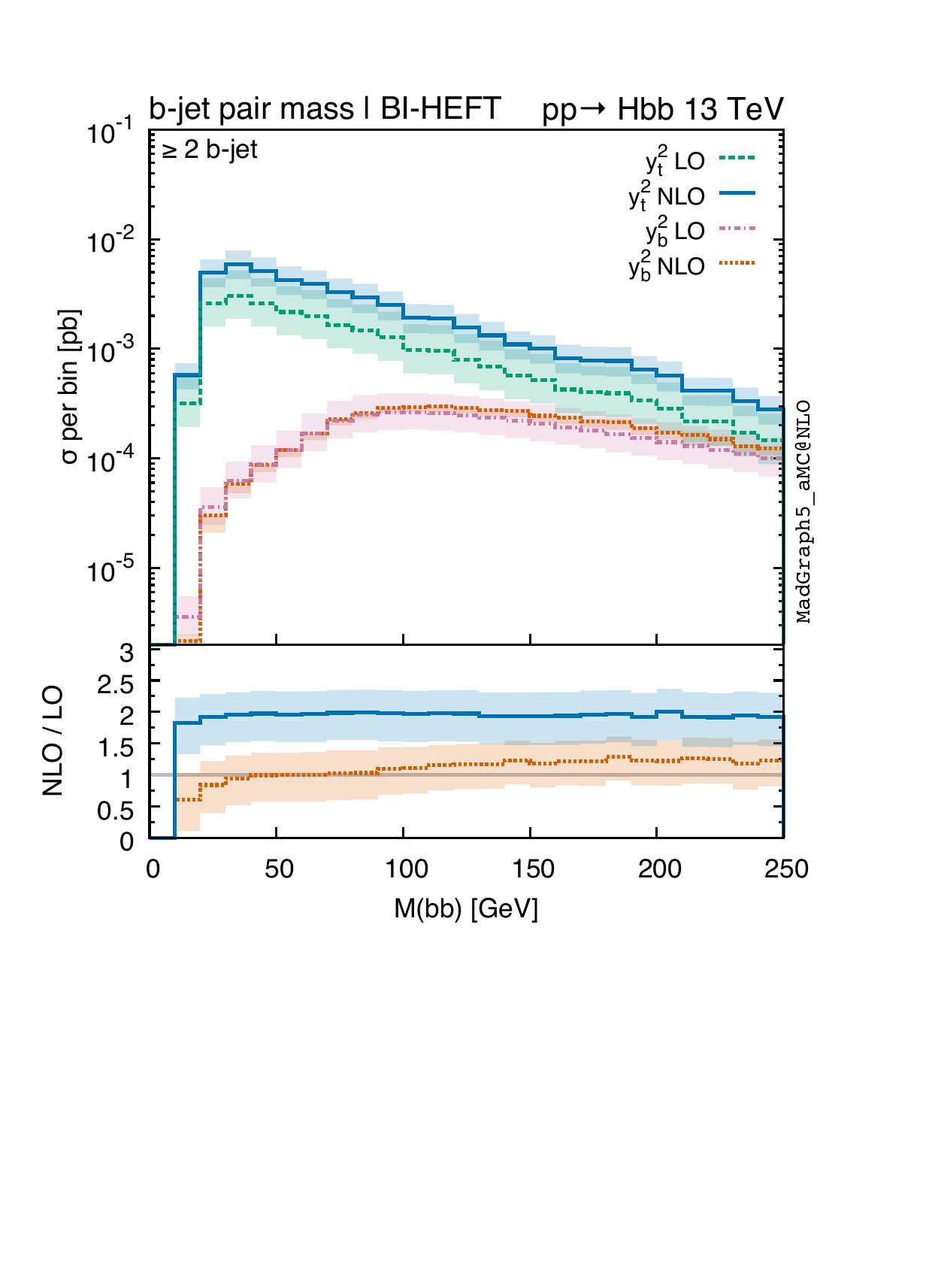}\label{fig:bb-m}}\hspace*{0.5cm}
\subfloat[]{\includegraphics[trim = 13mm 8.2cm 2cm 2cm, width=.33\textheight]{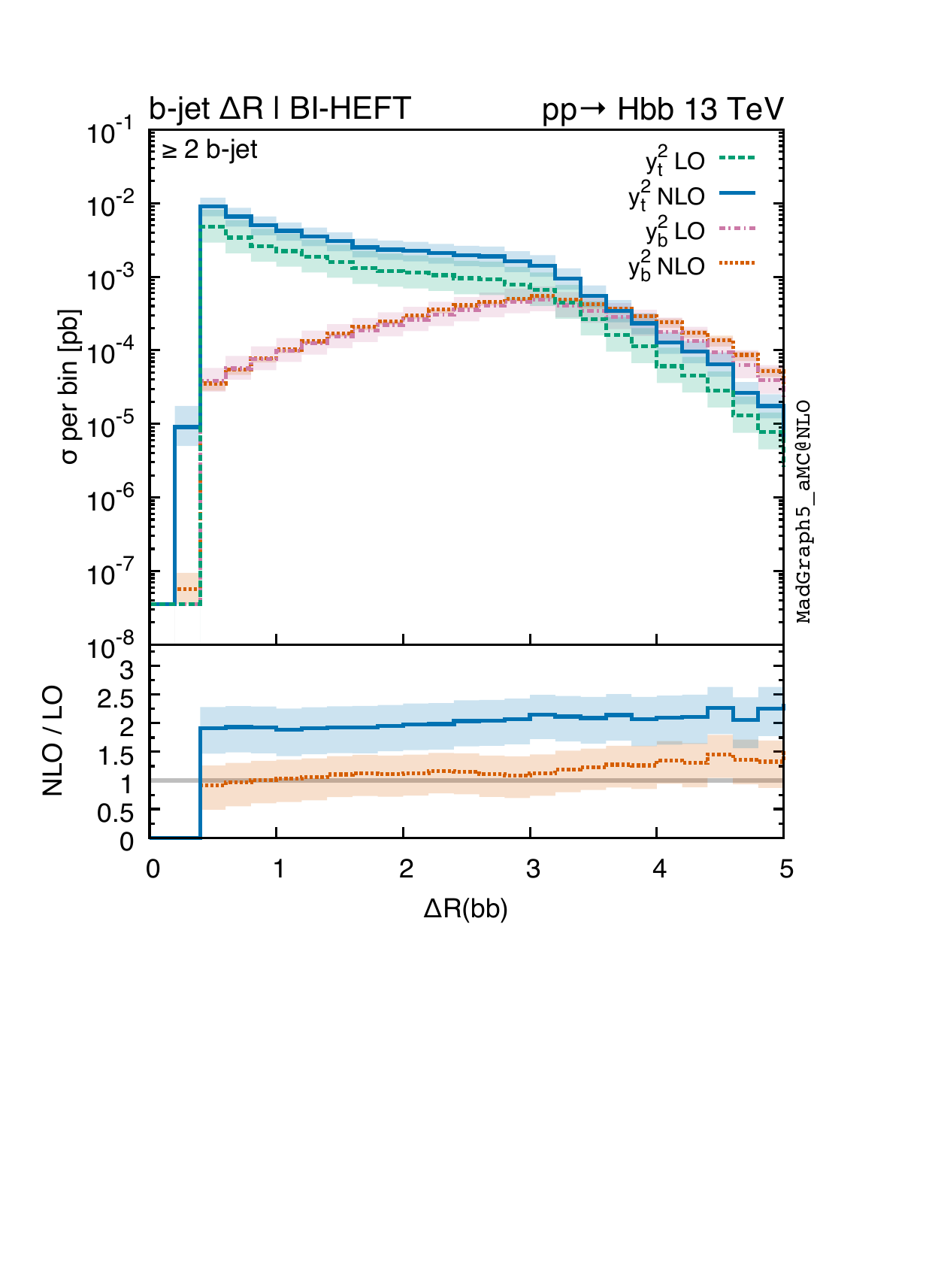} \label{fig:bb-dr}}
\caption[]{\label{fig:bb}{Same as in \fig{fig:pt}, for the invariant mass of the two $b$ jets (\ref{fig:bb-m}) and their distance in the $\eta$--$\phi$ plane (\ref{fig:bb-dr}).}}
\end{center}
\end{figure}

We now turn to discussing differential distributions. We first consider the NLO/LO $K$-factor of the different contributions
to the cross section in \figs{fig:pt} and \ref{fig:bb}. These figures are organised according
to the following pattern: there is a main frame, which shows histograms of
the LO \ytsq{} (green dashed),
NLO \ytsq{} (blue solid), LO \ybsq{} (purple dash-dotted), and NLO \ybsq{} (red dotted)
predictions as cross section per bin (namely, the sum of the values of the bins is equal to the
total cross section, possibly within cuts). In an inset we display the $K$-factor for each
contribution by taking the bin-by-bin ratio of the NLO histogram which appears in the
main frame over the LO one. The bands correspond to the residual uncertainties
estimated from scale variations according to \sct{sec:inputs}.

\Fig{fig:pt-h} shows the transverse momentum distribution of the Higgs boson. As expected from
the general hardness of the two different production processes leading to \ytsq{} and \ybsq{} (the Higgs
being radiated off a top-quark loop, and the Higgs boson being coupled to a bottom-quark line), the latter
features the significantly softer \pth{} spectrum. The $K$-factors for both contributions grow with the
value of \pth{} similarly, and become quite flat at large transverse momenta. However, as observed before,
the size of QCD corrections is larger for the \ytsq{} terms, ranging from $K\approx 2$ at small \pth{}
to $K\approx 3$ for $\pth{}\gtrsim 150\,\giga\electronvolt$. For \ybsq{} contributions they are $K\approx 1.5$ and
 $K\approx 2$ in the same regions.

Also the transverse-momentum distribution of the leading $b$ jet in \fig{fig:pt-b1} displays a harder spectrum for the \ytsq{} contributions.
Note that for the \ptbone{} distribution, as we select the leading $b$ jet,
the integral of the distribution
corresponds to the $\ge 1b$-jet rate. The behaviour of the $K$-factor is quite different in this case. While it is essentially flat and about $K=1.5$ for
the \ybsq{} terms, it is about $K=2.5$ for \ytsq{} at low \ptbone{}, decreases to $K\approx 2$ for $\ptbone{}= 150\,\giga\electronvolt$, and turns flat afterwards.

In \fig{fig:bb} we consider two observables which require the presence of at least two $b$ jets: the left panel,
\fig{fig:bb-m}, shows the
invariant-mass distribution of the $b$-jet pair, \Mbb{}, and the right panel, \fig{fig:bb-dr}, shows their distance in the $\eta$--$\phi$ plane, \Rbb{}.
It is interesting to notice the very different behaviour of \ybsq{} and \ytsq{} terms in the main frame of these two distributions.
While \ytsq{} clearly prefers small invariant-masses and small separations between the $b$ jets, \ybsq{} peaks around $\Mbb=100\,\giga\electronvolt$
and $\Rbb=3$. The reason is clear: the dominant contribution for \ytsq{} originates from the
$g\to b\bar{b}$ splittings, which
generate bottom-quark pairs that are hardly separated and, hence, also have a rather small invariant mass.
Looking at the $K$-factors in the lower inset, the one of the \ybsq{} terms turns out to be rather close to one for a large part
of the phase space. It slightly increases with both \Mbb{} and \Rbb{}. For \ytsq{}, on the other hand, the $K$-factor is around $K=2$, and shows an even milder increase with \Mbb{} and \Rbb{}.

The scale-uncertainty bands in all four plots show the same features as the scale uncertainties discussed in \tab{tab:rates}:
Their size decreases upon inclusion of higher-order corrections, but overall they are rather large even at NLO.
The \ytsq{} contributions feature a stronger scale dependence due to the logarithmic enhancement of $g\to b\bar{b}$ splittings.
By and large, LO and NLO at least have some overlap in most cases. Nevertheless, \ybsq{} contributions show the better converging
perturbative series in that respect, which of course is directly related to their smaller QCD corrections.

\subsection{Accessing the bottom-quark Yukawa coupling}\label{sec:yb}

\begin{figure}[tp]
\begin{center}
\subfloat[]{\includegraphics[trim = 13mm 7.2cm 2cm 2cm, width=.245\textheight]{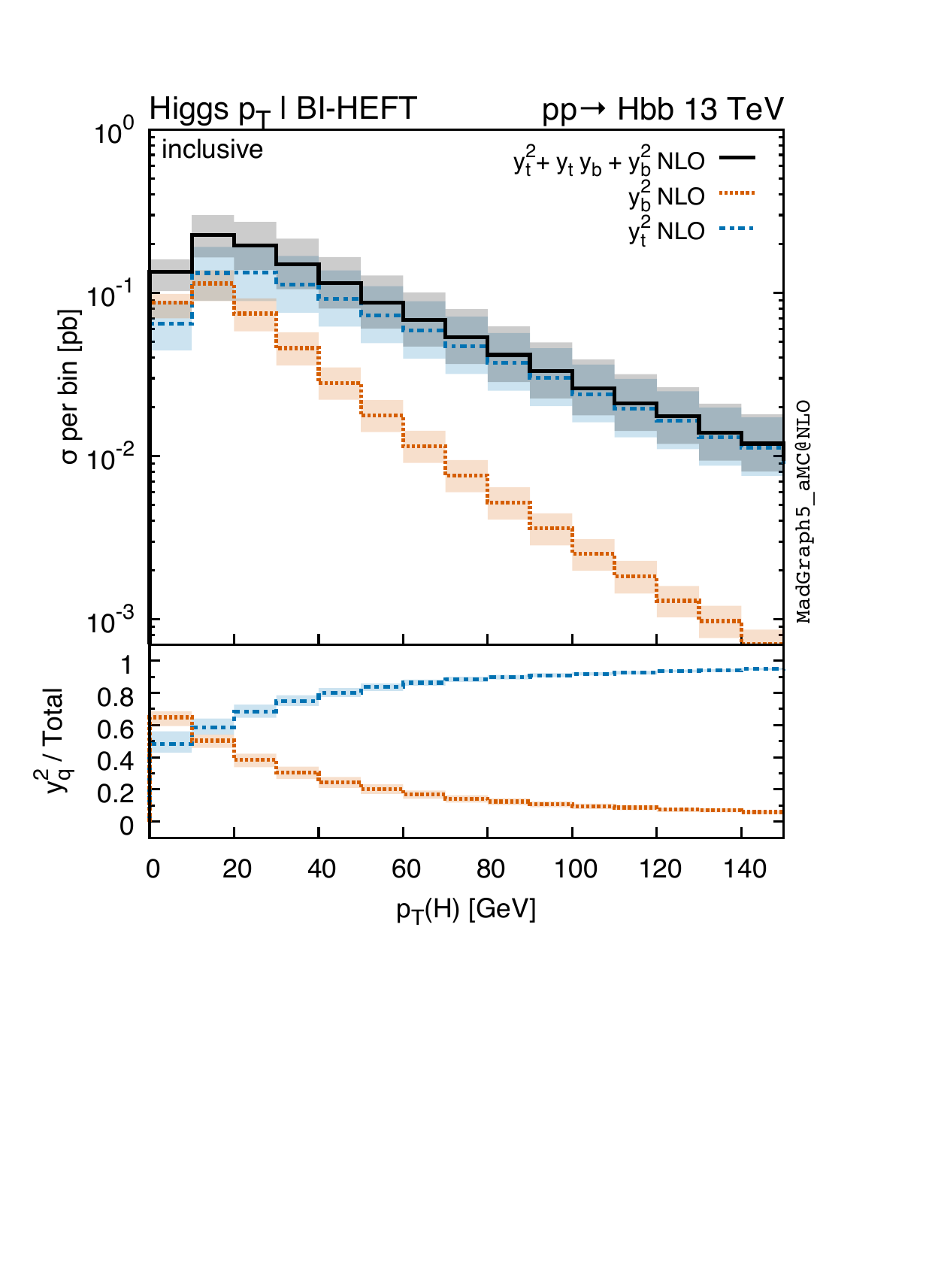}\label{fig:pthvs-nocut}}
\subfloat[]{\includegraphics[trim = 13mm 7.2cm 2cm 2cm, width=.245\textheight]{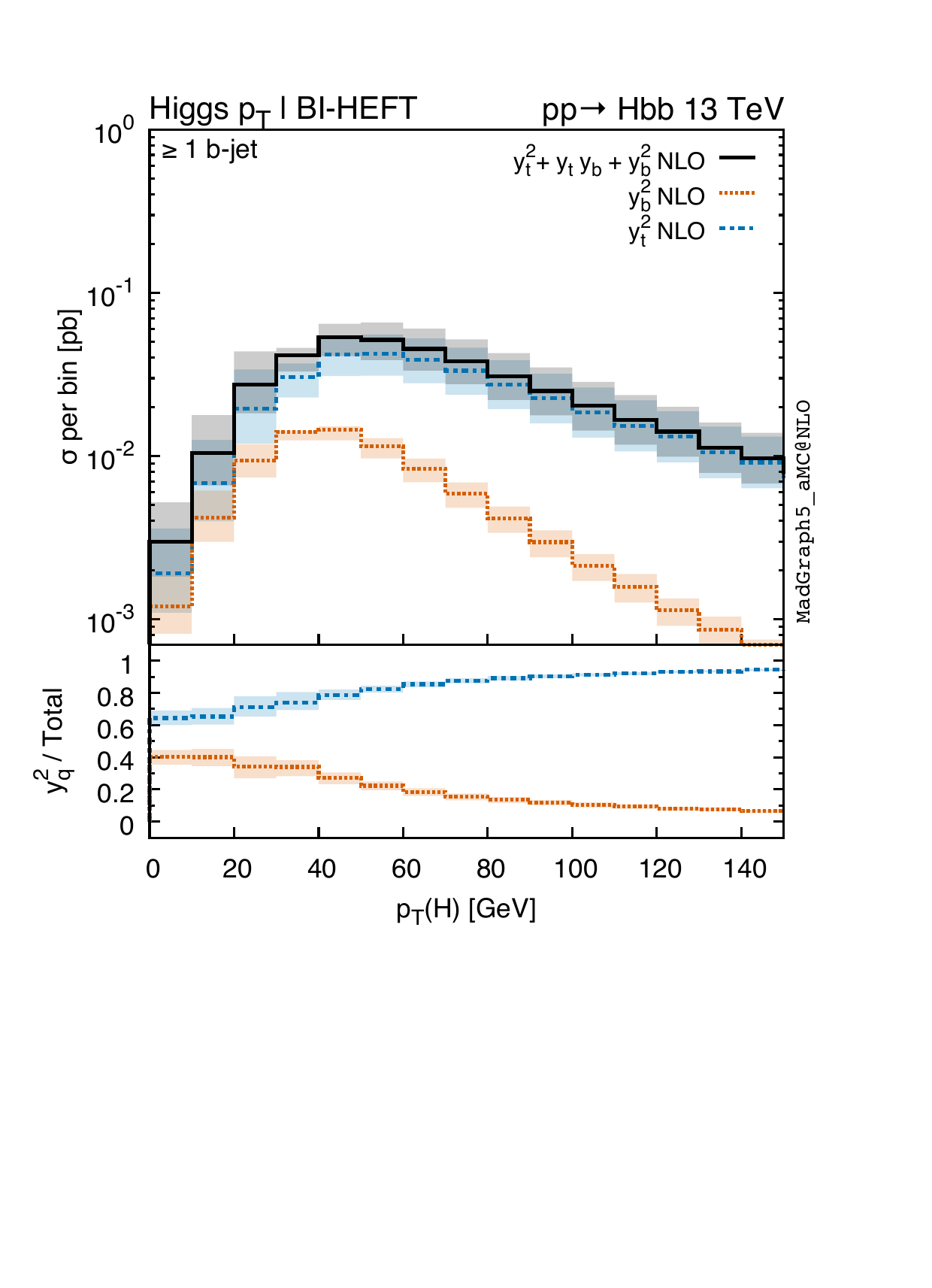}\label{fig:pthvs-1b}}
\subfloat[]{\includegraphics[trim = 13mm 7.2cm 2cm 2cm, width=.245\textheight]{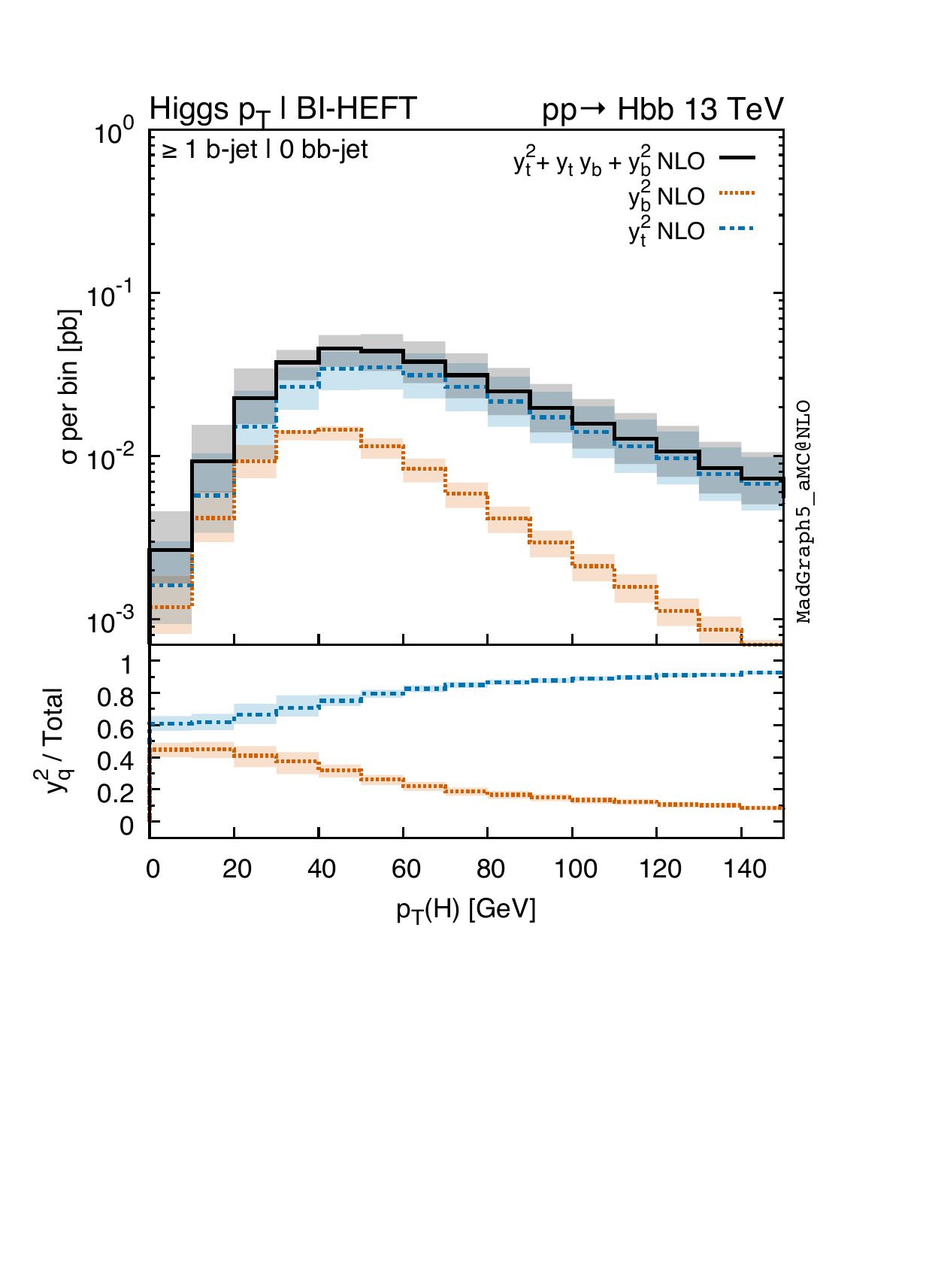}\label{fig:pthvs-1b0bb}}
\caption[]{\label{fig:pthvs}{Distributions in the transverse momentum of the Higgs boson
in three categories: inclusive (\ref{fig:pthvs-nocut}), $\ge 1b$-jet (\ref{fig:pthvs-1b}), and $\ge 1b$-jet $|$ $0\,bb$-jets (\ref{fig:pthvs-1b0bb}). See the text for details.}}
\end{center}
\end{figure}

\begin{figure}[tp]
\begin{center}
\subfloat[]{\includegraphics[trim = 13mm 8.2cm 2cm 2cm, width=.33\textheight]{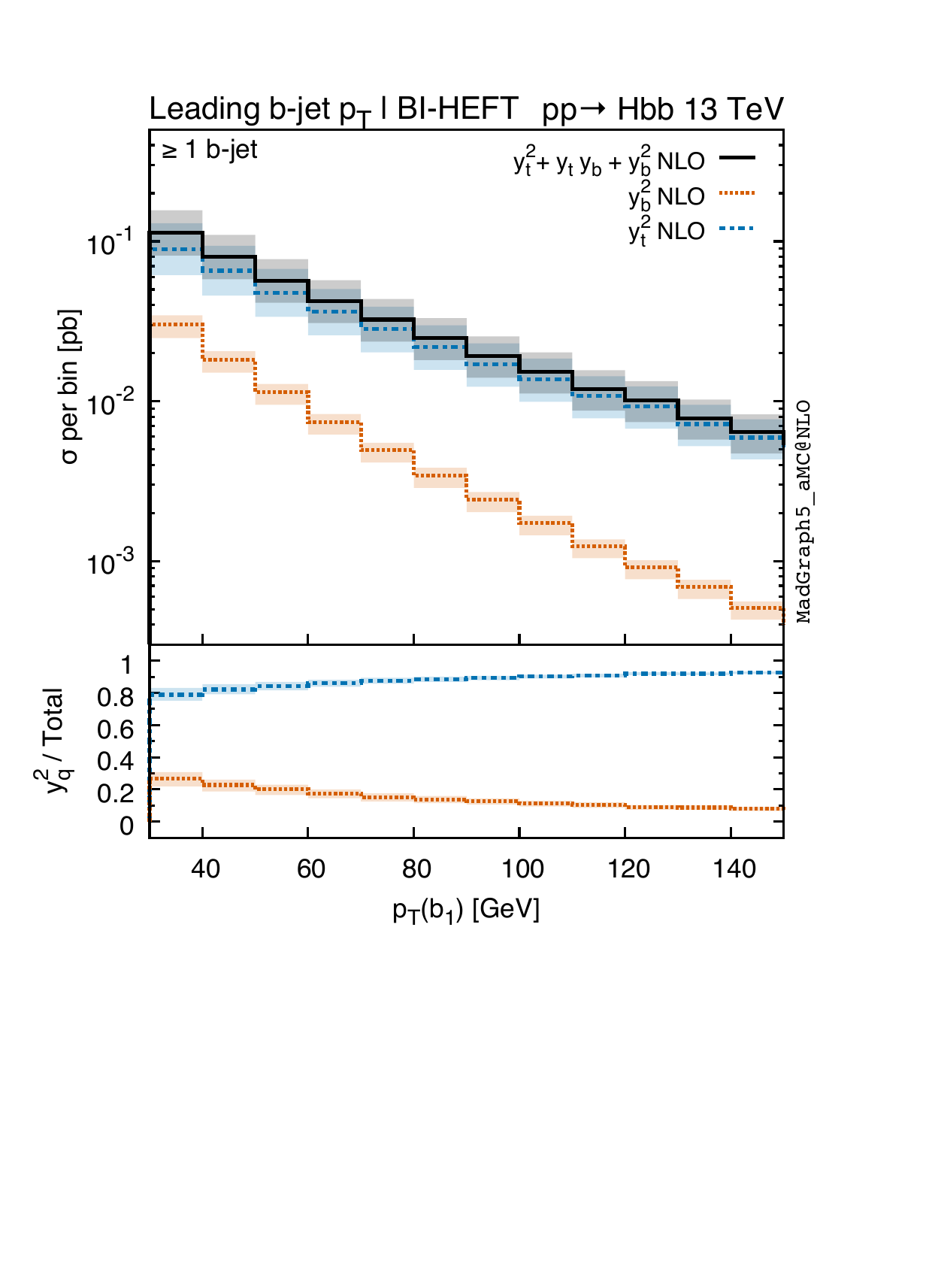}\label{fig:ptbvs-1b}}\hspace*{0.5cm}
\subfloat[]{\includegraphics[trim = 13mm 8.2cm 2cm 2cm, width=.33\textheight]{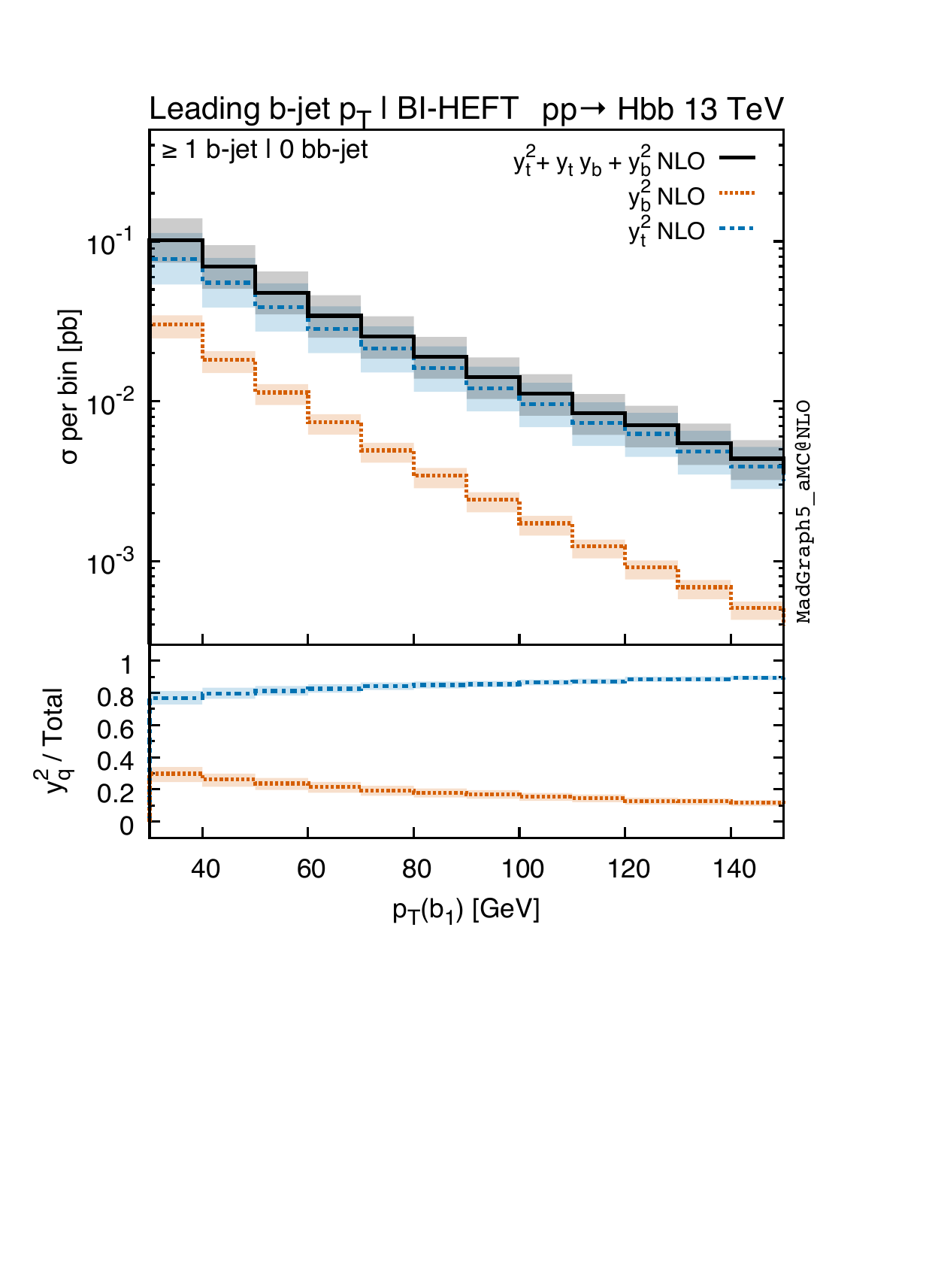}\label{fig:ptbvs-1b0bb}}
\caption[]{\label{fig:ptbvs}{Same as \fig{fig:pthvs}, for the transverse momentum of the hardest $b$ jet. Note that the inclusive and $\geq 1$ b-jet categories yield identical distributions and we show only the latter.}}
\end{center}
\end{figure}

\begin{figure}[h]
\vspace{1cm}
\begin{center}
\subfloat[]{\includegraphics[trim = 13mm 8.2cm 2cm 2cm, width=.33\textheight]{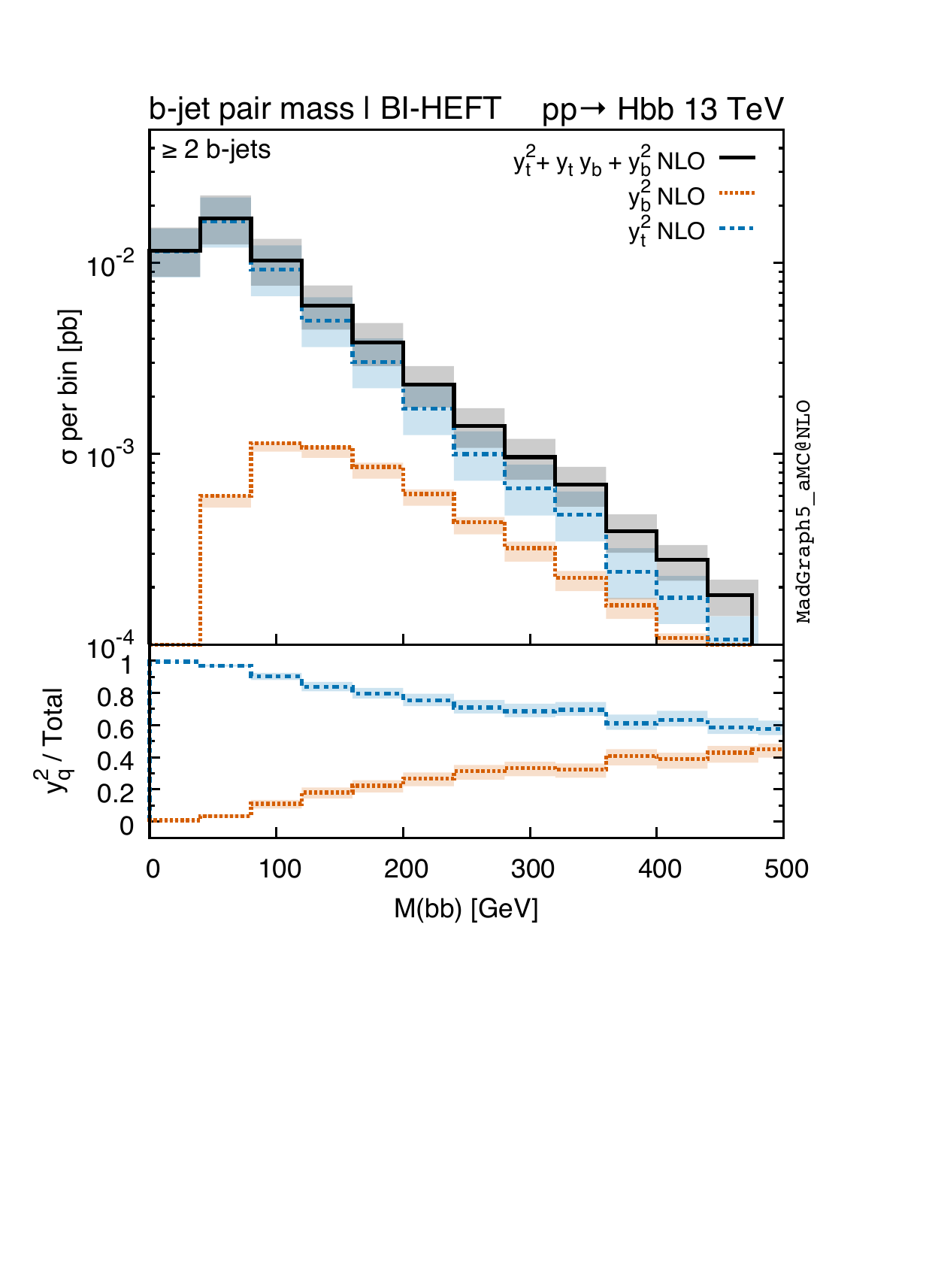}\label{fig:bbvs-mbbj}}\hspace*{0.5cm}
\subfloat[]{\includegraphics[trim = 13mm 8.2cm 2cm 2cm, width=.33\textheight]{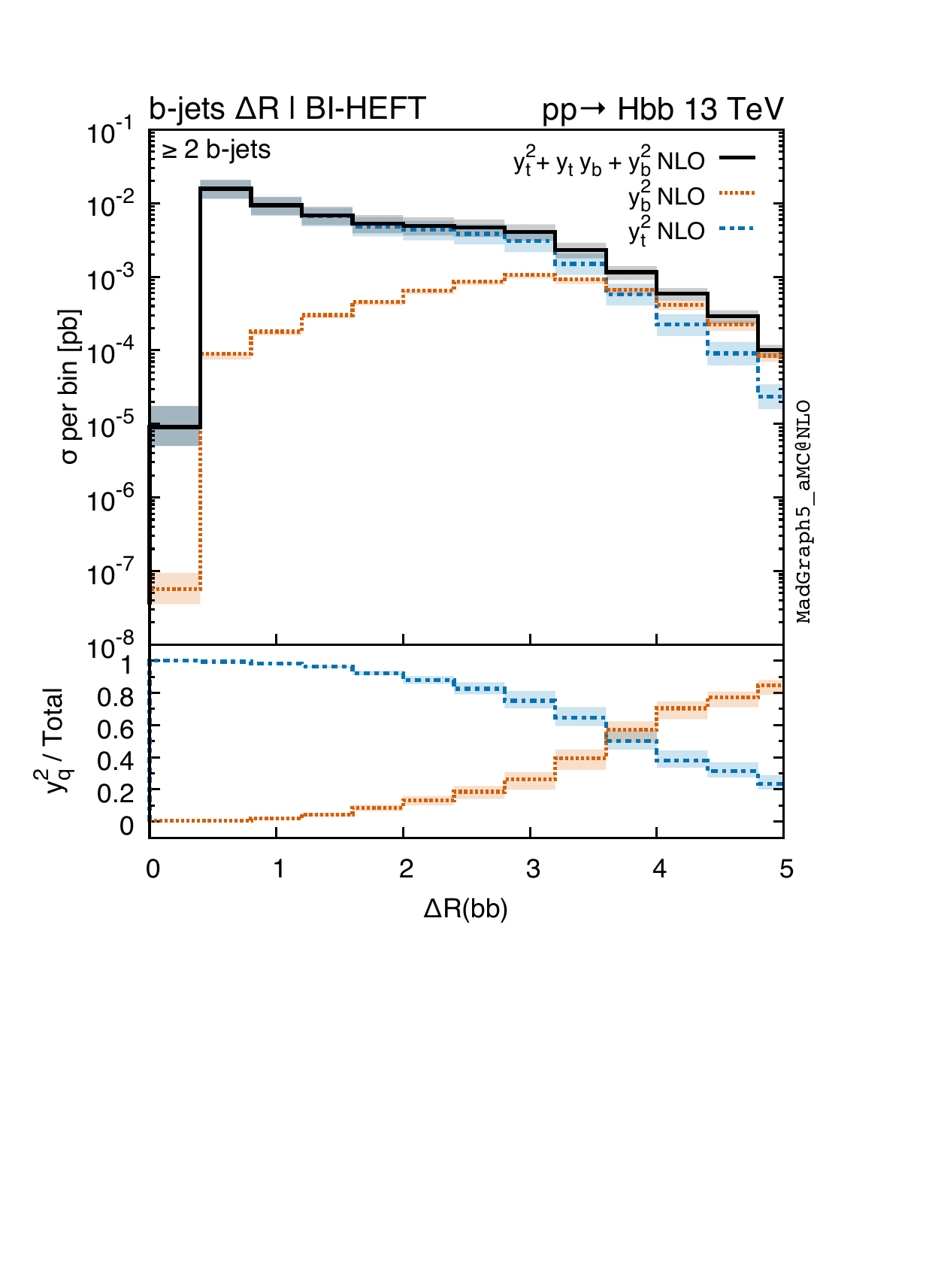}\label{fig:bbvs-drbbj}} \\
\subfloat[]{\includegraphics[trim = 13mm 8.2cm 2cm 2cm, width=.33\textheight]{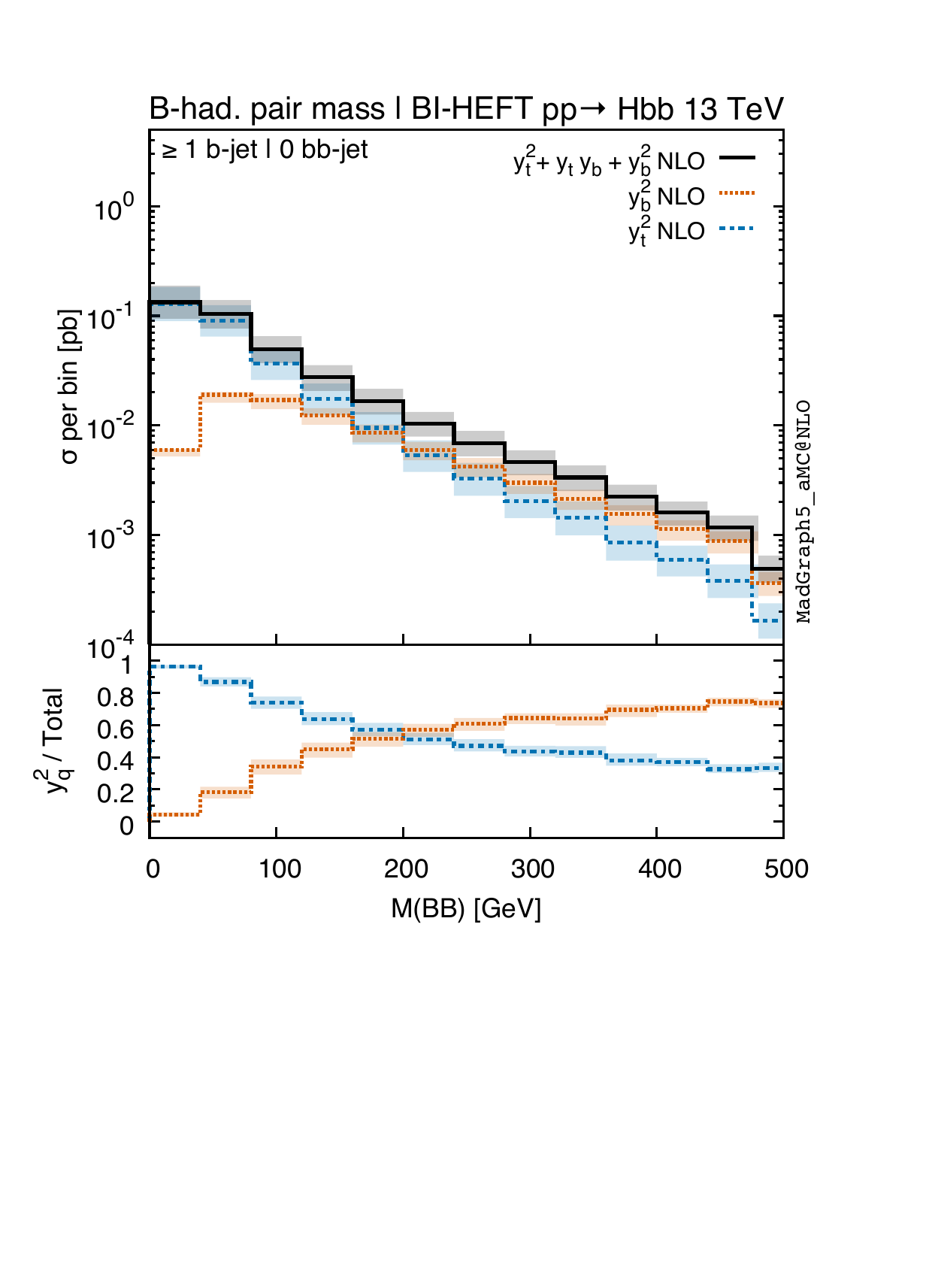}\label{fig:bbvs-mbbh}}\hspace*{0.5cm}
\subfloat[]{\includegraphics[trim = 13mm 8.2cm 2cm 2cm, width=.33\textheight]{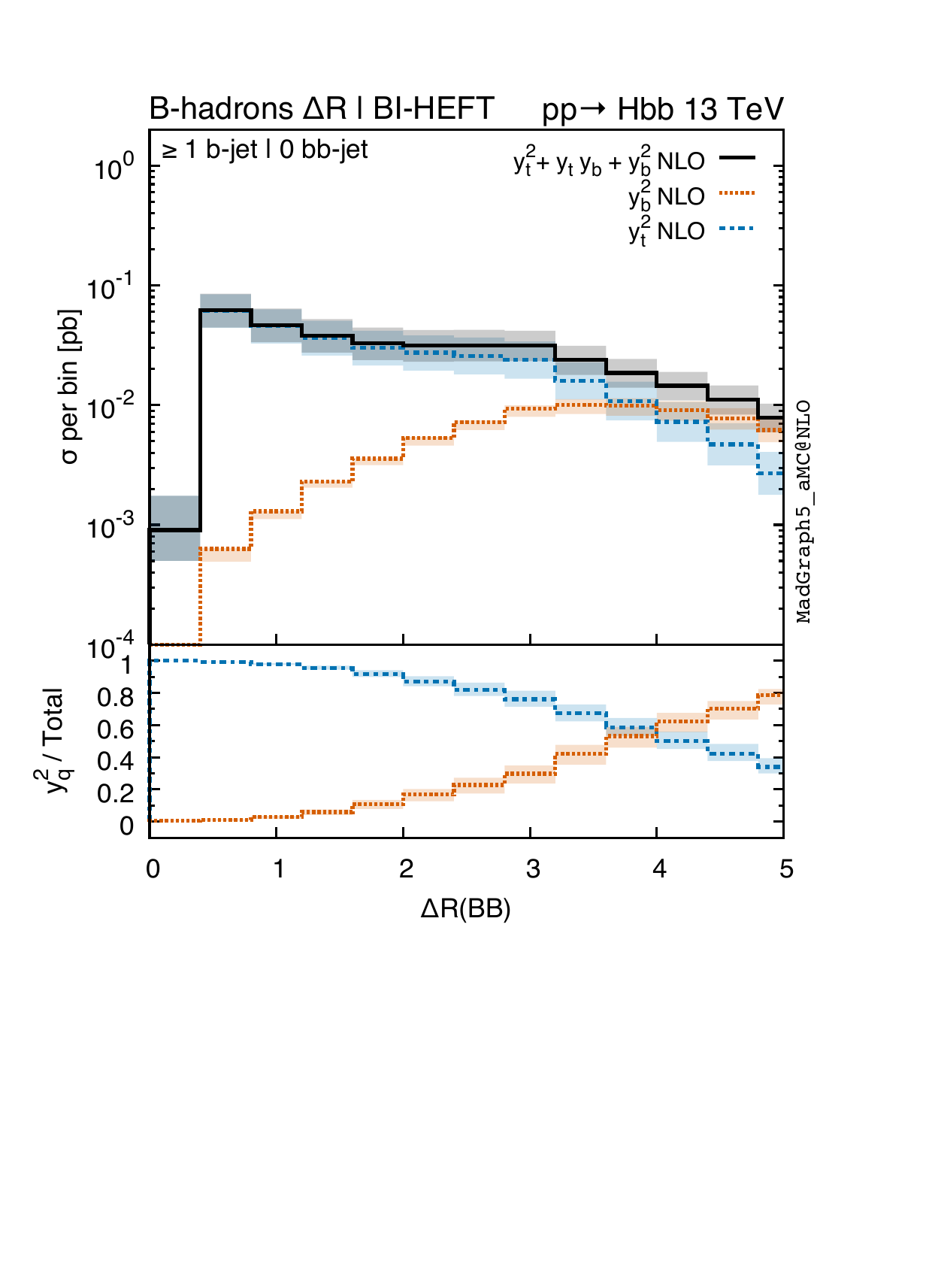}\label{fig:bbvs-drbbh}}
\caption[]{\label{fig:bbvs}{Same as \fig{fig:pthvs-1b0bb}, for the invariant mass of the two $b$ jets (\ref{fig:bbvs-mbbj}), their distance in the $\eta$--$\phi$ plane (\ref{fig:bbvs-mbbj}), and the same distributions for $B$ hadrons (\ref{fig:bbvs-mbbh} and \ref{fig:bbvs-drbbh}).}}
\end{center}
\vspace{0.8cm}
\end{figure}

We now return to the question of how to improve the sensitivity to the bottom-quark Yukawa coupling in \bbH{} production.
The goal is to increase the relative contribution of the \ybsq{} terms by suitable selections, while keeping the absolute value
of the cross section as large as possible. First, in order to be able to distinguish \bbH{} from inclusive Higgs
production, we require to observe at least one $b$ jet. Second, we have already noticed that by removing all $b$ jets containing a
pair of bottom quarks, we can decrease the \ytsq{} rate to some extent, with a negligible impact on the \ybsq{} rate.
The combination with additional phase-space requirements provides the most promising approach to further improve the sensitivity to the
bottom-quark Yukawa coupling and to recover \bbH{} production as the best process to measure \yb{} directly. To this end, we consider the relative contribution of
\ybsq{} and \ytsq{} terms to the \bbH{} cross section for various differential observables in \figs{fig:pthvs}--\ref{fig:bbvs}.
All the plots in these figures have a similar layout: the main frame shows NLO predictions for the \ytsq{} contribution (blue dash-dotted),
the \ybsq{} contribution (red dotted), and the sum of all contributions, including the interference (black solid). The ratio inset shows the relative contributions of
the \ybsq{} and \ytsq{} terms to \bbH{} cross section. The bands reflect the residual uncertainties
estimated from scale variations according to \sct{sec:inputs}.

We start in \fig{fig:pthvs} with the transverse-momentum distribution of the Higgs boson. The three panels show this distribution
with different requirements: \fig{fig:pthvs-nocut} displays the inclusive spectrum, \fig{fig:pthvs-1b} is in the $\ge 1b$-jet category, and \fig{fig:pthvs-1b0bb} is in the same category, but
vetoing $bb$ jets. This observable constitutes one of the strongest discriminators between
\ybsq{} and \ytsq{} terms. The reason is the significantly softer spectrum of the terms
proportional to \ybsq{}, which we already observed before. By looking at the
three plots in \fig{fig:pthvs} one can infer that the relative \ybsq{} contribution is maximal in the inclusive
case and at low Higgs transverse momentum, even exceeding 50\% in the lowest part of the spectrum. If
we require at least one $b$ jet the situation becomes worse, with the \ybsq{} term reaching at most 40\%, again
in the lowest transverse-momentum bins. If we require at least one $b$ jet and veto $bb$ jets, the relative contribution
 of \ybsq{} at low transverse momentum is mildly increased. In the three cases (inclusive, $\ge 1$ $b$ jet and
 $\ge 1$ $b$ jet without $bb$ jets) the relative \ybsq{} contribution quickly decreases with $\pth{}$, being less
 than 20\% already at $\pth{} = 50\,\giga\electronvolt$. At the level of the cross section, in the $\ge 1b-$jet category,
 the relative \ytsq{} and \ybsq{} contributions are respectively 81\% and 19\%.
 In the $\ge 1b-$jet and no $bb$-jet category their relative contributions
 become $\sim 77\%$ and $\sim 23\%$, respectively. All in all, the gain coming from vetoing $bb$ jets is moderate. Another strategy, which can be combined with
 the $bb$-jet veto, consists in discarding
 events with the Higgs transverse momentum larger than a given value. For example, with an upper cut on \pth{} at $50\,(100)\,\giga\electronvolt$, in the category with at least one $b$ jet
and no $bb$ jet,
we can increase the relative contribution of \ybsq{} terms to about $36\% \, (27\%)$, while keeping
about $50\% \, (90\%)$ of its rate.
Hence, restricting the phase space
to small \pth{} values allows us to increase the relative size of \ybsq{} terms, while the impact
on the rate is moderate due to the quite strong suppression at large \pth{}.

We continue in \fig{fig:ptbvs} with the transverse-momentum distribution of hardest $b$ jet.
The general features of the \ptbone{} spectrum are similar to the ones of \pth{}. However, as this observable
clearly does not help  very much in distinguishing between \ybsq{} and \ytsq{} contributions, we
do not suggest any additional cut on \ptbone{}. It becomes clear from these plots, though, that
a lower \ptb{} threshold used in the definition of $b$ jets would increase the relative
size of the \ybsq{} terms. In the present study jets are defined with a \ptj{} threshold of $30\,\giga\electronvolt$. A value of $25\,\giga\electronvolt$ or even $20\,\giga\electronvolt$  could be feasible at the LHC, and would further increase the sensitivity to the bottom-quark Yukawa coupling in \bbH{} production. We note that additional modifications of the $b$-jet definition,
for example the usage of a different jet radius (as shown in appendix~\ref{app:r1}), or of jet-substructure techniques, can
provide further handles to improve the discrimination of the \ybsq{} contribution.

Finally, we consider \fig{fig:bbvs}, where we show the
invariant-mass distributions of the two $b$ jets, \fig{fig:bbvs-mbbj}, and their distance $R$, \fig{fig:bbvs-drbbj}, and the corresponding distributions for $B$ hadrons, \figs{fig:bbvs-mbbh} and \ref{fig:bbvs-drbbh}. Note that the
\Mbb{} and \Rbb{} distributions by construction require the presence of two $b$ jets,
while \MBB{} and \RBB{} are shown with at least one $b$ jet and no $bb$ jet.
Clearly, all of these observables, especially those related to $B$ hadrons, could in principle provide information  to discriminate between \ybsq{} and \ytsq{} contributions. However, in practice, their usefulness is limited, due to two main reasons, both related to statistics. First, the two $b$-jet distributions
require the presence of at least two $b$ jets and the corresponding rate is significantly reduced (by roughly one order of magnitude) with respect to the $\ge 1b$-jet one. Second, the bulk of the $B$-hadron distributions feature $B$ hadrons which are quite soft, and
therefore possibly not accessible in the measurements. We therefore conclude that, despite showing
useful features, neither of these distributions can significantly help in obtaining additional
sensitivity to the bottom-quark Yukawa coupling.

\subsection{QCD corrections to $g\to b\bar{b}$ splitting}\label{sec:gbb}
Besides its phenomenological relevance for the extraction of the bottom-quark Yukawa or as a background to other Higgs production processes at the LHC, \bbH{} production induced by the top-quark Yukawa coupling offers a clean and simple theoretical setting to study the dynamics of $g\to b\bar{b}$ splitting in presence of a hard scale. Cases of interest at the LHC where such splitting plays an important role, and is in fact one of the main sources of uncertainties, are $t\bar t b \bar b$~\cite{Cascioli:2013era,Jezo:2018yaf} and $b \bar b Z$~\cite{Krauss:2016orf,Bagnaschi:2018dnh}, {\it i.e.}, irreducible backgrounds for Higgs production in the $t\bar t H$ and $ZH$ production modes, respectively. In these cases, however, the production of a pair of $b$ quarks proceeds through different mechanisms and it is difficult to study them independently.\footnote{One must bear in mind that besides applications in certain
LHC processes, proper modelling of $g\to b\bar{b}$ splitting plays an important
role also in the context of parton showers.}

In the previous sections, we have found that \bbH{} production is dominated by the top-quark Yukawa contribution.   At the lowest order of the \ytsq{} contribution the Higgs boson recoils against the $b \bar b$ pair coming from a gluon splitting, either in the initial or in the final state. When the splitting occurs in the initial state, a gluon typically produces a $b$ quark going forward and the other interacting at high $Q^2$, while when the splitting happens  in the final state, a gluon has already been scattered at high $Q^2$. Therefore, asking a pair of $b$ quarks at high $\pt{}$ mostly selects the mechanism of gluon splitting in the final state.

This LO picture is modified at NLO, in particular in presence of real radiation where additional configurations can appear: most importantly, the Higgs can recoil against a hard light parton, with the $b\bar b$ pair being soft/collinear. Such configurations can give a large contribution as they are possibly enhanced by soft and/or collinear logarithms. In fact, rather than treating them as NLO corrections to \bbH, such contributions can be thought as higher-order corrections to Higgs + jet production. In this case, they can be described either with a parton-shower or by employing a gluon fragmentation function and its evolution. Both approaches resum (with different accuracy) large logarithms of the form $\log(\pt{}/ m_b)$, with $\pt{}$ being the transverse momentum of the light parton.

The \ytsq{} contribution to \bbH{} production enables a direct assessment of the
importance of these configurations. This can be done by studying very simple observables. To this aim, we consider the fraction of energy (or equivalently of transverse momentum) of the $b$ jet which is carried by particles other than the $b$ quark. At NLO accuracy, where only one extra light parton, dubbed $g$, can be emitted, this fraction can be defined for the $i$-th ($b$) jet as
\begin{equation}
    z_g(j_i) = \frac{p^T_g}{p^T_{j_i}}\,.
\end{equation}
Jets featuring hard $b$ quarks and soft gluon emissions are characterised by $z_g \ll 1$, while $b$ jets
with a hard light parton and soft $b$ quarks yield $z_g \simeq 1$. In the latter case, one would rather consider the $b$ quark to be originated from the evolution of the light parton.  The limiting cases are easily identified: $z_g=0$ means that the jets are constituted only by one or two b quarks, while $z_g=1$ corresponds to light jets where $g$ is the only constituent.

\begin{figure}[tp]
\begin{center}
\hspace*{-0.17cm}
\includegraphics[trim = 13mm 8.2cm 2cm 2cm, width=.33\textheight]{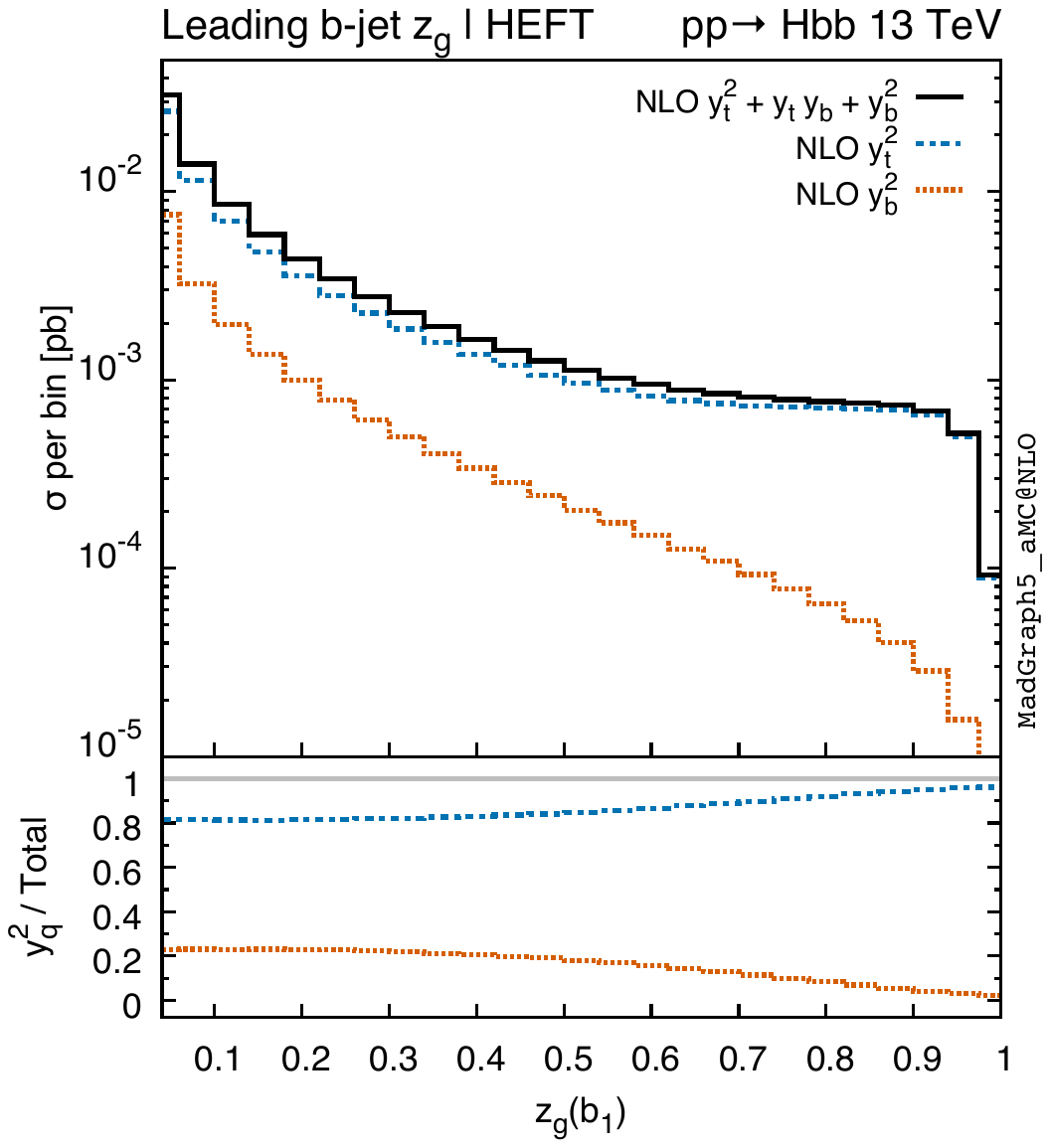}
\caption[]{\label{fig:zgybvseft}The momentum fraction of the first $b$ jet carried by light partons, $z_g^{b_1}$
in \bbH{} production.}
\end{center}\vspace{0.5cm}
\begin{center}
\subfloat[]{\includegraphics[trim = 13mm 8.2cm 2cm 2cm, width=.33\textheight]{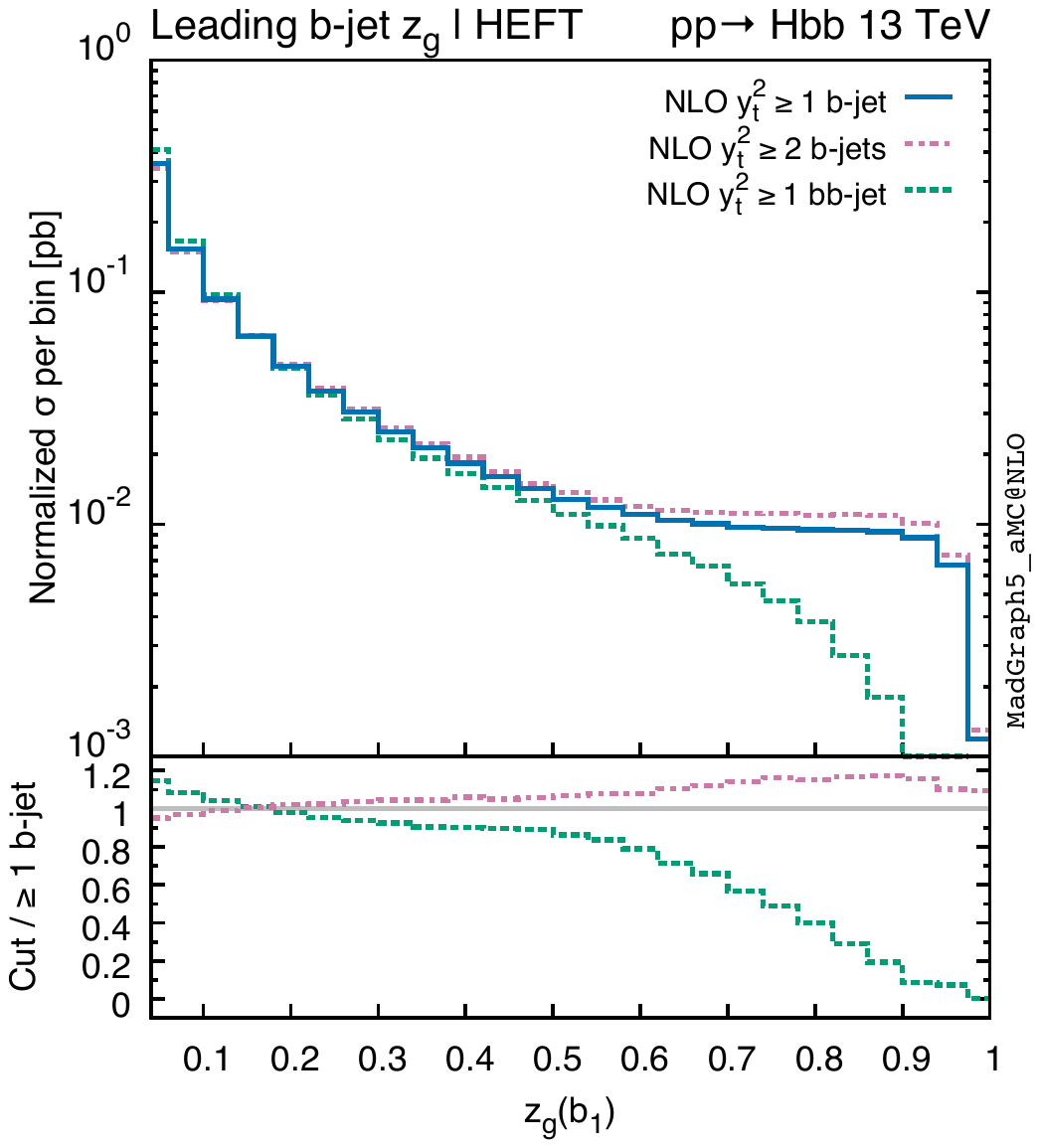}\label{fig:zgeftcuts-j}}\hspace*{0.5cm}
\subfloat[]{\includegraphics[trim = 13mm 8.2cm 2cm 2cm, width=.33\textheight]{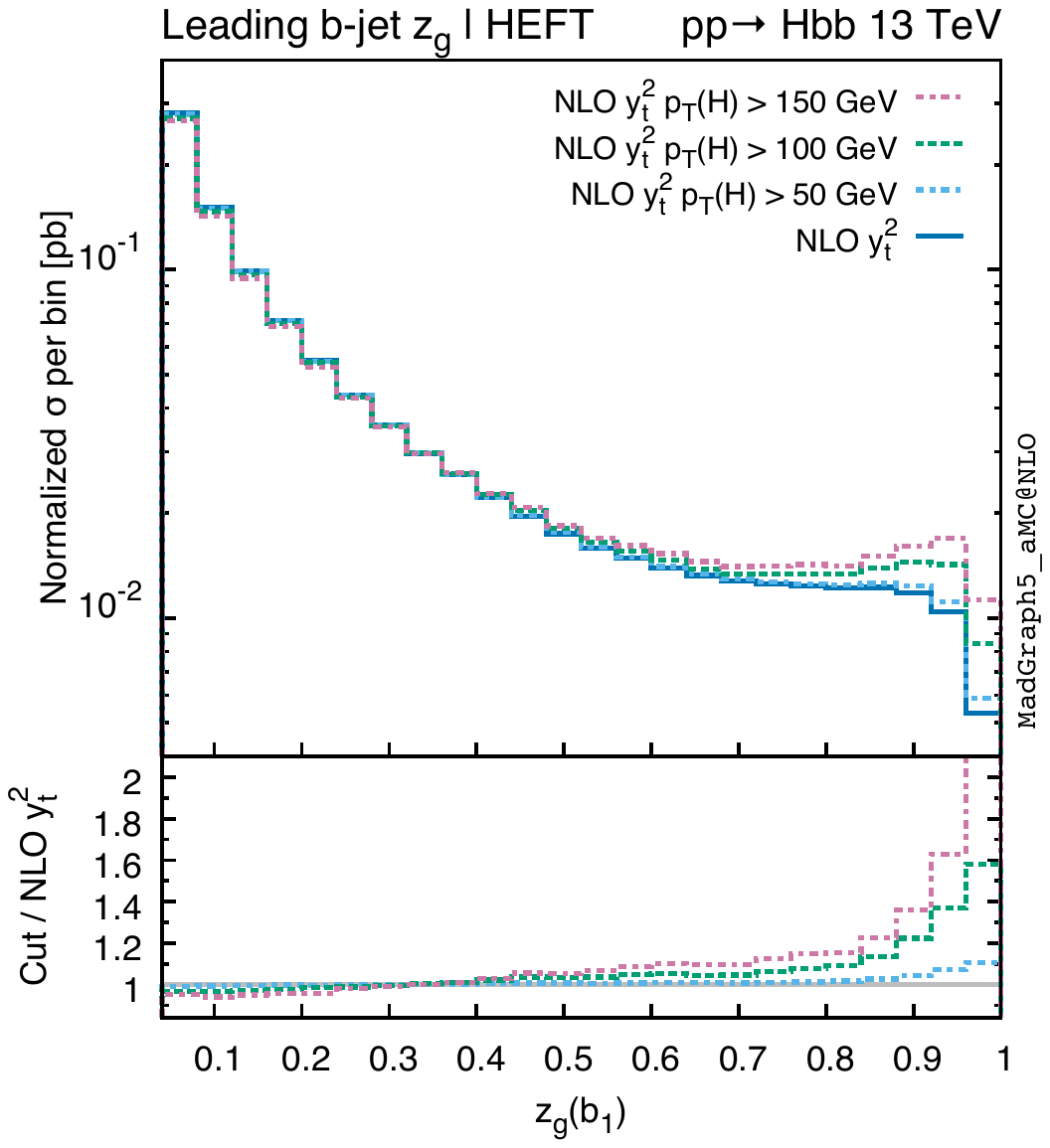}\label{fig:zgeftcuts-h}}
\caption[]{\label{fig:zgeftcuts}The momentum fraction of the first $b$ jet carried by light partons, $z_g^{b_1}$, for the \ytsq{} contribution to
\bbH{} production, for different $b$-jet acceptances (\ref{fig:zgeftcuts-j}) and different Higgs-$\pt$ cuts (\ref{fig:zgeftcuts-h}).}
\end{center}
\end{figure}

In the following, we consider the momentum fraction carried by light partons in the hardest $b$ jet,
$z_g(b_1)$. We start by showing,
in \fig{fig:zgybvseft}, the (normalised) $z_g(b_1)$ distribution for the $y_b^2$ (red dotted) and $y_t^2$ (blue dash-dotted) contributions
to \bbH{} production, as well as the complete \bbH{} cross section (black solid). The different behaviour in the two cases is manifest: for the $y_b^2$ contribution, $z_g(b_1)$
is monotonically decreasing, and configurations with a hard gluon inside the $b$ jet are very suppressed. On the contrary, the $y_t^2$ contribution shows a \emph{plateau} in the range  $0.6< z_g(b_1) < 0.95$. The integrated cross section for $ z_g(b_1) > 0.6$ amounts to 1.8\% of the $\ge 1$ b-jet one.

Next, we focus on the $y_t^2$ contribution, and consider the $z_g(b_1)$ distribution with different acceptance cuts. In \fig{fig:zgeftcuts-j}, we show the jet acceptances considered in
\sct{sec:predictions} and two $b$ jets while in \fig{fig:zgeftcuts-h} the Higgs transverse-momentum
cuts are shown. From the first figure, we conclude that requiring a $bb$ jet leads to a suppression of configurations with a hard light parton. This is reflected in the mild shape enhancement for $z_g(b_1)\simeq 1$ when a second, separate $b$ jet is required. In the second figure we appreciate how, at large Higgs transverse momenta, configurations with a hard light parton are more and more enhanced. When a minimum $\pth{}$ cut of 100 \giga\electronvolt\ is required, $z_g(b_1)$ even starts to increase as $z_g(b_1)> 0.8$. In this case, the fraction of cross section for $ z_g(b_1) > 0.6$ is 2.3\%, and it reaches 3\% for $\pth{}> 150\,\giga\electronvolt$. Although the relative importance of such configurations with respect of the total rate is marginal, the enhancement is manifest.

Being aware of the limitations of a fixed-order approach in this context, our findings could motivate a more complete study to explore the possibility of enhancing the $y_b^2$ contribution to \bbH{} production over the $y_t^2$ one by requiring, for instance, an energy threshold for the $B$ hadron inside the hardest $b$ jet, and/or vetoing double b-tagged jets. Alternatively, substructure techniques could be employed to reveal the internal details of jets and classify events more efficiently.

\section{Summary and outlook}\label{sec:summary}

The precise determination of the Higgs-boson couplings will be one of the main goals of the
LHC programme of the coming decades. In this work we have presented for the first time the computation of the contributions proportional to the top-quark Yukawa coupling to $\bbH{}$ production at NLO in QCD using the 4FS. Given that the exact NLO computation is beyond the reach of  the current multi-loop technology, we have employed an effective field theory approach, where the top quark is integrated out and the Higgs boson couples directly to gluon fields via a dimension-five operator. In order to reach NLO accuracy in the HEFT, in  addition to the usual tree-level, virtual and real terms,  we have also determined the finite $O(y_t)$ corrections to the bottom-quark Yukawa, by matching the HEFT to the full theory at two loops. Our results agree with previous
calculations available in literature~\cite{PhysRevLett.78.594,CHETYRKIN199719,Chetyrkin:1997un}. We have argued that the HEFT approximation is suitable to describe the phase space region where the bulk of the $\bbH{}$ cross section resides,   {\it i.e.}, for the Higgs boson,  up to $\pth{}\simeq 200\,\giga\electronvolt$.

The main result of our study is that the $\bbH{}$ final state is largely dominated by the top-quark Yukawa contributions in all regions of the phase space,
at least where the Born-improved HEFT approximation can be trusted. This is contrary to the common lore and intuition that the $\bbH{}$ final state gives direct access to $y_b^2$, as much as $\ttH{}$ gives access to  $y_t^2$. The failure of such simple-minded approach is mostly due to the large hierarchy between $y_t^2/y_b^2$ which makes up for the relative $\alpha_s^2$ and loop-squared suppression of the top-induced contribution. While being expected to some extent by existing LO computations of the $y_t^2$ terms, such conclusion becomes glaring (and robust) at NLO due to the very large $K$-factor (almost 3) associated to the $y_t^2$ contribution.
Apart from the decreased sensitivity to the bottom-quark Yukawa coupling,
this result also entails a larger cross section of associated \bbH{} production and
henceforth a better chance of measuring \bbH{} at the LHC.

We have then investigated in detail how the two main contributions, those proportional to $y_b^2$ and $y_t^2$, can be disentangled by using suitable observables and selection cuts.  By  systematically studying kinematical distributions of the final states at NLO accuracy, we have identified two main handles: first,  the largest relative contribution from $y_b^2$ terms resides at small Higgs transverse momentum. Second, the $y_t^2$ terms are strongly dominated by configurations with gluon splitting into a $b\bar b$ pair, appearing already at LO, often leading to a high-\pt{} jet consisting of two bottom quarks.  Our results indicate that the $y_t^2$ contribution will always provide a significant fraction of the irreducible background to the $y_b^2$ one, and therefore the $y_t^2$ contribution should be included in global fits, such as those based on the rescaling of SM couplings (kappa framework)~\cite{LHCHiggsCrossSectionWorkingGroup:2012nn}.

Finally, predictions for $\bbH{}$ total and differential rates at NLO in QCD have a wide range of phenomenological relevance and applications, which go beyond what was was discussed in this work and are worth exploring in the future.  The first natural extension of our work will be to promote the NLO results to fully exclusive ones by matching to parton showers. While technically straightforward, this step entails understanding and controlling the interplay between fixed-order and resummed radiation from the bottom quarks as well as the gluon-splitting mechanism, a topic which has been the subject of several recent investigations~\cite{Cascioli:2013era,Buonocore:2017lry,Jezo:2018yaf,Bagnaschi:2018dnh}. As briefly explored in this work, the $y_t^2$ contribution is very sensitive to $g\to b\bar{b}$ splittings and therefore could provide an optimal testing bench for further studies. The second extension will be a careful reassessment of $\bbH{}$ as potential background to other final states featuring the Higgs boson, in the SM measurements as well as
in Beyond the SM (BSM) searches, mostly due to the large enhancement of the cross section at NLO. For instance, in the search for $t\bar t H$ with $H\to \gamma \gamma$, only very weak constraints on additional ($b$-)jet activity beyond the two photons (on the Higgs mass shell) are required, and $\bbH{}$ production can contribute to the signal region. Similarly, $HH$ searches where at least one of the Higgs bosons decays into a $b\bar b$ pair will be affected by associated $\bbH{}$ production, as the $\bbH{}$ rate with $m_{bb}>100\,\giga\electronvolt$ is comparable to the $HH$ cross section and it is dominated by $y_t^2$ terms as soon as $H$ has a moderate \pt{}. Other channels, currently searched for in 3$b$-jet final states, such as associated heavy-quark  and double-scalar production in extensions of the SM like the 2HDM, could also be affected by a sizable $\bbH{}$ background.

%===============================================================================
%===============================================================================
\section*{Acknowledgments}
We thank the anonymous JHEP referee for the useful comments which helped us improving this work. We are grateful to Sasha Nikitenko and Nikos Romptis for discussions and information on the details of the
measurements. MW thanks Michael Spira for many useful insights and discussions on Higgs production in association with bottom quarks.  ND is thankful to Claude Duhr for his essential advice and guidance and many discussions on how to perform the two loop calculation reported in this work and to him and Falko Dulat for providing a version of $\textsc{PolylogTools}$. The work of ND has been supported by the ERC grants 637019 `MathAm` and 694712 `pertQCD`. ND further acknowledges the hospitality of the CERN TH department while this work was carried out. FM has received fundings from the European Union's Horizon 2020 research and innovation programme as part of the Marie Sk\l{}odowska-Curie Innovative Training Network MCnetITN3 (grant agreement no. 722104) and by the  F.R.S.-FNRS under the `Excellence of Science` EOS be.h project n. 30820817.
The work of MW is supported by the ERC Consolidator Grant 614577 HICCUP. The work of MZ is supported by the Netherlands National Organisation for Scientific Research (NWO).

\clearpage
\appendix
\section{Jet rates with $R=1$}
\label{app:r1}
In this appendix we show results for the jet categories considered in \sct{sec:results} using $R=1$ as jet-radius parameter. The choice of such a value is  motivated by boosted-Higgs searches. More in general, a larger jet-radius  enhances the $\ge 1 bb$-jet category, thus making a more efficient reduction of the $\ytsq{}$ contribution possible.
Apart from the jet-radius parameter, we employ exactly the same setup as described in \sct{sec:inputs}.
\begin{table}[!ht]
\begin{center}
    \begin{tabular}{cc| ccc}
         & $R=0.4 $  &  LO (acceptance) &  NLO (acceptance) & $K$ \\
 \hline\hline
\multirow{2}{*}{ $ \ge 1 b $ } 
 & ${y_t^2}_{\textrm{HEFT}}$  &  $ \phantom{-} 1.70 \cdot 10^{ -1 }\,\,{}^{+   72 \perc} _{ -39 \perc} $  (47\perc) &  $ \phantom{-} 3.67 \cdot 10^{ -1 }\,\,{}^{+   39 \perc} _{ -29 \perc} $ (42\perc) &   2.2  \\
 & $y_b^2$  &  $ \phantom{-} 6.02 \cdot 10^{ -2 }\,\,{}^{+   52 \perc} _{ -31 \perc} $  (23\perc) &  $ \phantom{-} 8.49 \cdot 10^{ -2 }\,\,{}^{+   13 \perc} _{ -16 \perc} $ (21\perc) &   1.4  \\

 \hline\hline
\multirow{2}{*}{ $ \ge 2 b $ } 
 & ${y_t^2}_{\textrm{HEFT}}$  &  $ \phantom{-} 2.48 \cdot 10^{ -2 }\,\,{}^{+   72 \perc} _{ -39 \perc} $  (6.9\perc) &  $ \phantom{-} 4.86 \cdot 10^{ -2 }\,\,{}^{+   33 \perc} _{ -27 \perc} $ (5.5\perc) &   2.0  \\
 & $y_b^2$  &  $ \phantom{-} 5.07 \cdot 10^{ -3 }\,\,{}^{+   51 \perc} _{ -31 \perc} $  (1.9\perc) &  $ \phantom{-} 5.92 \cdot 10^{ -3 }\,\,{}^{+    1 \perc} _{ -12 \perc} $ (1.5\perc) &   1.2  \\

 \hline\hline
\multirow{2}{*}{ $ \ge 1 bb $ } 
 & ${y_t^2}_{\textrm{HEFT}}$  &  $ \phantom{-} 3.84 \cdot 10^{ -2 }\,\,{}^{+   70 \perc} _{ -38 \perc} $  (11\perc) &  $ \phantom{-} 7.86 \cdot 10^{ -2 }\,\,{}^{+   36 \perc} _{ -28 \perc} $ (8.9\perc) &   2.0  \\
 & $y_b^2$  &  $ \phantom{-} 3.37 \cdot 10^{ -4 }\,\,{}^{+   57 \perc} _{ -34 \perc} $  (0.1\perc) &  $ \phantom{-} 2.53 \cdot 10^{ -4 }\,\,{}^{+    4 \perc} _{ -48 \perc} $ (0.1\perc) &   0.7  \\

    \end{tabular}
    \\[0.5cm]
    \begin{tabular}{cc| ccc}
        & $R=1$  &  LO (acceptance) &  NLO (acceptance) & $K$ \\
 \hline\hline
\multirow{2}{*}{ $ \ge 1 b $ } 
 & ${y_t^2}_{\textrm{HEFT}}$  &  $ \phantom{-} 1.80 \cdot 10^{ -1 }\,\,{}^{+   72 \perc} _{ -39 \perc} $  (50\perc) &  $ \phantom{-} 4.16 \cdot 10^{ -1 }\,\,{}^{+   43 \perc} _{ -30 \perc} $ (47\perc) &   2.3  \\
 & $y_b^2$  &  $ \phantom{-} 6.08 \cdot 10^{ -2 }\,\,{}^{+   52 \perc} _{ -31 \perc} $  (23\perc) &  $ \phantom{-} 9.23 \cdot 10^{ -2 }\,\,{}^{+   17 \perc} _{ -18 \perc} $ (23\perc) &   1.5  \\

 \hline\hline
\multirow{2}{*}{ $ \ge 2 b $ } 
 & ${y_t^2}_{\textrm{HEFT}}$  &  $ \phantom{-} 1.46 \cdot 10^{ -2 }\,\,{}^{+   73 \perc} _{ -39 \perc} $  (4.1\perc) &  $ \phantom{-} 3.32 \cdot 10^{ -2 }\,\,{}^{+   42 \perc} _{ -30 \perc} $ (3.8\perc) &   2.3  \\
 & $y_b^2$  &  $ \phantom{-} 4.89 \cdot 10^{ -3 }\,\,{}^{+   51 \perc} _{ -31 \perc} $  (1.9\perc) &  $ \phantom{-} 6.80 \cdot 10^{ -3 }\,\,{}^{+   12 \perc} _{ -16 \perc} $ (1.7\perc) &   1.4  \\

 \hline\hline
\multirow{2}{*}{ $ \ge 1 bb $ } 
 & ${y_t^2}_{\textrm{HEFT}}$  &  $ \phantom{-} 8.73 \cdot 10^{ -2 }\,\,{}^{+   71 \perc} _{ -38 \perc} $  (24\perc) &  $ \phantom{-} 1.91 \cdot 10^{ -1 }\,\,{}^{+   39 \perc} _{ -29 \perc} $ (22\perc) &   2.2  \\
 & $y_b^2$  &  $ \phantom{-} 2.22 \cdot 10^{ -3 }\,\,{}^{+   57 \perc} _{ -34 \perc} $  (0.8\perc) &  $ \phantom{-} 2.13 \cdot 10^{ -3 }\,\,{}^{+    1 \perc} _{ -19 \perc} $ (0.5\perc) &   1.0  \\

    \end{tabular}
    \caption{\label{tab:rates-r1} Cross sections (in pb) for different $b$-jet multiplicities with jet radius $R=0.4$ (top table) and $R=1$ (bottom table). For $R=0.4$, numbers are the same
    as in \tab{tab:rates}.}
\end{center}
\end{table}

Our results are shown in \tab{tab:rates-r1}. For convenience of the reader, we also quote in the upper part of the table the jet rates computed with $R=0.4$ (taking the results directly from \tab{tab:rates}), while the
bottom part of the table displays results with $R=1$. In order to keep the table minimal, we do not show the $y_b y_t$ interference and the
BI-HEFT contribution, nor the sum of all contributions to the total cross section. As displayed in
\tab{tab:rates}, the effect of including the full top-mass dependence in the jet rates is small, in particular on the acceptance.

The conclusions that can be drawn from the table are the following:
\begin{itemize}
    \item Choosing $R=1$ has the effect of mildly increasing the $\ge 1 b$-jet category, with a slightly more pronounced effect for the $\ytsq{}$ contribution than for the $\ybsq{}$ one. This  can be easily understood
        as $R=1$ makes it possible to cluster the radiation  more inclusively. Events with both $b$ quarks slightly below the jet-$\pt{}$ threshold may more easily lead to a $b$ jet with a larger jet radius. However, one
        should also keep in mind acceptance effects: if one of the $b$ quark is outside the jet-acceptance rapidity window, the resulting jet may not fall inside the acceptance, and hence be discarded. The
        fact that the cross section is larger with $R=1$ than with $R=0.4$ hints that acceptance effects should be less important than the inclusive clustering of the radiation. However, the former effects explain why
        the \emph{exclusive} one-$b$ jet category (obtained by subtracting the $\ge 2 b$ and $\ge 1 bb$ from the $\ge 1 b$ one) is larger for $R=0.4$ than for $R=1$, as it can be trivially computed from the numbers in the table.
    \item Concerning the $\ge 2b$-jet category, the larger jet radius leads to a relative $30-40\%$ reduction of the $\ytsq{}$ rate. The $\ybsq{}$ term rate is slightly reduced at LO while it is increased of
        about 10\% (relative) at NLO. Again, two competing effects should be considered in order to explain the behaviour: on the one hand, a larger jet radius requires the $b$ jets to be more separated, leading
        to a reduction of the $\ge 2b$-jet rate; on the other, it leads
        to a more inclusive clustering of the QCD radiation, thus instead giving an increase of the rate. Given the tendency of the two $b$ jets to
        lie closer in the $\ytsq{}$ than in the $\ybsq{}$ contribution to the cross section (see also \fig{fig:bb-dr}), the first
        effect will be more pronounced on the former contribution, and the second on the latter.
    \item Finally, the $\ge 1bb$ category is where the effect of using $R=1$ is most pronounced. The relative increase of the $\ybsq{}$ contribution is very large, more than a factor 6 (8) at LO (NLO). However, the
        absolute rate remains negligible to all practical purposes, with an acceptance below 1\%. For the  $\ytsq{}$ contribution the relative increase is still large, about a factor 2, and the acceptance now exceeds 20\%.
\end{itemize}

The above remarks support vetoing fat $bb$ jets as a way to further reduce the effect of the $\ytsq{}$ term and thus increase the sensitivity on the bottom-quark Yukawa. As we already mentioned in the conclusions of our work, the natural follow-up of these findings would be to perform a more detailed study based on fully exclusive final state, which includes matching with parton showers.

\section{Matching and renormalization schemes}
\label{app:renormalization}
The matching of the HEFT to the SM for the computation of higher-order corrections
requires a well-defined renormalization scheme in both theories. In practice, rather than considering the full SM, one can focus
only on the part relevant for the study of the $pp\to b\bar{b}H$ process, at the order under consideration. Therefore, as it is customary in the literature~\cite{PhysRevLett.78.594,CHETYRKIN199719}, we can restrict ourselves to a simplified model, featuring QCD with four massless and two massive quarks and a singlet scalar $H$ that couples to the massive quarks. Hence we do not have to deal with issues related to the breaking of $SU(2)$ gauge invariance or the relation between Yukawa couplings and masses. The Lagrangian of this model, which for brevity is denoted as SM, has the following form:
\begin{align}
\begin{split}
{\cal L}_\text{SM} &= -\frac{1}{4}G_{\mu\nu}G^{\mu\nu}+\frac{1}{2} \left(\partial_\mu H\partial^\mu H - m_H^2 H^2\right)-V(H)\\
&+ \sum_{\psi = u,d,c,s} i\bar{\psi} {\not} D \psi + \sum_{\Psi = b,t}\left[ i\bar{\Psi} \left({\not} D - m_\Psi\right) \Psi - \frac{y_\Psi}{\sqrt{2}} \bar{\Psi}\Psi H  \right]+{\cal L}_\text{gf},
\end{split}
\end{align}
where $G_{\mu\nu}$ is the gluon field strength tensor for the gluon field $G_\mu$. $H$ is the scalar Higgs field with mass $m_H$. The four light quarks $\psi\in\{u,d,c,s\}$ and 
the two massive quarks $\Psi\in\{b,t\}$ are labelled by the usual SM flavor symbols.
Their masses and Yukawa couplings are denoted by $m_b$, $m_t$ and $y_b$, $y_t$. 
Finally ${\cal L}_\text{gf}$ is the gauge-fixing and ghost Lagrangian of the gauge interaction. We denote the QCD coupling constant in this theory as $\alpha_s$.

The EFT to which we match our theory in the heavy-top quark limit is expressed by the following Lagrangian:
\begin{align}
\begin{split}
{\cal L}_\text{HEFT} &= -\frac{1}{4}\tilde G_{\mu\nu}\tilde G^{\mu\nu}+\frac{1}{2} \left(\partial_\mu \tilde H\partial^\mu \tilde H - \left(m_H^\text{HEFT}\right)^2 \tilde H^2\right)-V(\tilde H)\\
&+ \sum_{\tilde \psi = \tilde u,\tilde d,\tilde c,\tilde s} i\bar{\tilde \psi} {\not} D \tilde \psi + \left[ i\bar{\tilde b} \left({\not} D - m_b^\text{HEFT}\right) \tilde b - \frac{y_b^\text{HEFT}}{\sqrt{2}} \bar{\tilde b}\tilde b \tilde H  \right]+{\cal L}_\text{gf}\\
& -\frac{C_1}{4} H \tilde G_{\mu\nu}\tilde G^{\mu\nu} + \sum_{i=2}^5 C_i O_i,
\end{split}
\label{eq:HEFTfull}
\end{align}
where the fields with a tilde denote the low-energy analogue of the SM fields. The HEFT-labelled parameters have a SM equivalent to which they are matched at leading power. We consider next-to-leading power interactions, mediated by dimension-five interactions in the last line of  \eqn{eq:HEFTfull}, but show only the first of the five independent operators that were first listed in~\citere{Inami1983} as the others are not relevant for this paper (in particular they do not mix with $O_1$ under renormalization). While not used explicitly in  \eqn{eq:HEFTfull}, we refer to the strong coupling constant in this theory as $\alpha_s^\text{HEFT}$. The HEFT is equivalent to the original Lagrangian in the heavy-top quark limit if one can express the renormalized parameters of the HEFT and its renormalized fields in terms of those of the original theory such that amplitudes in the two theories are equal up to terms supressed by inverse powers of the top mass.

\begin{figure}
\centering
\subfloat[]{\includegraphics[height=2.2cm]{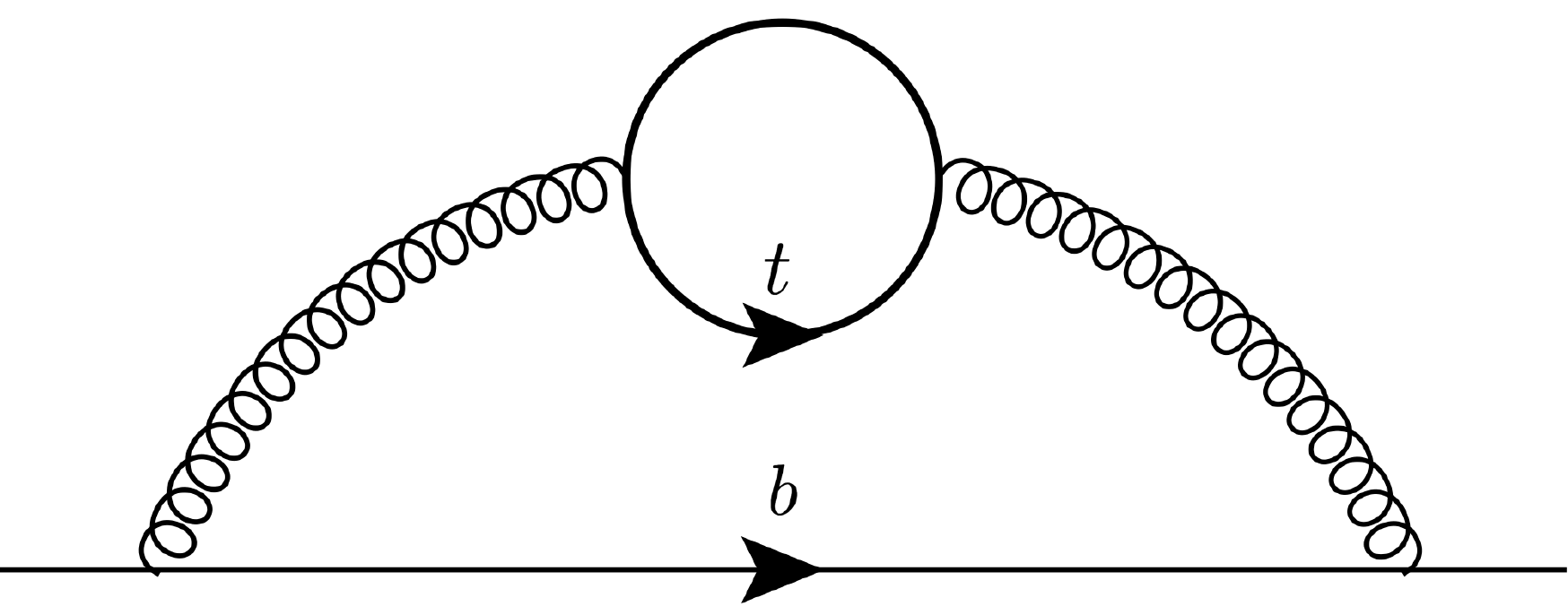}\label{fig:bmass_correction}}
\subfloat[]{\includegraphics[trim={21cm 0 1cm 0},clip,height=3.3cm]{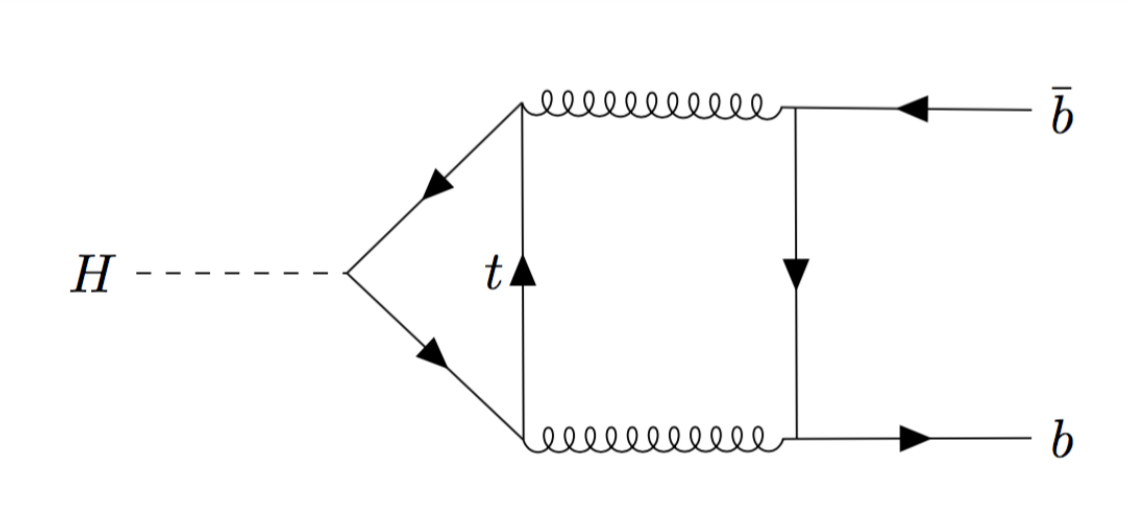}\label{fig:byuk_correction}}
\caption{Standard Model diagrams whose large top mass expansion will yield corrections to (a) the bottom-quark mass and (b) the bottom-quark Yukawa coupling.}
\end{figure}

In the case at hand it is established that a particularly suitable scheme is the so-called \textit{decoupling scheme}~\cite{Dawson:1990zj}, which is generally used in \texttt{Madgraph5\_aMC@NLO}.\footnote{See Appendix B of \citere{Hirschi:2011pa} for an explicit discussion.} In this scheme, fields and masses are renormalized on-shell, and couplings are renormalized in a mixed way, where the counterterms that cancels UV divergences from loops with light degrees of freedom are subtracted in the $\overline{\text{MS}}$ scheme, while those involving heavy quarks are renormalized on-shell. As 
a consequence, the running of the strong coupling is the same as in QCD with four flavors and $\overline{\text{MS}}$ renormalization. The matching condition for $\alpha_s$ and the gluon field are then trivial. In our calculation, we differ from the standard \texttt{Madgraph5\_aMC@NLO} scheme in that we renormalize the bottom-quark Yukawa coupling in the $\overline{\text{MS}}$ scheme. Finally, also for the gluon-Higgs operator coupling $\overline{\text{MS}}$ renormalization is used. Let us write the relevant renormalization equations for the processes considered in this paper. We label all bare quantities with a superscript $B$.

\begin{align}
\alpha_s^{\text{HEFT},B} &= \mu^{2 \epsilon} S_\epsilon^{-1}Z_\alpha \alpha_s^\text{HEFT}\,, \\
m_b^{\text{HEFT},B} & = Z_{m_b}m_b^\text{HEFT}\,,\\
y_b^{\text{HEFT},B} & = \mu^{\epsilon}S_\epsilon^{-1/2} Z_{y_b}y_b^\text{HEFT}\,,\\
C_1^B &= \mu^{\epsilon} S_\epsilon^{-1/2}  Z_{C} C_1\,,
\end{align}
where $S_\epsilon = \exp(-\gamma_E \epsilon) (4\pi)^{\epsilon}$ with $\gamma_E$ the Euler-Mascheroni constant and the renormalization constants take the following form:
\begin{align}
Z_\alpha &= 1- \frac{\alpha_s^\text{HEFT}}{4 \pi}\frac{1}{\epsilon} \left(\beta_0^{(4)} - \frac{2}{3} \left(\frac{\mu}{m_b^\text{HEFT}}\right)^{2\epsilon}\right)\,, \\
Z_{m_b} & = 1- \frac{\alpha_s^\text{HEFT}}{\pi}\left( \frac{1}{\epsilon}+\frac{4}{3}\right)\,,\\
Z_{y_b} & = 1- \frac{\alpha_s^\text{HEFT}}{\pi} \frac{1}{\epsilon}\,,\\
Z_C & = 1- \frac{\alpha_s^\text{HEFT}}{4 \pi}\frac{1}{\epsilon} \left(\beta_0^{(5)} \right)\,,\\
\end{align}
where $\beta_0^{(nf)} = 11-2/3n_f$.
The renormalized parameters in the HEFT are then expressed in terms of renormalized parameters of the full theory. The result of the matching is the following:
\begin{align}
C_1& =- \frac{\alpha_s}{3\pi}\left(\frac{y_t}{\sqrt{2}m_t}\right)\left(1 + \frac{\alpha_s}{\pi} \left(\frac{11}{4} + \frac{1}{6} \log\left(\frac{\mu^2}{m_t^2}\right)\right)\right) + {\cal O}\left(\alpha_s^3\right)\,,\\
\alpha_s^\text{HEFT}& = \alpha_s\,,\\
m_b^\text{HEFT} & = m_b + {\cal O}\left(\alpha_s^2\right)\,,\\
y_b^\text{HEFT} & = y_b+y_t\left(\frac{\alpha_s}{\pi}\right)^2 \frac{m_b}{m_t}C_F \left(\frac{5}{24}-\frac{1}{4}\log\left(\frac{\mu_R^2}{m_t^2}\right)\right)+ {\cal O}\left(\alpha_s^3\right).
\end{align}
In our ``limited" SM, we can keep the top-quark Yukawa coupling independent from the mass, which is useful to keep track of which terms of the EFT correspond to the expansion of the parts of the SM amplitudes we wish to consider in  \eqn{eq:hbbxsecheft}. As discussed in \citeres{PhysRevLett.78.594,CHETYRKIN199719}, there are ${\cal O}(\alpha_s^2)$ contributions to both the bottom-quark mass and Yukawa coupling, associated  to the SM diagrams shown in Figures~\ref{fig:bmass_correction} and~\ref{fig:byuk_correction}, respectively, and we have to consider only the latter at the perturbative order we are interested in. Indeed, in our SM picture, the bottom Yukawa correction contributes to the NLO QCD cross 
section of \bbH{} production that involves the top-quark Yukawa coupling  (\ytsq{} and \ybyt{}), while the mass correction would contribute only to the NNLO QCD cross section proportional to \ybsq{}, which is beyond the accuracy under consideration. A critical look at these two corrections shows that they are likely to have a similar impact and it might be surprising to 
include one, but not the other. It turns out, however, that the correction to the HEFT bottom Yukawa is extremely small: it yields a permille effect to the $\overline{\text{MS}}$ bottom-quark Yukawa, which contributes through an already subleading production channel. Thus, we could have ignored this correction to the EFT bottom-quark Yukawa without affecting our phenomenological results. Instead we have chosen to keep it in order to have an exact description of the leading power expansion of the SM amplitude at the perturbative order under consideration. Since the ${\cal O}\left(\alpha_s^2\right)$ corrections to the Yukawa coupling and the mass are of similar numerical size, only a large effect from the bottom Yukawa correction 
would have been a motivation to include also the mass correction. Since this is not the case, we refrain from including the latter.

\section{The HEFT bottom-quark Yukawa at $\calo{1/m_t}$}
\label{app:matching}
In this appendix, we rederive through a direct calculation the power-suppressed correction to the bottom-quark Yukawa coupling in the HEFT that was first obtained in \citere{PhysRevLett.78.594,CHETYRKIN199719}. This correction is required to compute the complete NLO coefficients $\Delta_{\ytsq}^{(1)}$ (and $\Delta_{\ybsq}^{(1)}$) in the heavy-top mass approximation. Indeed, as mentioned in \sct{sec:heft}, the matching of the HEFT to the SM requires not only the introduction of an effective point-like coupling between the Higgs boson and gluons, but also corrections, suppressed by inverse powers of the top-quark mass, to renormalisable couplings. In particular, as we shall see in the following, the bottom-quark Yukawa coupling is modified at $\calo{1/m_t}$ by a term scaling like $y_t\,\alpha_s^2\, m_b/m_t$. We obtain this correction by evaluating a form-factor contribution to the amplitude of the  $H\to b \bar b$ decay at $\calo{y_t\,\alpha_s^2}$, first in the HEFT and then as an expansion in inverse powers of the top-quark mass in the SM, and by matching the two results.

The amplitude for the $H\to b\bar b$ decay can be expressed as follows:
\begin{align}
\cala(H\to b\bar b) = \delta_{ii'}\bar u_\sigma^{h} \hat{\cala}_{\sigma \sigma'} v_{\sigma'}^{h'}\, ,
\end{align}
where $\bar u$ and $v$ are the spinors associated to the bottom and the anti-bottom quarks,  respectively, $\sigma$ and $\sigma'$ are their spin indices, $h$ and $h'$ are their helicities, and $i$ and $i'$ are color indices. We consider the following form factor
\begin{align}
{\cal M} &= \sum_{hh'}\delta_{ii'}\cala(H\to b\bar b) \bar u^h\cdot v^{h'}\\
&= \text{Tr}\left[({\not p}_b+m_b) \hat{\cala}({\not p}_{\bar b}-m_b) \right]\, ,
\end{align}
where $p_b$ and $p_{\bar b}$ denote the bottom and anti-bottom momenta, respectively, and $C_A$ is the $SU(N)$ adjoint Casimir ($C_A = N$).

\begin{figure}[!h]
  \centering
  \subfloat[]{\includegraphics[height=.2\textwidth]{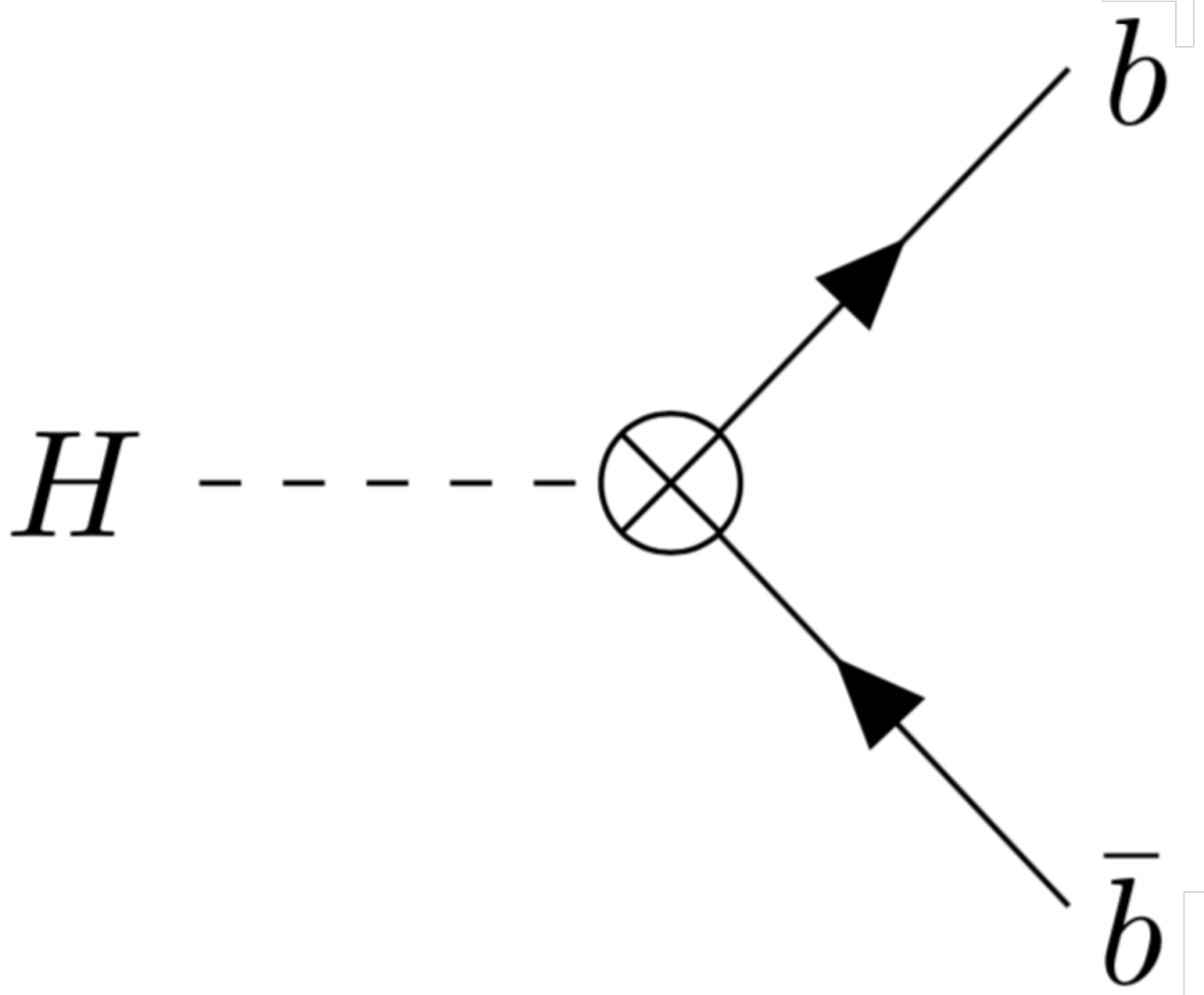}}\hspace{.2\textwidth}
  \subfloat[]{\includegraphics[height=.2\textwidth]{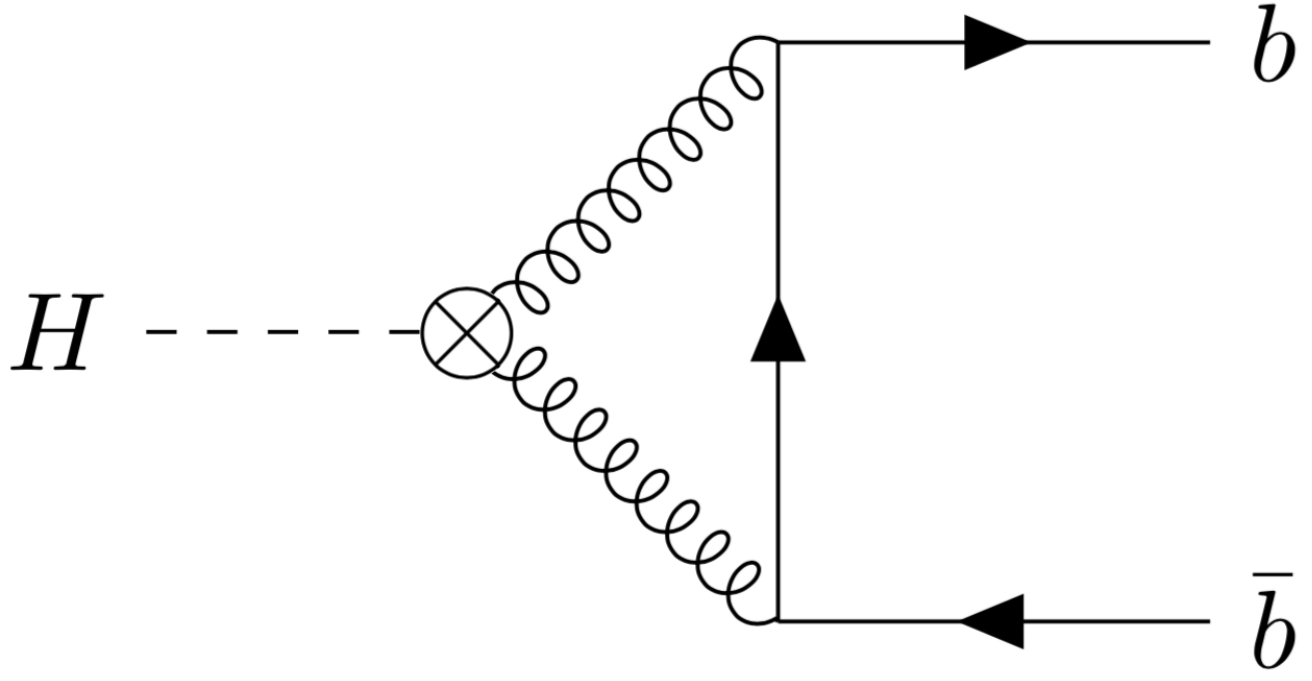}}
  \caption{Diagrams contributing to the $H\to b\bar b$ form factor in the HEFT at $\calo{y_t \alpha_s^2/m_t}$. In terms of HEFT couplings, these diagrams scale like ${\cal O}(\alpha_s C_1)$.
  \label{fig:hefthtobb}}
\end{figure}

In the HEFT, two types of diagrams contribute to ${\cal M}$ at $\calo{\alpha_s^2\, y_t/m_t}$, the contribution to the tree-level diagram from the bottom-quark Yukawa correction $\delta y_b$ and the one-loop diagram with an effective Higgs--gluon coupling, shown in \fig{fig:hefthtobb}.

The one loop diagram is UV-divergent and is rendered finite by the renormalisation of the bottom-quark Yukawa. Note that we should a priori renormalize the bottom-quark field as well. However, there is no bottom-quark propagator correction that is of order ${\cal O}\left(\alpha_s C_1 \right)$. Therefore, the field renormalization counterterm is not relveant in this calculation.

In the $\overline{\text{MS}}$ scheme, we find that we need to define the following counterterm for the HEFT bottom-quark Yukawa:
\begin{align}
\frac{\alpha_s}{\pi}\sqrt{2}\,m_b\, C_F\, \frac{3}{4}\, \frac{C_1}{\epsilon} = - y_t \left( \frac{\alpha_s}{\pi} \right)^2 \left( \frac{m_b}{m_t} \right) \frac{C_F}{4\epsilon}\, ,
\end{align}
where $C_F$ is the fundamental representation Casimir of $SU(N)$, $\epsilon$ is usual dimensional regularisation parameter in $d=4-2\epsilon$ spacetime dimensions. Note that the renormalization scheme chosen for other parameters is irrelevant to the matching procedure at hand, since it only has effects beyond accuracy.
We can therefore write the following expression for the bare HEFT bottom-quark Yukawa
\begin{align}
y_b^{\text{HEFT},B} = y_b^{\text{SM},B} + y_t \left(\frac{\alpha_s}{\pi} \right)^2 \left( \frac{m_b}{m_t} \right) \left[-\frac{C_F}{4\epsilon} + \Delta_F\right]\,,
\end{align}
where $\Delta_F$ is the finite contribution to $y_b^{\text{HEFT}}$ that we seek to obtain. 
Similarly to the renormalization, we should in principle allow for the possibility that the renormalized bottom-quark field in the HEFT ($\tilde b$) is matched to the SM field through a nontrivial finite counterterm. However, as was already pointed out in the previous Appendix, no SM bottom-quark propagator correction has the correct scaling in the top Yukawa to contribute to the $H\to b\bar{b}$ amplitude at the order of interest.

We find the following expression for the ${\cal O}(y_t \alpha_s^2)$ part of the renormalised form factor in the Euclidian region ($m_H^2>0$):
\begin{equation}
  \begin{array}{l}
    \cala_{H\to b\bar b}^{\text{HEFT}}\big|_{y_t \alpha_s^2} =   -i y_t \left( \dfrac{\alpha_s}{\pi} \right)^2  C_A C_F\left( \dfrac{m_b}{m_t} \right) m_b^2 \dfrac{(2 r+1)}{r (r+1)18 \sqrt{2}} \left(3 G(0,-1,r)+3 G(0,0,r)\vphantom{\dfrac{a}{b}}\right.\\
    \left.\vphantom{\dfrac{a}{b}}-3 G(-1,-1,r)-3 G(-1,0,r) +9 (1+2r) \log \left(\dfrac{m_b^2}{\mu ^2}\right)-24 r+2 \pi ^2-12\right) \\
   + i y_t \left( \dfrac{\alpha_s}{\pi} \right)^2 \left( \dfrac{m_b}{m_t} \right)  C_A \dfrac{m_b^2}{\sqrt{2} } \dfrac{ (2 r+1)^2}{r (r+1)} \Delta_F,
  \end{array}
\end{equation}
where $r=\dfrac{\sqrt{\tau} \sqrt{\tau+4}-\tau}{2 \tau}$, with $\tau=-\dfrac{m_H^2}{m_b^2}$. The form factor is expressed in terms of multiple polylogarithms $G$~\cite{Goncharov1998,Goncharov2001} defined iteratively by
\begin{align}
  G(a_1,\dots\,a_n; x ) = \int_0^x dt \frac{G(a_2,\dots,a_n;t)}{t-a_1},
\end{align}
where $G(;x)=1$ and we define the special case where all $a_i$ are 0 as
\begin{align}
  G(\underbrace{0,\dots,0}_n;x) = \frac{1}{n!}\log^n(x).
\end{align}

\begin{figure}[!t]
  \centering
  \includegraphics[width=.98\textwidth]{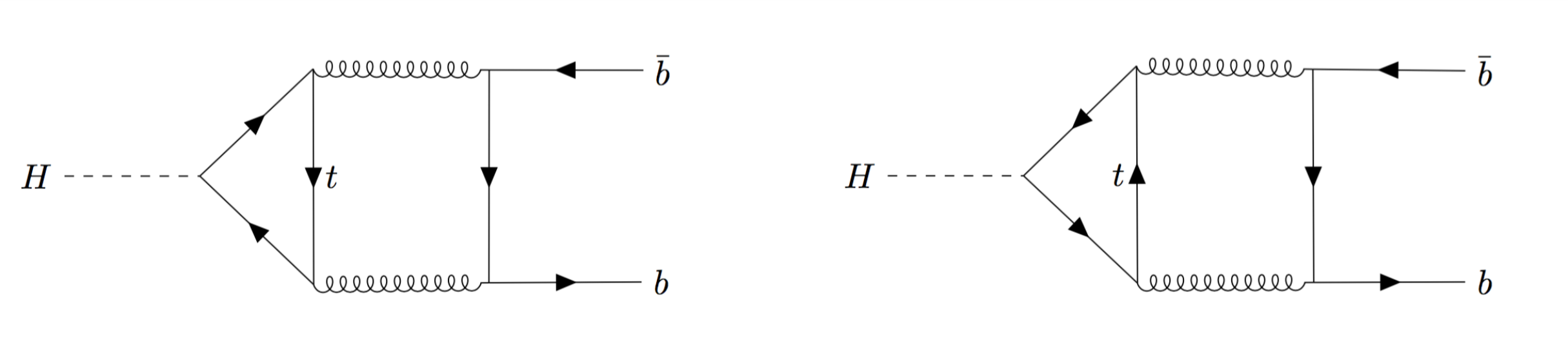}
  \caption{Diagrams contributing to $\cal M$ at $\calo {y_t \alpha_s^2}$ in the SM.}
  \label{fig:smhtobb}
\end{figure}

To derive $\Delta_F$, we need to compute the heavy-top quark limit (the leading term in $1/m_t$) of the SM expression for ${\cal M}$. At the perturbative order of interest two two-loop diagrams contribute, shown in \fig{fig:smhtobb}.

The SM form factor can be evaluated using modern multi-loop calculation techniques. We generate \textsc{FORM}~\cite{Kuipers2013} expressions for these form factors using our own \textsc{QGRAF}~\cite{Nogueira1993} interface. After the spinor and tensor algebra is performed with \textsc{FORM}, we obtain a scalar expression in terms of kinematic invariants involving the external momenta and the loop momenta. We define the following family of integrals:
\begin{align}
  J \left( n_1,\dots,n_7 \right) = \int \ddk{1}\ddk{2} \frac{1}{D_1^{n_1}\dots D_7^{n_7}}\,,
\end{align}
where the denominators are defined as
\begin{equation}
  \begin{array}{lll}
D_1 = k_2^2-m_b^2\,, & D_2 = (k_2-p_1)^2\,, & D_3 = (k_2+p_2)^2\,, \\
D_4 = (k_1-p_1)^2-m_t^2\,, & D_5 = (k_1+p_2)^2-m_t^2\,, & D_6 = (k_1-k_2)^2-m_t^2\,,\\
  & D_7 = (k_1+k_2)^2-m_t^2\,. &
\label{topoHEFT}
  \end{array}
\end{equation}
The form factor is expressed as a linear combination of integrals in this family using \textsc{Mathematica}. We reduce the set of integrals required for the expression of the amplitude to a basis of master integrals using \textsc{LiteRed}~\cite{Lee2014a}. We obtain a set of 23 master integrals shown in \eqn{eq:masters}, which we need expand in $1/m_t$:
\begin{equation}
  \begin{array}{cccc}
    J(0,0,0,1,0,1,0)\,, & J(1,0,0,1,0,0,0)\,, & J(0,0,1,1,0,1,0)\,, & J(0,0,2,1,0,1,0)\,, \\
    J(0,0,1,2,0,1,0)\,, & J(0,0,0,1,1,1,0)\,, & J(0,1,1,1,0,0,0)\,, & J(2,0,0,0,1,1,0)\,, \\
    J(1,0,0,1,0,1,0)\,, & J(2,0,0,1,0,1,0)\,, & J(1,0,0,1,1,0,0)\,, & J(0,1,1,1,1,0,0)\,, \\
    J(1,0,0,1,1,1,0)\,, & J(2,0,0,1,1,1,0)\,, & J(1,0,0,1,1,2,0)\,, & J(1,0,0,2,1,1,0)\,, \\
    J(1,0,1,1,0,1,0)\,, & J(2,0,1,1,0,1,0)\,, & J(1,0,1,2,0,1,0)\,, & J(1,1,1,1,0,0,0)\,, \\
    J(0,1,1,1,1,1,0)\,, & J(1,1,1,1,1,0,0)\,, & J(1,1,1,1,1,1,0)\,. &  \\
  \end{array}
  \label{eq:masters}
\end{equation}
These master integrals can be expressed in the parametric Feynman representation and their expansion for a heavy top-quark mass\footnote{This is technically achieved by rescaling the invariants with a spurious parameter $\rho$ as $m_b\to \rho\, m_b$ and $m_H\to \rho\, m_H$ and taking the limit $\rho\to 0$.} is done using the \textsc{Mathematica} package \textsc{ASY}~\cite{Jantzen2012} to perform an expansion by region~\cite{Smirnov:2002pj}. The expansion significantly simplifies the integrals and most can readily be identified with combinations of Euler $\Gamma$, $B$ integrals and logarithms, with the exception of $J(1,1,1,1,1,0,0)$ and $J(1,1,1,1,1,1,0)$.
These two integrals can be expressed in terms of multiple polylogarithms and can be straightforwardly evaluated using the algorithm described in Appendix C of \cite{Anastasiou2013}, whose application is significantly simplified by the package \textsc{PolylogTools}~\cite{PolyLogTools} which implements the reduction of multiple polylogarithms to the canonical form defined in~\citere{Anastasiou2013} and many useful automated tools to integrate multiple polylogarithms in canonical form. The expression for all master integrals in the Euclidean region ($m_H^2<0$) can be found in appendix~\ref{app:masters}. These integrals have been checked by comparing their evaluation with \textsc{GINAC}~\cite{Bauer:2000cp} to the numerical integration provided by \textsc{FIESTA}~\cite{Smirnov2015} in the Euclidean region, showing excellent agreement.

By insertion of the expanded master integrals into the form factor and expansion of their coefficients for
large top-quark masses, we obtain the expression for the SM version of $\cal M$ at order $1/m_t$.\footnote{We verified that the master integrals were all expanded to a sufficient order in $1/m_t$ by adding dummy higher order terms to their expressions and checking that they vanish in the $1/m_t$ term of the form factor.} The form factor is UV-finite in the SM at this order, and is therefore independent of the choice of the renormalization scheme. We ultimately obtain the following expression for the form factor in the SM:
\begin{equation}
  \begin{array}{rl}
  \cala_\text{SM} &=  -i y_t \left( \dfrac{\alpha_s}{\pi} \right) ^2 C_A C_F \left( \dfrac{m_b}{m_t}\right) \dfrac{m_b^2}{36 \sqrt{2} }  \dfrac{ (2 r+1)}{r (r+1)} \left( 6
     G(0,-1,r)+6 G(0,0,r) \vphantom{\dfrac{a}{b}}\right.\\
     &\left.-6 G(-1,-1,r)-6 G(-1,0,r)+36 \left( 2r+1 \right)  \log
     \left(\dfrac{m_b}{m_t}\right)-78 r+4 \pi ^2-39\right).
  \end{array}
\end{equation}

In the last step of our calculation we match the form factor in the two theories. We find that the polylogarithmic dependence of the two expressions is exactly the same, leaving us with the condition for the two renormalised expressions to be equal:
  \begin{equation}
    \Delta_F=\frac{C_F}{24}  \left(5-6 \log \left(\frac{\mu_R^2}{m_t^2}\right)\right)\, ,
  \end{equation}
which is in agreement with the result obtained in \citeres{PhysRevLett.78.594,CHETYRKIN199719} through low-energy theorems.

\section{Master integrals}
\label{app:masters}
In this appendix, we show the results for the 23 master integrals that appear in the calculation of the two-loop form factor expanded in the infinite top mass limit that we evluated in the previous appendix. The integrals showed here are defined with a $\overline{\text{MS}}$ prefactor
\begin{equation}
  J(n_1,\dots,n_7) = (4 \pi)^{-2 \epsilon} e^{2 \epsilon \gamma}m_t^{2 \sum n_i-2d}\int \frac{d^dk_1}{(2\pi)^d}\frac{d^dk_2}{(2\pi)^d} \frac{1}{D_1^{n_1}\dots D_7^{n_7}},
\end{equation}
where
\begin{equation}
  \begin{array}{lll}
D_1 = k_2^2-m_b^2, & D_2 = (k_2-p_1)^2, & D_3 = (k_2+p_2)^2, \\
D_4 = (k_1-p_1)^2-m_t^2, & D_5 = (k_1+p_2)^2-m_t^2, & D_6 = (k_1-k_2)^2-m_t^2,\\
  & D_7 = (k_1+k_2)^2-m_t^2. &
\end{array}
\end{equation}
We express our integrals in terms of $t=m_b^2/m_t^2$ and $r=\dfrac{\sqrt{\tau} \sqrt{\tau+4}-\tau}{2 \tau}$, with $\tau=-\dfrac{m_H^2}{m_b^2}$.
\clearpage
{\small
\begin{align*}
  &J(0,0,0,1,0,1,0) = \left(-\frac{1}{256 \pi ^4}\right) \frac{1}{\epsilon^2}+\left(-\frac{1}{128 \pi ^4}\right) \frac{1}{\epsilon}-\frac{1}{1536 \pi ^2}-\frac{3}{256
   \pi ^4}  \\&
  J(1,0,0,1,0,0,0)=\frac{1}{\epsilon^2} \left(t \left(-\frac{1}{256 \pi ^4}\right)\right)+\frac{1}{\epsilon} \left(t \frac{\log (t)-2}{256 \pi ^4}\right)+t
   \left(-\frac{3 \log ^2(t)-12 \log (t)+\pi ^2+18}{1536 \pi ^4}\right) \\&
J(0,0,1,1,0,1,0)= \left(-\frac{1}{256 \pi ^4}\right) \frac{1}{\epsilon^2}+\frac{1}{\epsilon} \left(t \left(-\frac{1}{1024 \pi ^4 r (r+1)}\right)-\frac{3}{256 \pi
   ^4}\right)\\&+\frac{1}{9216 \pi ^4 r^2 (r+1)^2} t^2+\frac{1}{2048 \pi ^4 r (r+1)} t-\frac{1}{1536 \pi ^2}-\frac{7}{256 \pi ^4} \\&
J(0,0,2,1,0,1,0)=\frac{1}{512 \pi ^4} \frac{1}{\epsilon^2}+\frac{1}{\epsilon} \left(t^2 \frac{1}{15360 \pi ^4 r^2 (r+1)^2}+t \left(-\frac{1}{1536 \pi ^4 r
   (r+1)}\right)-\frac{1}{512 \pi ^4}\right)\\&+\frac{17}{115200 \pi ^4 r^2 (r+1)^2} t^2+\frac{1}{9216 \pi ^4 r (r+1)} t+\frac{18+\pi ^2}{3072 \pi ^4}\\&
J(0,0,1,2,0,1,0)=\left(-\frac{1}{512 \pi ^4}\right) \frac{1}{\epsilon^2}+\left(-\frac{1}{512 \pi ^4}\right) \frac{1}{\epsilon}+\left(-\frac{1}{18432 \pi ^4 r^2
   (r+1)^2}\right) t^2\\&+\frac{1}{1024 \pi ^4 r (r+1)} t+\frac{-6-\pi ^2}{3072 \pi ^4}\\&
J(0,0,0,1,1,1,0)=\left(-\frac{1}{256 \pi ^4}\right) \frac{1}{\epsilon^2}+\frac{1}{\epsilon} \left(t^2 \left(-\frac{1}{15360 \pi ^4 r^2 (r+1)^2}\right)+t
   \frac{1}{1536 \pi ^4 r (r+1)}-\frac{1}{256 \pi ^4}\right)\\&+\left(-\frac{1}{7680 \pi ^4 r^2 (r+1)^2}\right) t^2+\frac{1}{1536 \pi ^4 r (r+1)}
   t+\frac{-6-\pi ^2}{1536 \pi ^4}\\&
J(0,1,1,1,0,0,0)=\left(-\frac{1}{256 \pi ^4}\right) \frac{1}{\epsilon^2}+\frac{1}{\epsilon} \frac{-\log (r (r+1))+\log (t)-3}{256 \pi ^4}\\&+\frac{2 \log (r (r+1))
   \log (t)-\log ^2(r (r+1))-6 \log (r (r+1))-\log ^2(t)+6 \log (t)-14}{512 \pi ^4}\\&
J(2,0,0,0,1,1,0)=\left(-\frac{1}{512 \pi ^4}\right) \frac{1}{\epsilon^2}+\frac{1}{\epsilon} \frac{2 \log (t)-1}{512 \pi ^4}+t^2 \frac{6 \log (t)-5}{9216 \pi
   ^4}\\&+\frac{-6 \log ^2(t)-\pi ^2+18}{3072 \pi ^4}+t \frac{2 \log (t)-3}{1024 \pi ^4}\\&
J(1,0,0,1,0,1,0)=\frac{1}{\epsilon^2} \left(t \left(-\frac{1}{512 \pi ^4}\right)-\frac{1}{256 \pi ^4}\right)+\frac{1}{\epsilon} \left(t^3 \left(-\frac{1}{3072 \pi
   ^4}\right)+t \frac{4 \log (t)-5}{1024 \pi ^4}-\frac{3}{256 \pi ^4}\right)\\&+t^3 \frac{10 \log (t)-21}{30720 \pi ^4}+t^2 \frac{6 \log (t)-11}{4608
   \pi ^4}+t \frac{-12 \log ^2(t)+24 \log (t)-2 \pi ^2+9}{6144 \pi ^4}+\frac{-42-\pi ^2}{1536 \pi ^4}\\&
J(2,0,0,1,0,1,0)=\left(-\frac{1}{512 \pi ^4}\right) \frac{1}{\epsilon^2}+\frac{1}{\epsilon} \frac{2 \log (t)-1}{512 \pi ^4}+t^2 \frac{6 \log (t)-5}{9216 \pi
   ^4}\\&+\frac{-6 \log ^2(t)-\pi ^2+18}{3072 \pi ^4}+t \frac{2 \log (t)-3}{1024 \pi ^4}\\&
J(1,0,0,1,1,0,0)=\frac{1}{\epsilon^2} \left(t \left(-\frac{1}{256 \pi ^4}\right)\right)\\&+\frac{1}{\epsilon} \left(t^3 \left(-\frac{1}{15360 \pi ^4 r^2
   (r+1)^2}\right)+t^2 \frac{1}{1536 \pi ^4 r (r+1)}+t \frac{\log (t)-1}{256 \pi ^4}\right)\\&+t^3 \frac{\log (t)-2}{15360 \pi ^4 r^2 (r+1)^2}+t^2
   \left(-\frac{\log (t)-1}{1536 \pi ^4 r (r+1)}\right)+t \left(-\frac{3 \log ^2(t)-6 \log (t)+\pi ^2+6}{1536 \pi ^4}\right)
\end{align*}

\begin{align*}
&J(0,1,1,1,1,0,0)=\left(-\frac{1}{256 \pi ^4}\right) \frac{1}{\epsilon^2}+\frac{1}{\epsilon} \left(\frac{-\log (r (r+1))+\log (t)-2}{256 \pi ^4}\right.\\&\left.+t^2
   \left(-\frac{1}{15360 \pi ^4 r^2 (r+1)^2}\right)+t \frac{1}{1536 \pi ^4 r (r+1)}\right)+t^2 \frac{-\log (r (r+1))+\log (t)-3}{15360 \pi ^4 r^2
   (r+1)^2}\\&+\frac{2 \log (r (r+1)) \log (t)-\log ^2(r (r+1))-4 \log (r (r+1))-\log ^2(t)+4 \log (t)-8}{512 \pi ^4}\\&+t \frac{\log (r (r+1))-\log
   (t)+2}{1536 \pi ^4 r (r+1)}\\&
J(1,0,0,1,1,1,0)= \left(-\frac{1}{512 \pi ^4}\right) \frac{1}{\epsilon^2}+\frac{1}{\epsilon} \left(t^2 \left(-\frac{1}{15360 \pi ^4 r^2 (r+1)^2}\right)+t
   \frac{1}{1536 \pi ^4 r (r+1)}\right.\\&\left.+t^3 \frac{30 r^4+60 r^3+23 r^2-7 r+2}{92160 \pi ^4 r^2 (r+1)^2}-\frac{1}{512 \pi ^4}\right)+t \frac{18 r^2-12
   (r+1) \log (t) r+18 r-1}{6144 \pi ^4 r (r+1)}\\&+t^3 \frac{78 r^4+156 r^3+61 r^2-17 r-\left(30 r^4+60 r^3+23 r^2-7 r+2\right) \log (t)+5}{92160 \pi
   ^4 r^2 (r+1)^2}\\&+t^2 \frac{300 r^4+600 r^3+215 r^2-90 \left(4 r^3+8 r^2+3 r-1\right) \log (t) r-85 r-43}{552960 \pi ^4 r^2 (r+1)^2}-\frac{1}{3072
   \pi ^2}-\frac{1}{512 \pi ^4}\\&
J(2,0,0,1,1,1,0)=t \left(-\frac{\left(6 r^2+6 r-1\right) (2 \log (t)-1)}{12288 \pi ^4 r (r+1)}\right)\\&+t^2 \left(-\frac{\left(30 r^4+60 r^3+25 r^2-5 r+1\right) (3
   \log (t)-1)}{138240 \pi ^4 r^2 (r+1)^2}\right)-\frac{\log (t)}{512 \pi ^4}+\frac{1}{512 \pi ^4}\\&
J(1,0,0,1,1,2,0)=\frac{1}{\epsilon} \left(t^3 \left(-\frac{90 r^4+180 r^3+77 r^2-13 r+2}{215040 \pi ^4 r^2 (r+1)^2}\right)\right)\\&+t \left(-\frac{2 r^2-12 (r+1)
   \log (t) r+2 r+5}{18432 \pi ^4 r (r+1)}\right)\\&+t^3 \left(-\frac{792 r^4+1584 r^3+686 r^2-106 r-3 \left(90 r^4+180 r^3+77 r^2-13 r+2\right) \log
   (t)+17}{645120 \pi ^4 r^2 (r+1)^2}\right)\\&+t^2 \left(-\frac{168 r^4+336 r^3+172 r^2-90 \left(8 r^3+16 r^2+7 r-1\right) \log (t) r+4 r-47}{1382400
   \pi ^4 r^2 (r+1)^2}\right)+\frac{1}{256 \pi ^4}\\&
J(1,0,0,2,1,1,0)=\frac{1}{\epsilon} \left(t^2 \frac{1}{15360 \pi ^4 r^2 (r+1)^2}+t \left(-\frac{1}{3072 \pi ^4 r (r+1)}\right)\right.\\&\left.+t^3 \left(-\frac{10 r^4+20 r^3+7
   r^2-3 r+1}{35840 \pi ^4 r^2 (r+1)^2}\right)+\frac{1}{512 \pi ^4}\right)+t \frac{3 r (r+1) \log (t)-2 \left(r^2+r+1\right)}{4608 \pi ^4 r
   (r+1)}\\&+t^3 \left(-\frac{176 r^4+352 r^3+126 r^2-50 r-6 \left(10 r^4+20 r^3+7 r^2-3 r+1\right) \log (t)+17}{215040 \pi ^4 r^2 (r+1)^2}\right)\\&+t^2
   \frac{-36 r^4-72 r^3-19 r^2+30 \left(3 r^3+6 r^2+2 r-1\right) \log (t) r+17 r+29}{230400 \pi ^4 r^2 (r+1)^2}\\&
J(1,0,1,1,0,1,0)=\left(-\frac{1}{512 \pi ^4}\right) \frac{1}{\epsilon^2}+\frac{1}{\epsilon} \left(\frac{\log (t)}{256 \pi ^4}-\frac{5}{512 \pi ^4}\right)\\&+t
   \frac{-26 r^2-26 r+6 \left(2 r^2+2 r-1\right) \log (t)+19}{18432 \pi ^4 r (r+1)}\\&+t^2 \left(-\frac{37 r^4+74 r^3+23 r^2-14 r-5 \left(6 r^4+12
   r^3+4 r^2-2 r+1\right) \log (t)+12}{230400 \pi ^4 r^2 (r+1)^2}\right)\\&-\frac{\log ^2(t)}{512 \pi ^4}+\frac{\log (t)}{128 \pi ^4}-\frac{1}{3072
   \pi ^2}-\frac{5}{512 \pi ^4}\\&
\end{align*}

\begin{align*}
  &J(2,0,1,1,0,1,0)=   \frac{1}{\epsilon^2} \left(\frac{1}{t} \left(-\frac{1}{512 \pi ^4}\right)\right)\\&+\frac{1}{\epsilon} \left(t \left(-\frac{1}{30720 \pi ^4 r^2
   (r+1)^2}\right)+t^2 \frac{10 r^4+20 r^3+7 r^2-3 r+1}{107520 \pi ^4 r^2 (r+1)^2}+\frac{1}{t} \frac{\log (t)}{512 \pi ^4}+\frac{1}{3072 \pi ^4 r
   (r+1)}\right)\\&+\frac{3110 r^4+6220 r^3+1967 r^2-1143 r+486}{22579200 \pi ^4 r^2 (r+1)^2} t^2\\&+t \left(-\frac{81 r^4+162 r^3+49 r^2-32 r-15 \left(6
   r^4+12 r^3+4 r^2-2 r+1\right) \log (t)+60}{460800 \pi ^4 r^2 (r+1)^2}\right)\\&-\frac{\log (t)}{3072 \pi ^4 r (r+1)}+\frac{1}{1536 \pi ^4 r
   (r+1)}+\frac{1}{t} \left(-\frac{3 \log ^2(t)+\pi ^2}{3072 \pi ^4}\right)\frac{\log (t)}{1536 \pi ^4}-\frac{5}{4608 \pi ^4}\\&
J(1,0,1,2,0,1,0)=t \frac{14 r^2+14 r-6 \left(2 r^2+2 r-1\right) \log (t)-13}{36864 \pi ^4 r (r+1)}\\&+t^2 \frac{44 r^4+88 r^3+26 r^2-18 r-10 \left(6 r^4+12 r^3+4 r^2-2
   r+1\right) \log (t)+19}{460800 \pi ^4 r^2 (r+1)^2}\\&-\frac{\log (t)}{512 \pi ^4}+\frac{3}{512 \pi ^4}\\&
J(1,1,1,1,0,0,0)=\\&\frac{1}{\epsilon} \left(\frac{1}{t} \frac{r (r+1) \left(-3 G(-1,-1,r)-3 G(-1,0,r)+3 G(0,-1,r)+3 G(0,0,r)+2 \pi ^2\right)}{768 \pi ^4 (2
   r+1)}\right)\\&+\frac{1}{t} \frac{1}{1536 \pi ^4 (2 r+1)}\left(r (r+1) \left(\pi ^2 G(-1,r)+8 \pi ^2 G\left(-\frac{1}{2},r\right)-\pi ^2 G(0,r)-6 G(-1,-1,r)\right.\right.\\&\left.\left.-6 G(-1,0,r)+6
   G(0,-1,r)+6 G(0,0,r)-6 G(-1,-1,-1,r)\right.\right.\\&\left.\left.-6 G(-1,-1,0,r)-6 G(-1,0,-1,r)-6 G(-1,0,0,r)\right.\right.\\&\left.\left.-12 G\left(-\frac{1}{2},-1,-1,r\right)-12
   G\left(-\frac{1}{2},-1,0,r\right)+12 G\left(-\frac{1}{2},0,-1,r\right)+12 G\left(-\frac{1}{2},0,0,r\right)\right.\right.\\&\left.\left.+6 G(0,-1,-1,r)+6 G(0,-1,0,r)+6
   G(0,0,-1,r)+6 G(0,0,0,r)+6 G(-1,-1,r) \log (t)\right.\right.\\&\left.\left.+6 G(-1,0,r) \log (t)-6 G(0,-1,r) \log (t)-6 G(0,0,r) \log (t)-4 \pi ^2 \log (t)+30 \zeta (3)+4
   \pi ^2\right)\right)\\&
J(0,1,1,1,1,1,0)=t^2 \frac{\log (r (r+1))-\log (t)+1}{46080 \pi ^4 r^2 (r+1)^2}+t \frac{-2 \log (r (r+1))+2 \log (t)-3}{12288 \pi ^4 r (r+1)}\\&+\frac{\log (r
   (r+1))}{512 \pi ^4}-\frac{\log (t)}{512 \pi ^4}+\frac{3}{512 \pi ^4}\\&
\end{align*}

\begin{align*}
&  J(1,1,1,1,1,0,0)=\\&\frac{1}{\epsilon} \left(\frac{1}{t} \frac{r (r+1) \left(-3 G(-1,-1,r)-3 G(-1,0,r)+3 G(0,-1,r)+3 G(0,0,r)+2 \pi ^2\right)}{768 \pi ^4 (2
     r+1)}\right)\\&+\frac{1}{t} \frac{1}{1536 \pi ^4 (2 r+1)}\left(r (r+1) \left(\pi ^2 G(-1,r)+8 \pi ^2 G\left(-\frac{1}{2},r\right)-\pi ^2 G(0,r)\right.\right.\\&\left.\left.-6 G(-1,-1,-1,r)-6 G(-1,-1,0,r)-6
     G(-1,0,-1,r)-6 G(-1,0,0,r)-12 G\left(-\frac{1}{2},-1,-1,r\right)\right.\right.\\&\left.\left.-12 G\left(-\frac{1}{2},-1,0,r\right)+12 G\left(-\frac{1}{2},0,-1,r\right)+12
     G\left(-\frac{1}{2},0,0,r\right)+6 G(0,-1,-1,r)\right.\right.\\&\left.\left.+6 G(0,-1,0,r)+6 G(0,0,-1,r)+6 G(0,0,0,r)+6 G(-1,-1,r) \log (t)\right.\right.\\&\left.\left.+6 G(-1,0,r) \log (t)-6 G(0,-1,r)
     \log (t)-6 G(0,0,r) \log (t)-4 \pi ^2 \log (t)+30 \zeta (3)\right)\right)\\&+t \frac{-3 G(-1,-1,r)-3 G(-1,0,r)+3 G(0,-1,r)+3
     G(0,0,r)+2 \pi ^2}{46080 \pi ^4 r (r+1) (2 r+1)}\\&+\frac{r G(-1,-1,r)}{1536 \pi ^4 (r+1) (2 r+1)}+\frac{r G(-1,0,r)}{1536 \pi ^4 (r+1) (2
     r+1)}-\frac{r G(0,-1,r)}{1536 \pi ^4 (r+1) (2 r+1)}\\&-\frac{r G(0,0,r)}{1536 \pi ^4 (r+1) (2 r+1)}+\frac{G(-1,-1,r)}{1536 \pi ^4 (r+1) (2
     r+1)}\\&+\frac{G(-1,0,r)}{1536 \pi ^4 (r+1) (2 r+1)}-\frac{G(0,-1,r)}{1536 \pi ^4 (r+1) (2 r+1)}-\frac{G(0,0,r)}{1536 \pi ^4 (r+1) (2
     r+1)}\\&-\frac{r}{2304 \pi ^2 (r+1) (2 r+1)}-\frac{1}{2304 \pi ^2 (r+1) (2 r+1)}\\
&J(1,1,1,1,1,1,0)=\frac{1}{\epsilon} \left(t \left(-\frac{r^2+r-1}{15360 \pi ^4 r (r+1)}\right)\right)\\&+\frac{1}{t} \left(-\frac{r (r+1) \left(-3 G(-1,-1,r)-3 G(-1,0,r)+3 G(0,-1,r)+3 G(0,0,r)+2 \pi ^2\right)}{1536 \pi ^4 (2
   r+1)}\right)\\&-\frac{G(-1,-1,r)}{6144 \pi ^4 (2 r+1)}-\frac{G(-1,0,r)}{6144 \pi ^4 (2 r+1)}+\frac{G(0,-1,r)}{6144 \pi ^4 (2
   r+1)}+\frac{G(0,0,r)}{6144 \pi ^4 (2 r+1)}\\&+\frac{-86
   r^2-86 r+91}{921600 \pi ^4 r (r+1)} t-\frac{r \log (t)}{1536 \pi ^4 (2 r+1)}-\frac{\log (t)}{3072 \pi ^4 (2 r+1)}+\frac{1}{9216 \pi ^2 (2
   r+1)}+\frac{1}{1152 \pi ^4}
\end{align*}
}

\bibliographystyle{JHEP}
\bibliography{content/biblio}

\providecommand{\href}[2]{#2}\begingroup\raggedright\begin{thebibliography}{100}

\bibitem{Aad:2012tfa}
{\bf ATLAS} Collaboration, G.~Aad et~al., {\it {Observation of a new particle
  in the search for the Standard Model Higgs boson with the ATLAS detector at
  the LHC}},  {\em Phys.Lett.} {\bf B716} (2012) 1--29,
  [\href{http://arxiv.org/abs/1207.7214}{{\tt arXiv:1207.7214}}].

\bibitem{Chatrchyan:2012ufa}
{\bf CMS} Collaboration, S.~Chatrchyan et~al., {\it {Observation of a new boson
  at a mass of 125 GeV with the CMS experiment at the LHC}},  {\em Phys.Lett.}
  {\bf B716} (2012) 30--61, [\href{http://arxiv.org/abs/1207.7235}{{\tt
  arXiv:1207.7235}}].

\bibitem{Englert:1964et}
F.~Englert and R.~Brout, {\it {Broken Symmetry and the Mass of Gauge Vector
  Mesons}},  {\em Phys.Rev.Lett.} {\bf 13} (1964) 321--323.

\bibitem{Higgs:1964pj}
P.~W. Higgs, {\it {Broken Symmetries and the Masses of Gauge Bosons}},  {\em
  Phys.Rev.Lett.} {\bf 13} (1964) 508--509.

\bibitem{Weinberg:1967tq}
S.~Weinberg, {\it {A Model of Leptons}},  {\em Phys.Rev.Lett.} {\bf 19} (1967)
  1264--1266.

\bibitem{Aad:2015gba}
{\bf ATLAS} Collaboration, G.~Aad et~al., {\it {Measurements of the Higgs boson
  production and decay rates and coupling strengths using pp collision data at
  $\sqrt{s}=7$ and 8 TeV in the ATLAS experiment}},  {\em Eur. Phys. J.} {\bf
  C76} (2016), no.~1 6, [\href{http://arxiv.org/abs/1507.04548}{{\tt
  arXiv:1507.04548}}].

\bibitem{Aad:2015zhl}
{\bf ATLAS, CMS} Collaboration, G.~Aad et~al., {\it {Combined Measurement of
  the Higgs Boson Mass in $pp$ Collisions at $\sqrt{s}=7$ and 8 TeV with the
  ATLAS and CMS Experiments}},  {\em Phys. Rev. Lett.} {\bf 114} (2015) 191803,
  [\href{http://arxiv.org/abs/1503.07589}{{\tt arXiv:1503.07589}}].

\bibitem{Khachatryan:2016vau}
{\bf ATLAS, CMS} Collaboration, G.~Aad et~al., {\it {Measurements of the Higgs
  boson production and decay rates and constraints on its couplings from a
  combined ATLAS and CMS analysis of the LHC pp collision data at $ \sqrt{s}=7
  $ and 8 TeV}},  {\em JHEP} {\bf 08} (2016) 045,
  [\href{http://arxiv.org/abs/1606.02266}{{\tt arXiv:1606.02266}}].

\bibitem{ATLAS:2018doi}
{\bf ATLAS} Collaboration, T.~A. collaboration, {\it {Combined measurements of
  Higgs boson production and decay using up to 80 fb$^{-1}$ of proton--proton
  collision data at $\sqrt{s}=$ 13 TeV collected with the ATLAS experiment}}, .

\bibitem{CMS:2018hhg}
{\bf CMS} Collaboration, C.~Collaboration, {\it {Combined measurement and
  interpretation of differential Higgs boson production cross sections at
  $\sqrt{s}$=13 TeV}}, .

\bibitem{Sirunyan:2018hoz}
{\bf CMS} Collaboration, A.~M. Sirunyan et~al., {\it {Observation of
  $\mathrm{t\overline{t}}$H production}},  {\em Phys. Rev. Lett.} {\bf 120}
  (2018), no.~23 231801, [\href{http://arxiv.org/abs/1804.02610}{{\tt
  arXiv:1804.02610}}].

\bibitem{Aaboud:2018urx}
{\bf ATLAS} Collaboration, M.~Aaboud et~al., {\it {Observation of Higgs boson
  production in association with a top quark pair at the LHC with the ATLAS
  detector}},  {\em Phys. Lett.} {\bf B784} (2018) 173--191,
  [\href{http://arxiv.org/abs/1806.00425}{{\tt arXiv:1806.00425}}].

\bibitem{Aad:2015vsa}
{\bf ATLAS} Collaboration, G.~Aad et~al., {\it {Evidence for the Higgs-boson
  Yukawa coupling to tau leptons with the ATLAS detector}},  {\em JHEP} {\bf
  04} (2015) 117, [\href{http://arxiv.org/abs/1501.04943}{{\tt
  arXiv:1501.04943}}].

\bibitem{Sirunyan:2017khh}
{\bf CMS} Collaboration, A.~M. Sirunyan et~al., {\it {Observation of the Higgs
  boson decay to a pair of $\tau$ leptons with the CMS detector}},  {\em Phys.
  Lett.} {\bf B779} (2018) 283--316,
  [\href{http://arxiv.org/abs/1708.00373}{{\tt arXiv:1708.00373}}].

\bibitem{ATLAS:2018nkp}
{\bf ATLAS} Collaboration, T.~A. collaboration, {\it {Observation of $H \to
  b\bar{b}$ decays and $VH$ production with the ATLAS detector}}, .

\bibitem{atlas2016}
{The ATLAS Collaboration} and {The CMS Collaboration}, {\it {Measurements of
  the Higgs boson production and decay rates and constraints on its couplings
  from a combined ATLAS and CMS analysis of the LHC pp collision data at
  {$\sqrt{s} = 7$} and {$8$} TeV}},  {\em Journal of High Energy Physics}
  (2016), no.~8 45, [\href{http://arxiv.org/abs/1507.04548}{{\tt
  arXiv:1507.04548}}].

\bibitem{TheCMSCollaboration2015}
{The CMS Collaboration}, {\it {Search for the standard model Higgs boson
  produced through vector boson fusion and decaying to bb}},  {\em Physical
  Review D} {\bf 92} (2015) 032008,
  [\href{http://arxiv.org/abs/1506.01010}{{\tt arXiv:1506.01010}}].

\bibitem{Aaboud2016}
{The ATLAS Collaboration}, {\it {Search for the Standard Model Higgs boson
  produced by vector-boson fusion and decaying to bottom quarks in
  {$\sqrt{s}=8$} TeV pp collisions with the ATLAS detector}},  {\em Journal of
  High Energy Physics} (2016), no.~11 112,
  [\href{http://arxiv.org/abs/1606.02181}{{\tt 1606.02181}}].

\bibitem{Chatrchyan2014}
{The CMS Collaboration}, {\it {Search for the standard model Higgs boson
  produced in association with a W or a Z boson and decaying to bottom
  quarks}},  {\em Physical Review D} {\bf 89} (2014) 012003,
  [\href{http://arxiv.org/abs/1310.3687}{{\tt arXiv:1310.3687}}].

\bibitem{Mantler2013}
H.~Mantler and M.~Wiesemann, {\it {Top- and bottom-mass effects in hadronic
  Higgs production at small transverse momenta through lo+nll}},  {\em European
  Physical Journal C} {\bf 73} (2013), no.~6 1--11,
  [\href{http://arxiv.org/abs/1210.8263}{{\tt arXiv:1210.8263}}].

\bibitem{Grazzini2013}
M.~Grazzini and H.~Sargsyan, {\it {Heavy-quark mass effects in Higgs boson
  production at the LHC}},  {\em Journal of High Energy Physics} {\bf 2013}
  (2013), no.~9 [\href{http://arxiv.org/abs/1306.4581}{{\tt arXiv:1306.4581}}].

\bibitem{Banfi2014}
A.~Banfi, P.~F. Monni, and G.~Zanderighi, {\it {Quark masses in Higgs
  production with a jet veto}},  {\em Journal of High Energy Physics} {\bf
  2014} (2014), no.~1 97, [\href{http://arxiv.org/abs/1308.4634}{{\tt
  arXiv:1308.4634}}].

\bibitem{Bagnaschi2012}
E.~Bagnaschi, G.~Degrassi, P.~Slavich, and A.~Vicini, {\it {Higgs production
  via gluon fusion in the POWHEG approach in the SM and in the MSSM}},  {\em
  Journal of High Energy Physics} {\bf 2012} (2012), no.~2 88,
  [\href{http://arxiv.org/abs/1111.2854}{{\tt arXiv:1111.2854}}].

\bibitem{Frederix:2016cnl}
R.~Frederix, S.~Frixione, E.~Vryonidou, and M.~Wiesemann, {\it {Heavy-quark
  mass effects in Higgs plus jets production}},  {\em JHEP} {\bf 08} (2016)
  006, [\href{http://arxiv.org/abs/1604.03017}{{\tt arXiv:1604.03017}}].

\bibitem{Bagnaschi:2015bop}
E.~Bagnaschi, R.~V. Harlander, H.~Mantler, A.~Vicini, and M.~Wiesemann, {\it
  {Resummation ambiguities in the Higgs transverse-momentum spectrum in the
  Standard Model and beyond}},  {\em JHEP} {\bf 01} (2016) 090,
  [\href{http://arxiv.org/abs/1510.08850}{{\tt arXiv:1510.08850}}].

\bibitem{Mantler:2015vba}
H.~Mantler and M.~Wiesemann, {\it {Hadronic Higgs production through NLO $+$ PS
  in the SM, the 2HDM and the MSSM}},  {\em Eur. Phys. J.} {\bf C75} (2015),
  no.~6 257, [\href{http://arxiv.org/abs/1504.06625}{{\tt arXiv:1504.06625}}].

\bibitem{Harlander:2014uea}
R.~V. Harlander, H.~Mantler, and M.~Wiesemann, {\it {Transverse momentum
  resummation for Higgs production via gluon fusion in the MSSM}},  {\em JHEP}
  {\bf 11} (2014) 116, [\href{http://arxiv.org/abs/1409.0531}{{\tt
  arXiv:1409.0531}}].

\bibitem{Raitio:1978pt}
R.~Raitio and W.~W. Wada, {\it {Higgs Boson Production at Large Transverse
  Momentum in {QCD}}},  {\em Phys. Rev.} {\bf D19} (1979) 941.

\bibitem{Rainwater2002}
D.~Rainwater, M.~Spira, and D.~Zeppenfeld, {\it {Higgs Boson Production at
  Hadron Colliders: Signal and Background Processes}},  in {\em Physics at TeV
  colliders. Proceedings, Euro Summer School, Les Houches.}, p.~15, mar, 2001.
\newblock \href{http://arxiv.org/abs/0203187}{{\tt 0203187}}.

\bibitem{Dittmaier2004}
S.~Dittmaier, M.~Kr{\"{a}}mer, and M.~Spira, {\it {Higgs radiation off bottom
  quarks at the Fermilab Tevatron and the CERN LHC}},  {\em Physical Review D}
  {\bf 70} (2004), no.~7 1--10, [\href{http://arxiv.org/abs/0309204}{{\tt
  0309204}}].

\bibitem{Wiesemann2015}
M.~Wiesemann, R.~Frederix, S.~Frixione, V.~Hirschi, F.~Maltoni, and
  P.~Torrielli, {\it {Higgs production in association with bottom quarks}},
  {\em Journal of High Energy Physics} {\bf 2015} (feb, 2015) 132,
  [\href{http://arxiv.org/abs/1409.5301}{{\tt arXiv:1409.5301}}].

\bibitem{Dittmaier:2003ej}
S.~Dittmaier, M.~Kr{\"a}mer, and M.~Spira, {\it {Higgs radiation off bottom
  quarks at the Tevatron and the CERN LHC}},  {\em Phys.Rev.} {\bf D70} (2004)
  074010, [\href{http://arxiv.org/abs/hep-ph/0309204}{{\tt hep-ph/0309204}}].

\bibitem{Dawson:2003kb}
S.~Dawson, C.~Jackson, L.~Reina, and D.~Wackeroth, {\it {Exclusive Higgs boson
  production with bottom quarks at hadron colliders}},  {\em Phys.Rev.} {\bf
  D69} (2004) 074027, [\href{http://arxiv.org/abs/hep-ph/0311067}{{\tt
  hep-ph/0311067}}].

\bibitem{Dawson:2005vi}
S.~Dawson, C.~Jackson, L.~Reina, and D.~Wackeroth, {\it {Higgs production in
  association with bottom quarks at hadron colliders}},  {\em Mod.Phys.Lett.}
  {\bf A21} (2006) 89--110, [\href{http://arxiv.org/abs/hep-ph/0508293}{{\tt
  hep-ph/0508293}}].

\bibitem{Liu:2012qu}
N.~Liu, L.~Wu, P.~W. Wu, and J.~M. Yang, {\it {Complete one-loop effects of
  SUSY QCD in $b\bar{b}h$ production at the LHC under current experimental
  constraints}},  {\em JHEP} {\bf 1301} (2013) 161,
  [\href{http://arxiv.org/abs/1208.3413}{{\tt arXiv:1208.3413}}].

\bibitem{Dittmaier:2014sva}
S.~Dittmaier, P.~H\"afliger, M.~Kr\"amer, M.~Spira, and M.~Walser, {\it
  {Neutral MSSM Higgs-boson production with heavy quarks: NLO supersymmetric
  QCD corrections}},  {\em Phys. Rev.} {\bf D90} (2014), no.~3 035010,
  [\href{http://arxiv.org/abs/1406.5307}{{\tt arXiv:1406.5307}}].

\bibitem{Zhang:2017mdz}
Y.~Zhang, {\it {NLO electroweak effects on the Higgs boson production in
  association with a bottom quark pair at the LHC}},  {\em Phys. Rev.} {\bf
  D96} (2017), no.~11 113009, [\href{http://arxiv.org/abs/1708.08790}{{\tt
  arXiv:1708.08790}}].

\bibitem{Dicus:1998hs}
D.~Dicus, T.~Stelzer, Z.~Sullivan, and S.~Willenbrock, {\it {Higgs boson
  production in association with bottom quarks at next-to-leading order}},
  {\em Phys.Rev.} {\bf D59} (1999) 094016,
  [\href{http://arxiv.org/abs/hep-ph/9811492}{{\tt hep-ph/9811492}}].

\bibitem{Balazs:1998sb}
C.~Balazs, H.-J. He, and C.~Yuan, {\it {QCD corrections to scalar production
  via heavy quark fusion at hadron colliders}},  {\em Phys.Rev.} {\bf D60}
  (1999) 114001, [\href{http://arxiv.org/abs/hep-ph/9812263}{{\tt
  hep-ph/9812263}}].

\bibitem{Harlander:2003ai}
R.~V. Harlander and W.~B. Kilgore, {\it {Higgs boson production in bottom quark
  fusion at next-to-next-to leading order}},  {\em Phys.Rev.} {\bf D68} (2003)
  013001, [\href{http://arxiv.org/abs/hep-ph/0304035}{{\tt hep-ph/0304035}}].

\bibitem{Campbell:2002zm}
J.~M. Campbell, R.~K. Ellis, F.~Maltoni, and S.~Willenbrock, {\it {Higgs-Boson
  production in association with a single bottom quark}},  {\em Phys.Rev.} {\bf
  D67} (2003) 095002, [\href{http://arxiv.org/abs/hep-ph/0204093}{{\tt
  hep-ph/0204093}}].

\bibitem{Harlander:2010cz}
R.~V. Harlander, K.~J. Ozeren, and M.~Wiesemann, {\it {Higgs plus jet
  production in bottom quark annihilation at next-to-leading order}},  {\em
  Phys. Lett.} {\bf B693} (2010) 269--273,
  [\href{http://arxiv.org/abs/1007.5411}{{\tt arXiv:1007.5411}}].

\bibitem{Harlander:2011fx}
R.~Harlander and M.~Wiesemann, {\it {Jet-veto in bottom-quark induced Higgs
  production at next-to-next-to-leading order}},  {\em JHEP} {\bf 1204} (2012)
  066, [\href{http://arxiv.org/abs/1111.2182}{{\tt arXiv:1111.2182}}].

\bibitem{Buehler:2012cu}
S.~Buhler, F.~Herzog, A.~Lazopoulos, and R.~Muller, {\it {The fully
  differential hadronic production of a Higgs boson via bottom quark fusion at
  NNLO}},  {\em JHEP} {\bf 1207} (2012) 115,
  [\href{http://arxiv.org/abs/1204.4415}{{\tt arXiv:1204.4415}}].

\bibitem{Ozeren:2010qp}
K.~J. Ozeren, {\it {Analytic Results for Higgs Production in Bottom Fusion}},
  {\em JHEP} {\bf 1011} (2010) 084, [\href{http://arxiv.org/abs/1010.2977}{{\tt
  arXiv:1010.2977}}].

\bibitem{Belyaev:2005bs}
A.~Belyaev, P.~M. Nadolsky, and C.-P. Yuan, {\it {Transverse momentum
  resummation for Higgs boson produced via b anti-b fusion at hadron
  colliders}},  {\em JHEP} {\bf 0604} (2006) 004,
  [\href{http://arxiv.org/abs/hep-ph/0509100}{{\tt hep-ph/0509100}}].

\bibitem{Harlander:2014hya}
R.~V. Harlander, A.~Tripathi, and M.~Wiesemann, {\it {Higgs production in
  bottom quark annihilation: Transverse momentum distribution at NNLO$+$NNLL}},
   {\em Phys. Rev.} {\bf D90} (2014), no.~1 015017,
  [\href{http://arxiv.org/abs/1403.7196}{{\tt arXiv:1403.7196}}].

\bibitem{Ahmed:2014pka}
T.~Ahmed, M.~Mahakhud, P.~Mathews, N.~Rana, and V.~Ravindran, {\it {Two-loop
  QCD corrections to Higgs $\to b+\overline{b}+g$ amplitude}},  {\em JHEP} {\bf
  1408} (2014) 075, [\href{http://arxiv.org/abs/1405.2324}{{\tt
  arXiv:1405.2324}}].

\bibitem{Gehrmann:2014vha}
T.~Gehrmann and D.~Kara, {\it {The $Hb\bar{b}$ form factor to three loops in
  QCD}},  {\em JHEP} {\bf 09} (2014) 174,
  [\href{http://arxiv.org/abs/1407.8114}{{\tt arXiv:1407.8114}}].

\bibitem{Jager:2015hka}
B.~Jager, L.~Reina, and D.~Wackeroth, {\it {Higgs boson production in
  association with b jets in the POWHEG BOX}},  {\em Phys. Rev.} {\bf D93}
  (2016), no.~1 014030, [\href{http://arxiv.org/abs/1509.05843}{{\tt
  arXiv:1509.05843}}].

\bibitem{Krauss:2016orf}
F.~Krauss, D.~Napoletano, and S.~Schumann, {\it {Simulating $b$-associated
  production of $Z$ and Higgs bosons with the SHERPA event generator}},  {\em
  Phys. Rev.} {\bf D95} (2017), no.~3 036012,
  [\href{http://arxiv.org/abs/1612.04640}{{\tt arXiv:1612.04640}}].

\bibitem{Maltoni:2012pa}
F.~Maltoni, G.~Ridolfi, and M.~Ubiali, {\it {b-initiated processes at the LHC:
  a reappraisal}},  {\em JHEP} {\bf 1207} (2012) 022,
  [\href{http://arxiv.org/abs/1203.6393}{{\tt arXiv:1203.6393}}].

\bibitem{Lim:2016wjo}
M.~Lim, F.~Maltoni, G.~Ridolfi, and M.~Ubiali, {\it {Anatomy of double
  heavy-quark initiated processes}},  {\em JHEP} {\bf 09} (2016) 132,
  [\href{http://arxiv.org/abs/1605.09411}{{\tt arXiv:1605.09411}}].

\bibitem{Forte:2015hba}
S.~Forte, D.~Napoletano, and M.~Ubiali, {\it {Higgs production in bottom-quark
  fusion in a matched scheme}},  {\em Phys. Lett.} {\bf B751} (2015) 331--337,
  [\href{http://arxiv.org/abs/1508.01529}{{\tt arXiv:1508.01529}}].

\bibitem{Forte:2016sja}
S.~Forte, D.~Napoletano, and M.~Ubiali, {\it {Higgs production in bottom-quark
  fusion: matching beyond leading order}},  {\em Phys. Lett.} {\bf B763} (2016)
  190--196, [\href{http://arxiv.org/abs/1607.00389}{{\tt arXiv:1607.00389}}].

\bibitem{Bonvini:2015pxa}
M.~Bonvini, A.~S. Papanastasiou, and F.~J. Tackmann, {\it {Resummation and
  matching of b-quark mass effects in $ b\overline{b}H $ production}},  {\em
  JHEP} {\bf 11} (2015) 196, [\href{http://arxiv.org/abs/1508.03288}{{\tt
  arXiv:1508.03288}}].

\bibitem{Bonvini:2016fgf}
M.~Bonvini, A.~S. Papanastasiou, and F.~J. Tackmann, {\it {Matched predictions
  for the $ b\overline{b}H $ cross section at the 13 TeV LHC}},  {\em JHEP}
  {\bf 10} (2016) 053, [\href{http://arxiv.org/abs/1605.01733}{{\tt
  arXiv:1605.01733}}].

\bibitem{Anastasiou:2002yz}
C.~Anastasiou and K.~Melnikov, {\it {Higgs boson production at hadron colliders
  in NNLO QCD}},  {\em Nucl. Phys.} {\bf B646} (2002) 220--256,
  [\href{http://arxiv.org/abs/hep-ph/0207004}{{\tt hep-ph/0207004}}].

\bibitem{Anastasiou:2015ema}
C.~Anastasiou, C.~Duhr, F.~Dulat, F.~Herzog, and B.~Mistlberger, {\it {Higgs
  Boson Gluon-Fusion Production in QCD at Three Loops}},  {\em Phys. Rev.
  Lett.} {\bf 114} (2015) 212001, [\href{http://arxiv.org/abs/1503.06056}{{\tt
  arXiv:1503.06056}}].

\bibitem{Gerlach:2018hen}
M.~Gerlach, F.~Herren, and M.~Steinhauser, {\it {Wilson coefficients for Higgs
  boson production and decoupling relations to $
  \mathcal{O}\left({\alpha}_s^4\right) $}},  {\em JHEP} {\bf 11} (2018) 141,
  [\href{http://arxiv.org/abs/1809.06787}{{\tt arXiv:1809.06787}}].

\bibitem{Mistlberger:2018etf}
B.~Mistlberger, {\it {Higgs boson production at hadron colliders at N$^{3}$LO
  in QCD}},  {\em JHEP} {\bf 05} (2018) 028,
  [\href{http://arxiv.org/abs/1802.00833}{{\tt arXiv:1802.00833}}].

\bibitem{Anastasiou:2016cez}
C.~Anastasiou, C.~Duhr, F.~Dulat, E.~Furlan, T.~Gehrmann, F.~Herzog,
  A.~Lazopoulos, and B.~Mistlberger, {\it {High precision determination of the
  gluon fusion Higgs boson cross-section at the LHC}},  {\em JHEP} {\bf 05}
  (2016) 058, [\href{http://arxiv.org/abs/1602.00695}{{\tt arXiv:1602.00695}}].

\bibitem{Harlander:2012hf}
R.~V. Harlander, T.~Neumann, K.~J. Ozeren, and M.~Wiesemann, {\it {Top-mass
  effects in differential Higgs production through gluon fusion at order
  $\alpha_s^4$}},  {\em JHEP} {\bf 08} (2012) 139,
  [\href{http://arxiv.org/abs/1206.0157}{{\tt arXiv:1206.0157}}].

\bibitem{Neumann:2014nha}
T.~Neumann and M.~Wiesemann, {\it {Finite top-mass effects in gluon-induced
  Higgs production with a jet-veto at NNLO}},  {\em JHEP} {\bf 11} (2014) 150,
  [\href{http://arxiv.org/abs/1408.6836}{{\tt arXiv:1408.6836}}].

\bibitem{Artoisenet:2013puc}
P.~Artoisenet et~al., {\it {A framework for Higgs characterisation}},  {\em
  JHEP} {\bf 11} (2013) 043, [\href{http://arxiv.org/abs/1306.6464}{{\tt
  arXiv:1306.6464}}].

\bibitem{Demartin:2014fia}
F.~Demartin, F.~Maltoni, K.~Mawatari, B.~Page, and M.~Zaro, {\it {Higgs
  characterisation at NLO in QCD: CP properties of the top-quark Yukawa
  interaction}},  {\em Eur. Phys. J.} {\bf C74} (2014), no.~9 3065,
  [\href{http://arxiv.org/abs/1407.5089}{{\tt arXiv:1407.5089}}].

\bibitem{Alwall:2014hca}
J.~Alwall, R.~Frederix, S.~Frixione, V.~Hirschi, F.~Maltoni, et~al., {\it {The
  automated computation of tree-level and next-to-leading order differential
  cross sections, and their matching to parton shower simulations}},  {\em
  JHEP} {\bf 1407} (2014) 079, [\href{http://arxiv.org/abs/1405.0301}{{\tt
  arXiv:1405.0301}}].

\bibitem{PhysRevLett.78.594}
K.~G. Chetyrkin, B.~A. Kniehl, and M.~Steinhauser, {\it Virtual top-quark
  effects on the
  $\mathit{H}\ensuremath{\rightarrow}\mathit{b}\overline{\mathit{b}}$ decay at
  next-to-leading order in qcd},  {\em Phys. Rev. Lett.} {\bf 78} (Jan, 1997)
  594--597.

\bibitem{CHETYRKIN199719}
K.~Chetyrkin, B.~Kniehl, and M.~Steinhauser, {\it Three-loop ${\cal
  o}(\alpha_s^2 g_f m_t^2)$ corrections to hadronic higgs decays},  {\em
  Nuclear Physics B} {\bf 490} (1997), no.~1 19 -- 39.

\bibitem{Chetyrkin:1997un}
K.~G. Chetyrkin, B.~A. Kniehl, and M.~Steinhauser, {\it {Decoupling relations
  to O (alpha-s**3) and their connection to low-energy theorems}},  {\em Nucl.
  Phys.} {\bf B510} (1998) 61--87,
  [\href{http://arxiv.org/abs/hep-ph/9708255}{{\tt hep-ph/9708255}}].

\bibitem{Frederix:2018nkq}
R.~Frederix, S.~Frixione, V.~Hirschi, D.~Pagani, H.~S. Shao, and M.~Zaro, {\it
  {The automation of next-to-leading order electroweak calculations}},  {\em
  JHEP} {\bf 07} (2018) 185, [\href{http://arxiv.org/abs/1804.10017}{{\tt
  arXiv:1804.10017}}].

\bibitem{Degrande:2016hyf}
C.~Degrande, R.~Frederix, V.~Hirschi, M.~Ubiali, M.~Wiesemann, and M.~Zaro,
  {\it {Accurate predictions for charged Higgs production: Closing the
  $m_{H^{\pm}}\sim m_t$ window}},  {\em Phys. Lett.} {\bf B772} (2017) 87--92,
  [\href{http://arxiv.org/abs/1607.05291}{{\tt arXiv:1607.05291}}].

\bibitem{Dittmaier:2006cz}
S.~Dittmaier, M.~Kr{\"a}mer, A.~Muck, and T.~Schluter, {\it {MSSM Higgs-boson
  production in bottom-quark fusion: Electroweak radiative corrections}},  {\em
  JHEP} {\bf 0703} (2007) 114, [\href{http://arxiv.org/abs/hep-ph/0611353}{{\tt
  hep-ph/0611353}}].

\bibitem{Dawson:2011pe}
S.~Dawson, C.~Jackson, and P.~Jaiswal, {\it {SUSY QCD Corrections to Higgs-b
  Production : Is the $\Delta_b$ Approximation Accurate?}},  {\em Phys.Rev.}
  {\bf D83} (2011) 115007, [\href{http://arxiv.org/abs/1104.1631}{{\tt
  arXiv:1104.1631}}].

\bibitem{Ball:2017nwa}
{\bf NNPDF} Collaboration, R.~D. Ball et~al., {\it {Parton distributions from
  high-precision collider data}},  {\em Eur. Phys. J.} {\bf C77} (2017), no.~10
  663, [\href{http://arxiv.org/abs/1706.00428}{{\tt arXiv:1706.00428}}].

\bibitem{Buckley:2014ana}
A.~Buckley, J.~Ferrando, S.~Lloyd, K.~Nordstrm, B.~Page, M.~Rfenacht,
  M.~Schnherr, and G.~Watt, {\it {LHAPDF6: parton density access in the LHC
  precision era}},  {\em Eur. Phys. J.} {\bf C75} (2015) 132,
  [\href{http://arxiv.org/abs/1412.7420}{{\tt arXiv:1412.7420}}].

\bibitem{Frederix:2011ss}
R.~Frederix, S.~Frixione, V.~Hirschi, F.~Maltoni, R.~Pittau, et~al., {\it
  {Four-lepton production at hadron colliders: aMC@NLO predictions with
  theoretical uncertainties}},  {\em JHEP} {\bf 1202} (2012) 099,
  [\href{http://arxiv.org/abs/1110.4738}{{\tt arXiv:1110.4738}}].

\bibitem{Marquard:2015qpa}
P.~Marquard, A.~V. Smirnov, V.~A. Smirnov, and M.~Steinhauser, {\it {Quark Mass
  Relations to Four-Loop Order in Perturbative QCD}},  {\em Phys. Rev. Lett.}
  {\bf 114} (2015), no.~14 142002, [\href{http://arxiv.org/abs/1502.01030}{{\tt
  arXiv:1502.01030}}].

\bibitem{Kataev:2015gvt}
A.~L. Kataev and V.~S. Molokoedov, {\it {On the flavour dependence of the
  $\mathcal{O}(\alpha_s^4)$ correction to the relation between running and pole
  heavy quark masses}},  {\em Eur. Phys. J. Plus} {\bf 131} (2016), no.~8 271,
  [\href{http://arxiv.org/abs/1511.06898}{{\tt arXiv:1511.06898}}].

\bibitem{deFlorian:2016spz}
{\bf LHC Higgs Cross Section Working Group} Collaboration, D.~de~Florian
  et~al., {\it {Handbook of LHC Higgs Cross Sections: 4. Deciphering the Nature
  of the Higgs Sector}},  \href{http://arxiv.org/abs/1610.07922}{{\tt
  arXiv:1610.07922}}.

\bibitem{Cacciari:2008gp}
M.~Cacciari, G.~P. Salam, and G.~Soyez, {\it {The anti-$k_t$ jet clustering
  algorithm}},  {\em JHEP} {\bf 0804} (2008) 063,
  [\href{http://arxiv.org/abs/0802.1189}{{\tt arXiv:0802.1189}}].

\bibitem{Cacciari:2011ma}
M.~Cacciari, G.~P. Salam, and G.~Soyez, {\it {FastJet User Manual}},  {\em
  Eur.Phys.J.} {\bf C72} (2012) 1896,
  [\href{http://arxiv.org/abs/1111.6097}{{\tt arXiv:1111.6097}}].

\bibitem{Bishara:2016jga}
F.~Bishara, U.~Haisch, P.~F. Monni, and E.~Re, {\it {Constraining Light-Quark
  Yukawa Couplings from Higgs Distributions}},  {\em Phys. Rev. Lett.} {\bf
  118} (2017), no.~12 121801, [\href{http://arxiv.org/abs/1606.09253}{{\tt
  arXiv:1606.09253}}].

\bibitem{Cascioli:2013era}
{Cascioli, Fabio and Maierh\"ofer, Philipp and Moretti, Niccolo and Pozzorini,
  Stefano and Siegert, Frank}, {\it {NLO matching for $t\bar t b \bar b$
  production with massive $b$-quarks}},  {\em Phys. Lett.} {\bf B734} (2014)
  210--214, [\href{http://arxiv.org/abs/1309.5912}{{\tt arXiv:1309.5912}}].

\bibitem{Jezo:2018yaf}
T.~Je\v{z}o, J.~M. Lindert, N.~Moretti, and S.~Pozzorini, {\it {New NLOPS
  predictions for $\boldsymbol{t \bar{t} +b}$ -jet production at the LHC}},
  {\em Eur. Phys. J.} {\bf C78} (2018), no.~6 502,
  [\href{http://arxiv.org/abs/1802.00426}{{\tt arXiv:1802.00426}}].

\bibitem{Bagnaschi:2018dnh}
E.~Bagnaschi, F.~Maltoni, A.~Vicini, and M.~Zaro, {\it {Lepton-pair production
  in association with a $ b\overline{b} $ pair and the determination of the $W$
  boson mass}},  {\em JHEP} {\bf 07} (2018) 101,
  [\href{http://arxiv.org/abs/1803.04336}{{\tt arXiv:1803.04336}}].

\bibitem{LHCHiggsCrossSectionWorkingGroup:2012nn}
{\bf LHC Higgs Cross Section Working Group} Collaboration, A.~David, A.~Denner,
  M.~Duehrssen, M.~Grazzini, C.~Grojean, G.~Passarino, M.~Schumacher, M.~Spira,
  G.~Weiglein, and M.~Zanetti, {\it {LHC HXSWG interim recommendations to
  explore the coupling structure of a Higgs-like particle}},
  \href{http://arxiv.org/abs/1209.0040}{{\tt arXiv:1209.0040}}.

\bibitem{Buonocore:2017lry}
L.~Buonocore, P.~Nason, and F.~Tramontano, {\it {Heavy quark radiation in
  NLO+PS POWHEG generators}},  {\em Eur. Phys. J.} {\bf C78} (2018), no.~2 151,
  [\href{http://arxiv.org/abs/1711.06281}{{\tt arXiv:1711.06281}}].

\bibitem{Inami1983}
T.~Inami, T.~Kubota, and Y.~Okada, {\it Effective gauge theory and the effect
  of heavy quarks in higgs boson decays},  {\em Zeitschrift f{\"u}r Physik C
  Particles and Fields} {\bf 18} (Mar, 1983) 69--80.

\bibitem{Dawson:1990zj}
S.~Dawson, {\it {Radiative corrections to Higgs boson production}},  {\em Nucl.
  Phys.} {\bf B359} (1991) 283--300.

\bibitem{Hirschi:2011pa}
V.~Hirschi, R.~Frederix, S.~Frixione, M.~V. Garzelli, F.~Maltoni, et~al., {\it
  {Automation of one-loop QCD corrections}},  {\em JHEP} {\bf 1105} (2011) 044,
  [\href{http://arxiv.org/abs/1103.0621}{{\tt arXiv:1103.0621}}].

\bibitem{Goncharov1998}
A.~B. Goncharov, {\it {Multiple polylogarithms, cyclotomy and modular
  complexes}},  {\em Math Res. Letters} {\bf 5} (1998), no.~3 497--516,
  [\href{http://arxiv.org/abs/1105.2076}{{\tt arXiv:1105.2076}}].

\bibitem{Goncharov2001}
A.~B. Goncharov, {\it {Multiple polylogarithms and mixed Tate motives}},
  \href{http://arxiv.org/abs/0103059}{{\tt 0103059}}.

\bibitem{Kuipers2013}
J.~Kuipers, T.~Ueda, J.~a.~M. Vermaseren, and J.~Vollinga, {\it {FORM version
  4.0}},  {\em Computer Physics Communications} {\bf 184} (2013), no.~5
  1453--1467, [\href{http://arxiv.org/abs/1203.6543}{{\tt arXiv:1203.6543}}].

\bibitem{Nogueira1993}
P.~Nogueira, {\it {Automatic Feynman Graph Generation}},  {\em Journal of
  Computational Physics} {\bf 105} (1993), no.~2 279--289.

\bibitem{Lee2014a}
R.~N. Lee, {\it {LiteRed 1.4: a powerful tool for reduction of multiloop
  integrals}},  {\em Journal of Physics: Conference Series} {\bf 523} (jun,
  2014) 012059, [\href{http://arxiv.org/abs/1310.1145}{{\tt arXiv:1310.1145}}].

\bibitem{Jantzen2012}
B.~Jantzen, A.~V. Smirnov, and V.~a. Smirnov, {\it {Expansion by regions:
  Revealing potential and Glauber regions automatically}},  {\em European
  Physical Journal C} {\bf 72} (2012), no.~9 1--14,
  [\href{http://arxiv.org/abs/1206.0546}{{\tt arXiv:1206.0546}}].

\bibitem{Smirnov:2002pj}
V.~A. Smirnov, {\it {Applied asymptotic expansions in momenta and masses}},
  {\em Springer Tracts Mod. Phys.} {\bf 177} (2002) 1--262.

\bibitem{Anastasiou2013}
C.~Anastasiou, C.~Duhr, F.~Dulat, and B.~Mistlberger, {\it {Soft triple-real
  radiation for Higgs production at N3LO}},  {\em Journal of High Energy
  Physics} {\bf 2013} (2013), no.~7 [\href{http://arxiv.org/abs/1302.4379}{{\tt
  arXiv:1302.4379}}].

\bibitem{PolyLogTools}
C.~Duhr and F.~Dulat, ``Polylogtools.'' private code, 2014.

\bibitem{Bauer:2000cp}
C.~W. Bauer, A.~Frink, and R.~Kreckel, {\it {Introduction to the GiNaC
  framework for symbolic computation within the C++ programming language}},
  {\em J. Symb. Comput.} {\bf 33} (2000) 1,
  [\href{http://arxiv.org/abs/cs/0004015}{{\tt cs/0004015}}].

\bibitem{Smirnov2015}
A.~V. Smirnov, {\it {FIESTA 4: optimized Feynman integral calculations with GPU
  support}},  {\em Computer Physics Communications} {\bf 204} (2016) 189--199,
  [\href{http://arxiv.org/abs/1511.03614}{{\tt arXiv:1511.03614}}].

\end{thebibliography}\endgroup

\end{document}